\documentclass[11pt,a4paper,twoside]{scrartcl}
\DeclareOldFontCommand{\it}{\normalfont\itshape}{\mathit}
\usepackage[square,comma,sort&compress,numbers]{natbib}
\usepackage[version=4]{mhchem}
\usepackage{amsmath,amssymb,amstext,amsfonts}
\usepackage{parskip}
\usepackage[utf8]{inputenc}
\usepackage[T1]{fontenc}
\usepackage{titleref}
\usepackage{setspace}
\usepackage{pdfpages}
\usepackage[colorlinks=false,linkcolor=black]{hyperref}
\usepackage{epigraph}
\usepackage{graphicx}
\usepackage{floatflt}
\usepackage{sectsty}
\usepackage{siunitx}
\usepackage[top=1.5cm,bottom=1.0cm,includeheadfoot]{geometry}
\usepackage[margin=11pt,labelfont=bf,hangindent=-.0cm]{caption} 
\usepackage{sectsty}
\sectionfont{\Huge}
\subsectionfont{\Large}
\subsubsectionfont{\large}
\paragraphfont{\normalsize\textbf}
\subparagraphfont{\normalsize\normalfont\textbf}
\usepackage{tabularx}
\usepackage{wrapfig}
\usepackage{rotating}
\usepackage{array}

\usepackage{longtable}
\author{Hans Kleemann}
\usepackage{color}
\usepackage[english]{babel}
\RedeclareSectionCommands[afterskip=1sp]{paragraph,subparagraph}

\usepackage{fancyhdr}
\usepackage{afterpage}

\renewcommand{\headrulewidth}{1pt}

\usepackage{textcomp}
\begin{document}
\begin{titlepage}
\doublespacing
\begin{center}
\vspace{1cm}
Dresden Integrated Center for Applied Physics and Photonic Materials (IAPP) \\
Fakult\"at Physik \\
Bereich Mathematik und Naturwissenschaften \\ 
Technische Universit\"at Dresden \\
\vspace{2cm}
\begin{Large}
\textbf{Novel Concepts for Organic Transistors:\\ Physics, Device Design, and Applications} \\
\end{Large}
\vspace{2cm}
\begin{large}
	Habilitationsschrift \\[1cm]
	vorgelegt von Herrn \\
\end{large}
\begin{large}
\textbf{Dr. rer. nat. Hans Kleemann}\\
\end{large}
geboren am \\
10.10.1983 in S\"ommerda \\
\vspace{2cm}
\begin{figure}[hb]
\centering
\includegraphics[scale=0.65]{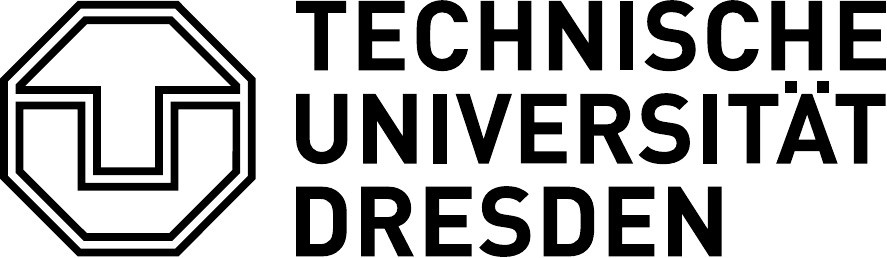}
\end{figure}
\vspace{1cm}
\date{\today}
Die Habilitationsschrift wurde in der Zeit von 10/2018 bis
11/2020 im Institut für Angewandte Physik angefertigt.

Dresden, 2020

\newpage

\begin{flushleft}
 \doublespacing
\vspace*{10cm}
Dr. rer. nat. Hans Kleemann: \textit{Novel Concepts for Organic Transistors: Physics, Device Design, and Applications}, Habilitation,\\
E-mail: \textit{hans.kleemann1@tu-dresden.de}\\
\vspace{1cm}
Die Habilitationsschrift habe ich selbst und ohne andere als die darin angegebenen Hilfsmittel angefertigt. Die w\"{o}rtlich oder inhaltlich \"{u}bernommenen Stellen wurden als solche gekennzeichnet. Bei gemeinschaftlichen Arbeiten erfolgte die Angabe, worauf sich meine Mitarbeit erstreckte. Es wurden keine fr\"{u}heren Habilitationsgesuche an anderen Hochschulen gestellt.\\
\vspace{1cm}
Dresden den: 12.11.2020\\
\vspace{1.5cm}
Hans Kleemann
\vspace{1.5cm}
\thispagestyle{empty}
\end{flushleft}

\newpage
\end{center}
\end{titlepage}

\thispagestyle{empty}
\newpage
\vspace*{3cm}
\begin{center}
\begin{huge}\textbf{Abstract}\end{huge}
\vspace{1.5cm}
\end{center}
\thispagestyle{empty}
In the first wave of commercialization of organic electronics, about ten years ago, active-matrix organic light-emitting diode (AMOLED) displays became the first large scale industrial application of organic electronic devices. The victory continues and AMOLED displays attain an ever-increasing market share in the global display industry. In the second wave, organic solar cells are about to enter the mass-production stage, and the possibility for low-cost production on flexible substrates will revolutionize the solar industry. The third wave will be the implementation of organic thin-film transistors for truly flexible, printed, large-area circuits. However, there is a multitude of challenges with regard to device physics, material, and process engineering which need to be overcome to make organic thin-film transistors fit for the step into industrial fabrication.\\
The focus of this thesis is at organic thin-film transistors, covering the whole spectrum of device physics, design principles, and the exploration of new applications. In particular, charge carrier transport and injection in  vertical organic transistors with ultra-short channel length are investigated in order to derive device architectures suitable for high and ultra-high frequency operation. Self-heating and a strongly thermally activated charge carrier transport at high current densities are identified as the limiting factors for high-frequency operation on low thermal conductivity, flexible substrates. Besides fundamental questions on charge carrier transport, this thesis also addresses questions related to the device fabrication. In particular, new fabrication methods for vertical organic transistors are proposed enabling reliable and stable device operation and integration of ultra-short channel length devices without using costly high-resolution patterning techniques.\\
Beyond conventional organic thin-film transistors, this thesis explores possible paths for the fourth wave of organic electronics. In this context, mixed ionic-electronic conductors and organic electro-chemical transistors (OECTs) are identified as highly promising approaches for electronic bio-interfaces enabling ultra-sensitive detection of biological signals. Furthermore, these systems show fundamental properties of biological synapses, namely the synaptic plasticity, which renders the possibility to build brain-inspired, neuromorphic networks enabling highly efficient computing. In particular, the combination of OECTs acting as sensor units and self-learning neural networks at once enables the development of intelligent tags for medical applications.\\
Overall, this thesis adds substantially new insight into the field of organic electronics and draws a vision towards further research and applications. The advancements in the field of vertical organic transistors open new perspectives for the implementation of organic transistors in high-resolution AMOLED displays or radio-frequency identification tags. Furthermore, the exploration of OECTs for neuromorphic computing will create a whole new research field across the disciplines of physics, material, and computer science.

\cleardoublepage
    \setcounter{secnumdepth}{5}
    \setcounter{tocdepth}{4}
\tableofcontents
\cleardoublepage

\pagestyle{fancy}
\fancyhead{}
\fancyhead[R]{Introduction} 
\fancyhead[L]{\scshape \thepage}
\fancyfoot{}
\renewcommand{\headrulewidth}{1pt}
\setcounter{section}{1}
\section*{Introduction}
\addcontentsline{toc}{section}{1 \hspace{0.1cm} Introduction}
\thispagestyle{empty}
\vspace{2cm}
\setlength{\epigraphwidth}{.5\textwidth} 
\setlength{\epigraphrule}{1pt} 
\epigraph{\large{\textit{'If future generations are to remember us more with gratitude than sorrow, we must achieve more than just the miracles of technology. We must also leave them a glimpse of the world as it was created, not just as it looked when we got through with it.'}}}{\footnotesize{Lyndon B. Johnson, 36$^\mathrm{th}$ president of the United States}}
\vspace{2cm}
For several decades, the field of organic electronics is an inspiring area of research fruitfully connecting innovations across the disciplines of synthetic chemistry, material science, physics, and electrical engineering. In the last 15 years, the technology related to organic electronic devices substantially matured helping to implement the vision of low-cost, large-area, flexible electronics. In particular, organic light-emitting diodes (OLEDs) and solar cells successfully entered the mass-production market due to their economic attractiveness caused by continuously improved device efficiency and reduction of fabrication costs compared to competing technologies. However, for more complex flexible devices based on organic electronics, powerful organic thin-film transistors (OTFTs) are indispensable. In this context, it remains one of the biggest challenges to develop organic thin-film transistors that can be fabricated using low-cost processes and at the same time fulfill the demanding performance targets required for flexible electronic circuits in e.g., active-matrix OLED (AMOLED) displays, liquid crystal displays (LCD), or smart labels including radio-frequency identification tags (RFID).\\
Besides its attractiveness for low-cost, large-area electronics, there is another striking advantage of organic electronic technology over conventional silicon-based approaches. Since human society is facing the limits of our planet all at once, it is absolutely essential to realign large parts of the global economy on technologies that possess a low-carbon footprint, low toxicity, and true sustainability. In this context, organic electronics might revolutionize the entire field of micro-electronics due to low-temperature processing and possible biodegradability of many organic semiconductor materials and substrates \cite{Vladu2012, Wang2016a, Li2018a}. Additionally, organic electronic devices render the possibility to emulate the behavior of biological synapses and neurons using organic electro-chemical transistors (OECTs). As recently demonstrated, such devices enable highly efficient brain-inspired, hardware-based computing which can even outperform leading-edge inorganic approaches \cite{Burgt2017, Fuller2019}. Astoundingly, brain-inspired computing based on OECTs might even help to resolve the problem of the ever-increasing power demand of silicon-based electronics.\\
The focus of this thesis lies on organic thin-film transistors - their device physics, design principles, and the exploration of new applications. Two device structures are described in this thesis. Firstly, vertical organic transistors are studied due to the enormous capabilities in the field of high-frequency operation. Secondly, the potential of organic electro-chemical transistors for neuromorphic networks as well as chemical sensors is explored.  

\subsection{Organic Field-Effect Transistors}
The vision of large-area, flexible electronics is to fabricate electronic devices on virtually any kind of substrate at low process temperature and low production costs. It is a revolutionary type of electronics since it breaks with the paradigm of wafer-to-wafer production by adapting highly cost-effective methods such as roll-to-roll fabrication. Furthermore, the fabrication on flexible substrates enables exciting new applications such as wearable electronics \cite{Zhou2006, Harding2017}, or artificial skin \cite{Bryant2017}.
Several material systems have been proposed as active semiconductors in this new kind of electronics, including organic semiconductors (OSC), oxide semiconductors, carbon nanotubes, and printed silicon \cite{Myny2018}. Although outstanding performance has been demonstrated for the two latter material systems \cite{Tang2018, Shimoda2006}, organic and oxide semiconductors have attracted more interest because they allow for low process temperatures, which is an essential prerequisite for processing on flexible polymer substrates. Oxide semiconductors such as amorphous indium-gallium-zinc-oxide (a-IGZO) nowadays have been adopted as a leading transistor technology for large-scale and high-resolution displays \cite{Ha2015}. Although encouraging developments regarding low-temperature processing have been reported \cite{Kim2012}, there is still a trade-off regarding the transistor performance, i.e., operational stability. In particular, sufficiently stable operation can only be achieved for a processing temperature of $\ge$250$^\circ$C, which is too high for low-cost, flexible substrates (require temperature $\le$\,150$^\circ$C). Organic semiconductors offer the possibility to be processed at temperature $\le$100$^\circ$C on virtually any kind of substrate. Moreover, with values for charge carrier mobility approaching 10\,cm$^2$/(Vs), OSCs outperform amorphous silicon in terms of mobility, opening various possibilities for electronic applications of organic thin-film transistors. However, state-of-the-art OTFTs suffer from various parasitic effects, which strongly limit their performance in circuits. In particular, contact resistance and electrode overlap capacitance severely restrict the dynamic response of OTFTs prevents them from being used in flexible RFID-tags or AMOLED displays up to now.\\ 
In this thesis, the vertical organic transistor concept is introduced. It enables the fabrication of ultra-short channel length devices without employing costly, high-resolution patterning techniques. These vertical transistor structures possess lower overlap capacitances than comparable lateral transistors, and the ultra-short channel length enables current densities up to 1\,kA/cm$^2$. Overall, vertical organic transistors combine the ease of processing of lateral OTFTs and the advantages of short-channel structures, thereby offering operation at frequencies above 100\,MHz.
\subsection{Organic Electro-Chemical Transistors}
In contrast to organic thin-film transistors described above, where the frequency of operation is an important figure in terms of applications, organic electro-chemical transistors are not conceived for traditional circuitry in the first place. OECTs operate in an electrolyte and employ ions to dope or dedope the channel of the transistor. Being restricted by the ion movement, OECTs are intrinsically slow. However, they act as transducers for ionic and electronic signals, which renders the possibility to use OECTs as highly efficient chemical sensors. Furthermore, the large double-layer capacitance of OECTs enables them to operate in the milli-volt or even micro-volt regime, which is ideal for interfacing OECTs with biological systems such as neurons \cite{Keene2020}. Even beyond the bio-interface, OECTs can mimic the behavior of biological neurons and synapses, offering the possibility to employ OECTs in neuromorphic networks for power-efficient computing. The field of neuromorphic computing using OECTs just recently emerged and there is a multitude of open questions related to the general behavior of OECT synapses, the integration into complex networks, and the connection to traditional hardware components.
In this thesis, a new method is proposed for the tunable growth of OECT synapses with adjustable synaptic properties. This method allows for seamless integration of OECT synapses into complex network structures and hence, sets a milestone towards the realization of brain-inspired, neuromorphic networks.
\subsection{Outline of this Thesis}
This thesis is divided into two parts. The first part focuses on vertical organic thin-film transistors as an alternative device concept enabling fully integrated devices for high-frequency operation. The chapter begins with a summary about physics, material, and processing related aspects to organic thin-film transistors. Additionally, the shortcomings of traditional lateral TFT geometries are discussed. In order to overcome these limitations, two concepts for vertical organic transistors are introduced - the organic permeable base transistor and the vertical organic field-effect transistor. Both device concepts are analyzed regarding the physics of device operation and the potential for high-frequency operation. Based on this thorough understanding, optimizations of device, layout, and process properties are proposed in order to push the performance of these devices to their limits. In particular, device heating is identified as a major obstacle restricting the high-frequency operation on flexible substrates.\\ 
Beyond these optimizations, new applications for vertical transistor structures are explored by adding new functionalities to the devices. In particular, vertical organic light-emitting transistors show great promise for power-efficient displays, and double-gated vertical organic transistors enable compact integration and high-frequency switching of logic gates.\\
The second part of this thesis is devoted to organic electro-chemical transistors and their use as ion-sensors and synapses for neuromorphic networks. The chapter begins with a summary of the device physics of OECTs and a brief review of their applications. In the second part of this chapter, a new fabrication method denoted as field-directed electropolymerization (FDP) is introduced, which allows for the on-demand growth of OECTs with controllable electrical properties. Furthermore, this method is employed to build complex networks of electro-chemical transistors serving as an essential building-block for neuromorphic networks. In particular, it is be shown that OECTs possess inherent properties of biological synapses (synaptic plasticity) and that the synaptic properties can be adjusted on demand by the FDP growth conditions. Hence, this new technique allows for the fabrication of complex synaptic networks with adjustable short- and long-term plasticity which is exploited to demonstrate Pavlovian conditioning and an example of pattern recognition. Furthermore, OECT-based synaptic networks show a highly non-linear electrical response, which is a prerequisite for the use in so-called echo-state or reservoir networks.\\
Finally, the potential of OECTs to be used as highly efficient sensors is discussed. In particular, its potential is be shown using only electrodes and a single OECT to achieve selective detection of different kinds of ions with exceptional resolution.
\newpage

\cleardoublepage

\pagestyle{fancy}
\fancyhead{}
\fancyhead[LE]{\scshape \thepage} 
\fancyhead[RE]{\scshape \rightmark}
\fancyhead[LO]{\scshape \rightmark} 
\fancyhead[RO]{\scshape \thepage}
\fancyfoot{}
\renewcommand{\headrulewidth}{1pt}
\setcounter{section}{2}
\setcounter{subsection}{0}
\section*{Advanced Architectures for Organic Transistors}
\addcontentsline{toc}{section}{2 \hspace{0.1cm} Novel Concepts for Organic Transistors}
\thispagestyle{empty}
\vspace{2cm}
\setlength{\epigraphwidth}{.5\textwidth} 
\setlength{\epigraphrule}{1pt} 
\epigraph{\large{\textit{'Es gibt zum Optimismus keine vernünftige Alternative.'}}}{\footnotesize{Karl Popper, philosopher}}
\vspace{2cm}
Thin-film transistors based on organic semiconductors are an essential building block for the kind of flexible electronics envisioned for the future. Still, a manifold of practical problems is to be solved that are related to the fabrication on flexible substrates (e.g., increased thermal expansion, reduced heat conductivity, limited chemical resistivity, etc.). On the other hand, the desire to have ever faster, more power-efficient, and versatile electronic devices requires advancing the performance of OTFTs. In particular, OTFTs need to improve in transconductance, reduced parasitic device capacitance, operation voltage, balancing of hole and electron mobility for complementary circuits, increased operation stability, and device uniformity.\\
This chapter is devoted to new architectures for organic transistors aiming for higher device transconductance, reduced parasitic capacitance, and high-frequency operation. To begin with, a summary of the operation of thin-film transistors is provided, followed by a review on the performance of state-of-the-art OTFTs, considering materials as well as processing related aspects. Moreover, an overview of possible application scenarios for OTFTs is given. Later, recent approaches towards high-performance organic transistors are highlighted, focusing on vertical organic transistors and doped organic transistors. In the fourth part of this chapter, the achievements of my research in this field are presented, which span from high-frequency vertical organic transistors, over organic permeable base transistors, to self-aligned OTFTs.\\
The overall goal of my research in this field is to increase the operation frequency of organic transistors as well as to demonstrate integrated and scalable high-performance transistors. As shown, reducing the channel length and the parasitic overlap of electrodes is the more promising strategy to reach this goal rather than focusing solely on the charge carrier mobility. For this reason, the main motive of this chapter is to find suitable approaches for the fabrication of ultra-short channel length organic transistors without using advanced and costly high-resolution structuring methods.
\subsection{Operation of Organic Thin-Film Transistors}
\label{subsec:OTFT_operation}
\begin{figure}[htb]
	\begin{center}
		\includegraphics[width=.99\textwidth,clip]{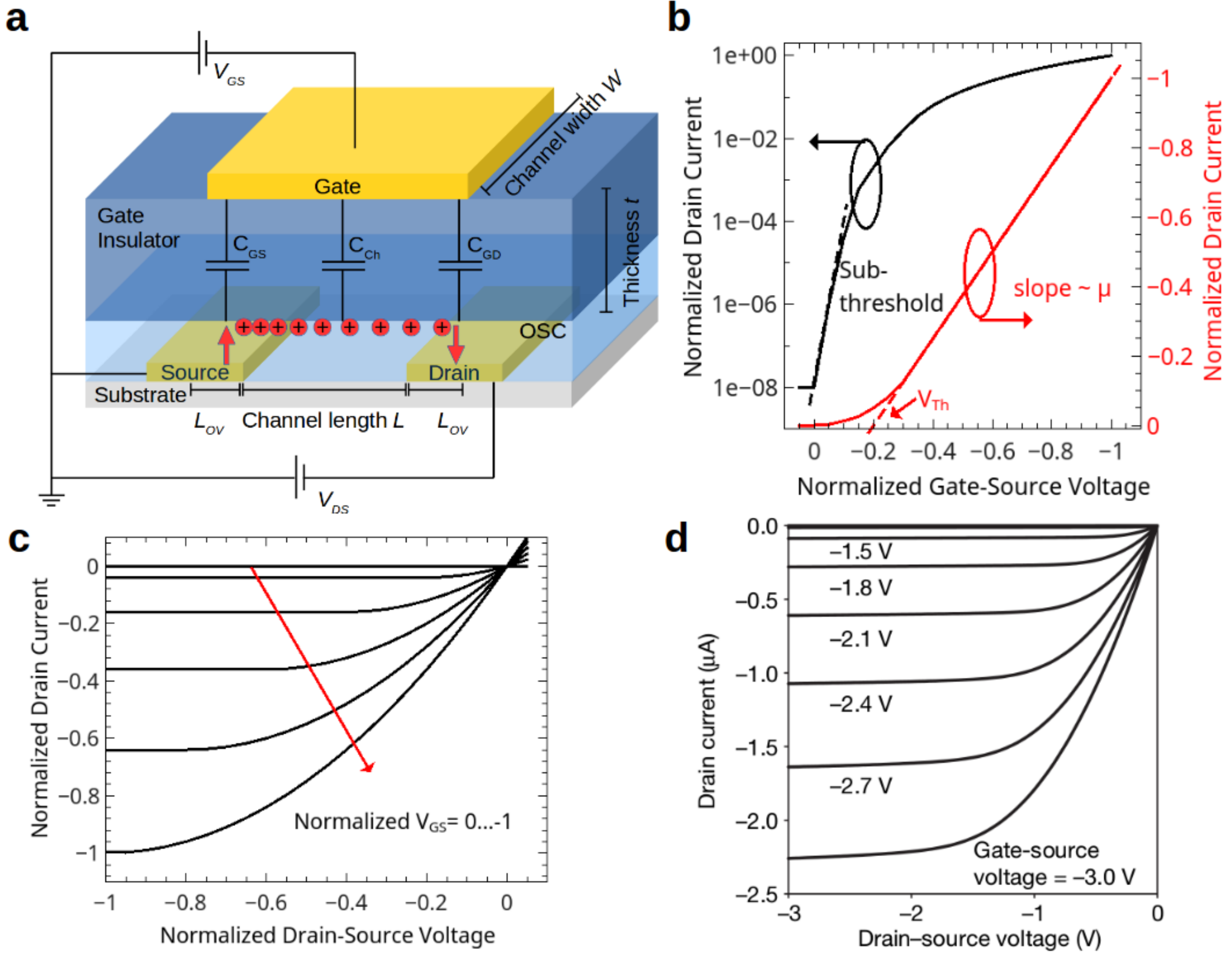}
	\end{center}
	\caption[]{\textbf{a} Schematic of a staggered OTFT in top-gate configuration. 
		The accumulation of holes at the semiconductor-insulator interface is shown. \textbf{b} Normalized transfer and \textbf{c} output characteristics of an ideal TFT according to the gradual channel approximation. \textbf{d} Experimental output curve of a pentacene-based OTFT showing an almost ideal transistor behavior \cite{Klauk2007}. Figure \textbf{d} reprinted from \cite{Klauk2007} with permission by Springer Nature.}
	\label{fig:OTFT_Intro}
\end{figure}

The essential building block of every field-effect and thin-film transistor is the metal-insulator-semiconductor (MIS) capacitor (cf. scheme of a staggered TFT in Figure \ref{fig:OTFT_Intro}a). 
By applying voltage $V_\mathrm{GS}$ between the metal electrode (gate) and the semiconductor, mobile charge carriers within the semiconductor are either pulled towards the insulator-semiconductor interface by the electric field or pushed away. 
In principle, three different situations with regard to the mobile charge carrier density at the interfacial layer (known as the channel) can occur 
upon applying an electrical field - accumulation, depletion, or inversion. Even though inversion has been demonstrated in OTFTs \cite{Luessem2013}, 
the vast majority of organic transistors operates in the depletion or accumulation mode, respectively. 
Whether electrons (n-type) or holes (p-type) are majority charge carriers in OTFTs depends on the energy of the 
frontier molecular orbitals but also to a large extent on the selectivity of source and drain electrode used for charge injection and extraction, respectively. 
The number of mobile charges in the channel is approximated by a simple plate-capacitor model which yields for the density of electrons per area $\mathrm{\tilde{n}}$, 
$\mathrm{\tilde{n}}=\mathrm{\tilde{C}}\times(\mathrm{V}_\mathrm{GS}-\mathrm{V}_\mathrm{th})/\mathrm{e}$, using the elementary charge $\mathrm{e}$, the specific capacitance of the insulator $\mathrm{\tilde{C}}$  defined as 
$\mathrm{\tilde{C}}=\mathrm{\epsilon}_\mathrm{0} \mathrm{\epsilon}_\mathrm{r}/\mathrm{t}$ ($\mathrm{t}$ and $\mathrm{\epsilon}_\mathrm{r}$ are the thickness and relative permittivity of the insulator, respectively), and $\mathrm{V}_\mathrm{th}$ the threshold voltage. Upon applying a voltage between source and drain $\mathrm{V}_\mathrm{DS}$, charge carriers move along the channel yielding a net drain current $\mathrm{I}_\mathrm{D}$.\\
Using the plate-capacitor model and assuming a field and charge carrier density independent mobility (gradual channel approximation), the drain current in the linear and saturation regime obeys
\begin{align}
\mathrm{I}_\mathrm{D}^\mathrm{lin}&=\text{\textmu}_\mathrm{FET}\frac{\mathrm{W}}{\mathrm{L}}\mathrm{\tilde{C}}((\mathrm{V}_\mathrm{GS}-\mathrm{V}_\mathrm{th})\mathrm{V}_\mathrm{DS}-\frac{1}{2} \mathrm{V}_\mathrm{DS}^2)&;&\,\,\, |\mathrm{V}_\mathrm{DS}|\le |\mathrm{V}_\mathrm{GS}-\mathrm{V}_\mathrm{th}|,\,\,\mathrm{linear\,regime} \label{eq:TFT_IV_1}\\
\mathrm{I}_\mathrm{D}^\mathrm{sat}&=\text{\textmu}_\mathrm{FET}\frac{\mathrm{W}}{2\mathrm{L}}\mathrm{\tilde{C}}(\mathrm{V}_\mathrm{GS}-\mathrm{V}_\mathrm{th})^2&;&\,\,\, |\mathrm{V}_\mathrm{DS}|\ge |\mathrm{V}_\mathrm{GS}-\mathrm{V}_\mathrm{th}|,\,\,\mathrm{saturation\,regime}
\label{eq:TFT_IV_2}
\end{align}
where $\mathrm{\mu}_\mathrm{FET}$ is the majority charge carrier field-effect mobility and $\mathrm{L}$ and $\mathrm{W}$ are channel length and width, respectively (cf. Fig.\,\ref{fig:OTFT_Intro}b and c for ideal current-voltage curves according to Eq. \ref{eq:TFT_IV_1} and \ref{eq:TFT_IV_2}). This ideal behavior is commonly observed
in OTFTs \cite{Klauk2007, Yi2016, Yamamura2018}(cf. Fig.\,\ref{fig:OTFT_Intro}d). 
Hence, the definition of the transconductance $\mathrm{g}_\mathrm{m}=\left.\frac{\partial \mathrm{I}_\mathrm{D}}{\partial \mathrm{V}_\mathrm{GS}}\right\rvert_{|\mathrm{V}_\mathrm{DS}|=const.}$ 
can be utilized to derive the FET mobility \textmu$_\mathrm{FET}$, which is proportional to the slope of the transfer curve (cf. Fig.\,\ref{fig:OTFT_Intro}b).
The FET mobility often provides an adequate approximation of the charge carrier mobility in the high charge carrier density limit ($|\mathrm{V}_\mathrm{GS}|>>|\mathrm{V}_\mathrm{th}|$).
However, for non-ideal contact properties as well as for low charge carrier concentrations, more complex mobility models are needed to describe the current-voltage characteristics correctly \cite{Vissenberg1998}. Furthermore, the upper set of equations fails to explain the current-voltage behavior for $|\mathrm{V}_\mathrm{GS}|<|\mathrm{V}_\mathrm{th}|$. This so-called subthreshold region is governed by an exponential current rise. Its steepness, the subthreshold slope $\mathrm{S}$ defined by
\begin{align}
\mathrm{S}=\left.\left(\frac{\partial \mathrm{log}_{10}|\mathrm{I}_\mathrm{D}|}{\partial |\mathrm{V}_\mathrm{GS}|}\right\rvert_{|\mathrm{V}_\mathrm{GS}|<|\mathrm{V}_\mathrm{th}|}\right)^{-1}
\label{eq:TFT_SS}
\end{align}
is a measure of how effective interface states can be occupied by mobile charge carriers. In particular, the analysis of $\mathrm{S}$ can be used to 
reconstruct the density of states at the semiconductor-insulator interface \cite{Kalb2010}. Within the Boltzmann approximation, the lower limit of $\mathrm{S}$ is $\simeq60~\mathrm{meV/dec}$ at room temperature, but often much larger for OTFTs. For an ideal, non-leaking gate insulator, the off-state of the transistor (cf. Fig.\,\ref{fig:OTFT_Intro}b) is governed 
by the bulk conductivity of the semiconductor, which is determined by the energy gap of the semiconductor, background doping, density of trap states, and temperature. Altogether, in order to find a mathematical description that fits the entire current-voltage range of an organic transistor, in-depth understanding of the microscopic charge carrier transport processes and generation/ recombination mechanisms is required. However, from a circuit design perspective, compact models based on empirical or generic functions, such as Eq. \ref{eq:TFT_IV_1}, \ref{eq:TFT_IV_2}, \ref{eq:TFT_SS} are often sufficient to describe the TFT behavior in a circuit \cite{Marinov2009, Fischer2017}.
\subsubsection{High-Mobility Organic Semiconductors and Gate Dielectrics}
\begin{figure}[!htb]
	\begin{center}
		\includegraphics[width=.99\textwidth,clip]{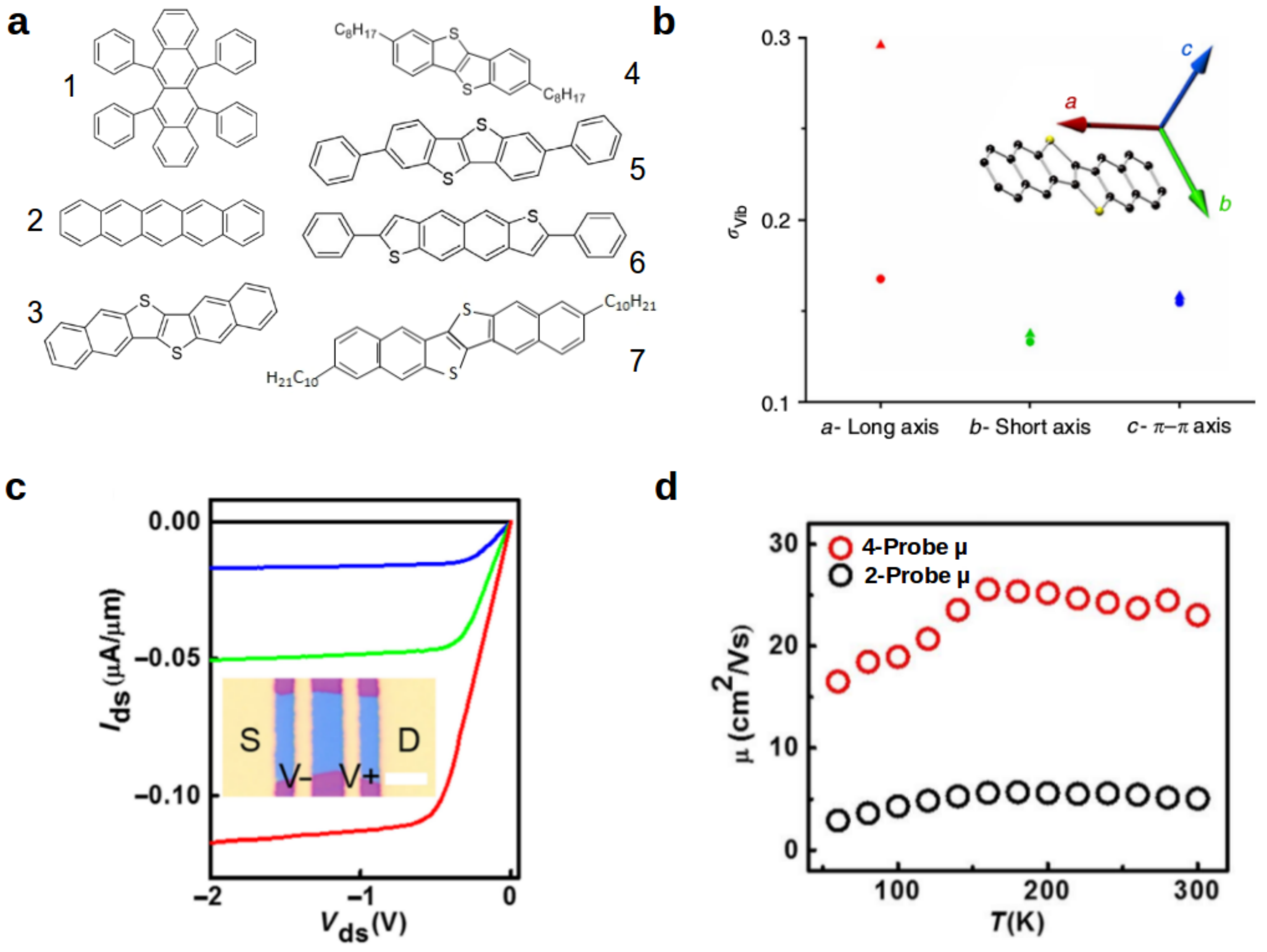}
	\end{center}
	\caption[]{\textbf{a} Selected small molecules typically used in vacuum processed OTFTs: (1) rubrene (2) pentacene, (3) DNTT, (4) C$_8$-BTBT, (5) DPh-BTBT, 
		(6) DPh-anti-NDT (7) C$_{10}$-DNTT 
		\textbf{b} Thermal amplitudes along different axis a, b and c for DNTT (triangles) and C$_{10}$-DNTT (circles) (reprinted from \cite{Illig2016}, published under CC BY-NC license for free non-commercial reuse). 
		The large amplitude along the a-axis causes the larger dynamic disorder in DNTT compared to C$_{10}$-DNTT. 
		\textbf{c} Output characteristics of a single layer C$_8$-BTBT OTFT (single-crystal) on hexagonal boron nitride (hBN) and \textbf{d} temperature dependent 2- and 4-probe mobility obtained from an OTFT as shown in \textbf{c} \cite{He2017}$*$. (Reprinted from \cite{He2017}, published under CC BY-NC license for free non-commercial reuse)}
	\label{fig:OTFT_materials}
\end{figure}

\begin{figure}[!htb]
	\begin{center}
		\includegraphics[width=.99\textwidth,clip]{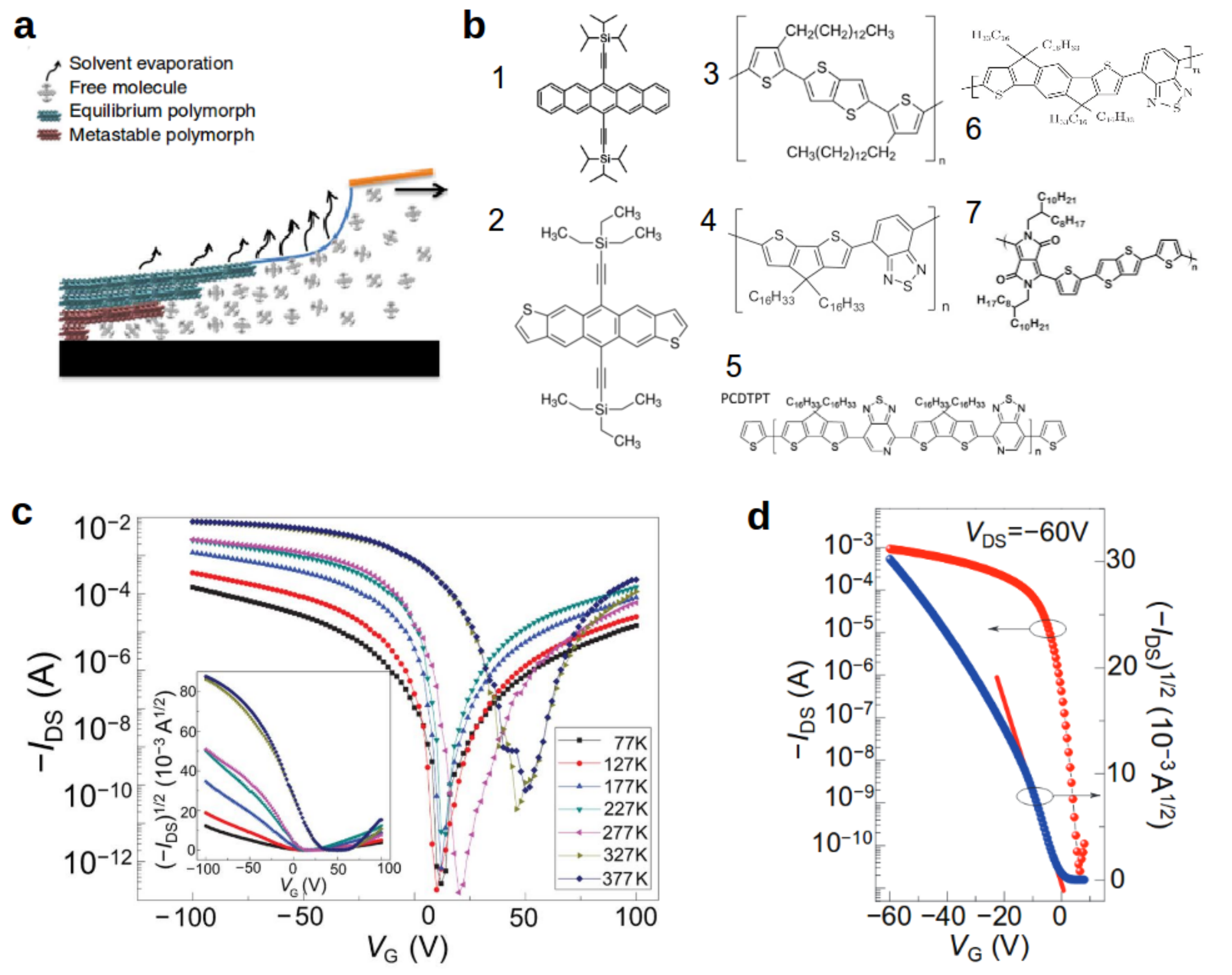}
	\end{center}
	\caption[]{\textbf{a} Illustration of solution shearing method (reprinted from \cite{Giri2013}$*$, published under CC BY-NC license for free non-commercial reuse). Initially, 
		the equilibrium polymorph (blue unit cells) is formed at the solution-air interface and grows downwards towards the 
		substrate. Afterwards, the metastable polymorph (red unit cells) is formed. \textbf{b} Selected small molecules and polymers typically 
		used for OTFTs processed from solution: (1) TIPS-pentacene (2) TES-ADT, (3) PBTTT, (4) CDT-BTZ, (5) PCDTPT, (6) IDTBT (7) DPh-DTT 
		\textbf{c} Transfer curve of an ambipolar OTFT composed of DPP-DTT as active 
		semiconductor material \cite{Li2012}
		\textbf{d} Transfer curve of an OTFTs comprising DPP-DTT  as an example for the overestimation of mobility for small $\mathrm{V}_\mathrm{GS}$ (reprinted from \cite{Li2012}$*$, published under CC BY-NC license for free non-commercial reuse.)}
	\label{fig:OTFT_materials_2}
\end{figure}

The primary target of the semiconductor material development for OTFTs is the improvement of mobility and operational stability. 
The charge carrier mobility, which directly relates to the transconductance, is an important figure of merit for the TFT operation. 
A high value is desirable for circuit design because it enables engineers to develop circuits with high integration density, 
low driving voltage, and more functionality. Since transistors are interface devices, the development of new semiconductor materials
needs to go hand in hand with the evaluation of optimized deposition techniques as well as tailored gate insulator materials. Hence, 
improving the charge carrier mobility in OTFTs is a versatile topic connecting theory, molecular design, film growth, and interface engineering.\\
Among high-mobility organic semiconductors (OSCs), small molecules such as rubrene or pentacene have traditionally been preferred since they can be grown as bulk single-crystals under vacuum conditions. Such crystals are ideal to study the intrinsic transport in the semiconductor since it is only limited by the degree of dynamic disorder (due to inter- and intramolecular vibrations)  \cite{Hasegawa2009, Yi2016, Tsurumi2017}. 
The perfection of these single-crystals allows for mobility in excess of 20-40\,cm$^2$/(Vs) as shown, e.g., for rubrene \cite{Hasegawa2009}.\\ 
Bulk single-crystals, however, are of limited use for electronic circuits. For single- and poly-crystalline thin-films, as favored for OTFTs in electronic circuits, the class of heteroacenes and oligobenzothiophenes \cite{Ebata2007, Takimiya2011, Takimiya2016} (cf. Fig.\,\ref{fig:OTFT_materials}a) has gained large popularity. Besides their potential for high charge carrier mobility, important molecular parameters and film properties (e.g., ionization potential, optical gap, thermal stability, solubility, etc.) can be adjusted by the conjugation length and side-chain substitutions \cite{Yokota2013, Takimiya2014}. Furthermore, they are perfect model systems to investigate the influence of dynamic disorder on charge transport, due to the rod-like shape of these molecules. In particular, it has been shown that the strength of the dynamic disorder is determined by the amplitude of molecular
vibrations along the long axis of these molecules. Such modes can be effectively suppressed using flexible alkyl side-chains along the long axis of oligobenzothiophenes like in C$_8$-BTBT \cite{Illig2016} (cf. Fig.\,\ref{fig:OTFT_materials}b).\\
In the following, recent works which have proven excellent charge carrier transport in organic semiconductor thin-films are reviewed. However, it should be emphasized that many reports claiming charge carrier mobility records suffer from incorrect determination of this quantity \cite{Bittle2016a, Choi2017a, Uemura2016a}, which will be discussed in more detail in subsection \ref{subsec:Electric_Properties_of_State_of_the_Art_OTFTs}. In the following, all reports where this incorrect method was employed, will be highlighted by a $*$.\\
The superior transport in oligobenzothiophenes with alkyl side-chains has recently been demonstrated by record mobility values $\ge$\,20\,cm$^2$/(Vs)$^{-1}$ in crystalline single-layer and few-layer films of C$_8$-BTBT deposited in vacuum \cite{He2017, Zhang2016}$*$ (cf. Fig.\,\ref{fig:OTFT_materials}c and d). Unfortunately, the vacuum growth of uniform crystalline thin-films on large areas represents a substantial challenge since the growth kinetics of competing crystalline facets are difficult to control. In this context, the deposition of semiconductor films from solution provides an alternative offering several possibilities to manipulate the molecular alignment during film growth \cite{Li2014}$*$. In particular, for deposition techniques such as off-center spin-coating \cite{Yuan2014}$*$ or meniscus-guided solution-shearing \cite{Giri2011, Gu2018}$*$, the crystal growth occurs far from thermal equilibrium, facilitating metastable polymorphs to grow, which are not obtained in films prepared in vacuum (cf. Figure \ref{fig:OTFT_materials_2}a). 
These metastable polymorphs can offer a more efficient charge transport than the equilibrium structures \cite{Giri2011, Giri2013}$*$ yielding higher mobility values. Furthermore, using meniscus-guided solution-shearing, centimeter-sized crystalline films can be prepared, which are ideal for large area OTFT arrays \cite{Yamamura2018, Rocha2018}. The second advantage of the deposition from solution is that a wider range of materials is available, i.e., molecules with a large molecular mass and polymers. The most important representatives among these materials are (selected structures displayed in Fig.\,\ref{fig:OTFT_materials_2}b): benzothiophenes with alkyl side-chains \cite{Ebata2007}, silylethyne-substituted heteroacenes and benzothiophenes such as TIPS-pentacene or TES-ADT \cite{Anthony2001,Payne2005, Jia2018}, polymers such as PBTTT \cite{Venkateshvaran2014}, donor-acceptor copolymers such as IDTBT \cite{Venkateshvaran2014}, DPP-DTT \cite{Li2012}$*$, or CDT-BTZ \cite{Yamashita2016}$*$. In particular, donor-acceptor copolymers have gained a lot of interest because the high charge carrier mobility observed in these materials (record values of $\ge$\,50\,cm$^2$/(Vs) \cite{Lee2016a}$*$) does not originate from a high degree of crystallinity within the film. It rather relies on the highly effective charge transport along the rigid conjugated polymer backbone, stabilized by the donor-acceptor interaction \cite{Sirringhaus2014}. Furthermore, due to the small energy gap of many donor-acceptor copolymer systems, ambipolar charge carrier transport (cf. Figure \ref{fig:OTFT_materials_2}c) is frequently observed and the balance between hole and electron transport can be adjusted by the acceptor and donor strength, respectively \cite{Yuen2011}$*$.\\
In conclusion, innovative material concepts for high-mobility OSCs along with advanced deposition techniques 
such as solution-shearing facilitate high-performance p-type OTFTs which can be a perfect match to high-mobility n-type transparent semiconducting oxides for complementary circuits. However, it should be emphasized that all record mobility values of $\ge10\, \mathrm{cm^2/(Vs)}$ have been
reported either for long channel TFTs (channel length $\ge20\,$\textmu m, cf. e.g. \cite{Paterson2018, Rocha2018, Yamamura2018}) or the reports suffer from an incorrect mobility extraction \cite{Bittle2016a, Choi2017a, Uemura2016a, Lee2016, Lee2016a, Li2012, He2017, Zhang2016, Yuan2014, Giri2011}* (cf. Figure \ref{fig:OTFT_materials_2}d). Both aspects are related to the increasing contribution of contact resistance in short-channel OTFTs which will be discussed in more detail in the next section.\\
In order to obtain a high charge carrier mobility, low driving voltage, and sufficient operational stability of the transistors, 
the semiconductor needs to be combined with a tailored dielectric material. Properties of an ideal dielectric film for TFTs are: high capacitance, high breakdown field, low interface trap density, and scalability of the deposition process. The best materials to obtain high capacitance and high breakdown strength using scalable processes are inorganic oxides such as Al$_2$O$_3$, HfO$_2$, or SiO$_2$ (specific capacitance in excess of $1\,$\textmu$\mathrm{F}/\mathrm{cm^2}$ with a breakdown strength of $\ge$\,5\,$\mathrm{MV/cm}$  \cite{Klauk2007}). Unfortunately, these materials often possess an unacceptably high surface trap density in organic transistors limiting the charge carrier transport. Organic insulator materials such as polymers \cite{Kalb2007, Ji2018}, self-assembled monolayers (SAM) \cite{Klauk2007, Kraft2015}, or small molecules \cite{Xiang2016a} have the potential to result in almost ideal, trap-free surfaces due to the lack of dangling bonds and the weak interaction with the OSC. Unfortunately, these materials suffer from low breakdown strength, e.g., related to pinholes in the film. As a consequence, polymer dielectrics often require a film thickness of $\ge$\,$500\,\mathrm{nm}$, resulting in a high driving voltage \cite{Venkateshvaran2014}. A solution to this problem is the application of hybrid stacks of inorganic oxides (e.g., Al$_2$O$_3$) and low trap-density polymers \cite{Jia2018} or self-assembled monolayers \cite{Klauk2007}. In particular, the combination of the fluoropolymer CYTOP (Asahi Glass Corporation) and robust Al$_2$O$_3$ films seems to be a valuable approach to achieve low driving voltages and operational stability even comparable to crystalline silicon TFTs \cite{Jia2018}.
%
%
\subsubsection{Electric Properties of State-of-the-Art OTFTs}
\label{subsec:Electric_Properties_of_State_of_the_Art_OTFTs}
\begin{figure}[!htb]
	\begin{center}
		\includegraphics[width=.99\textwidth,clip]{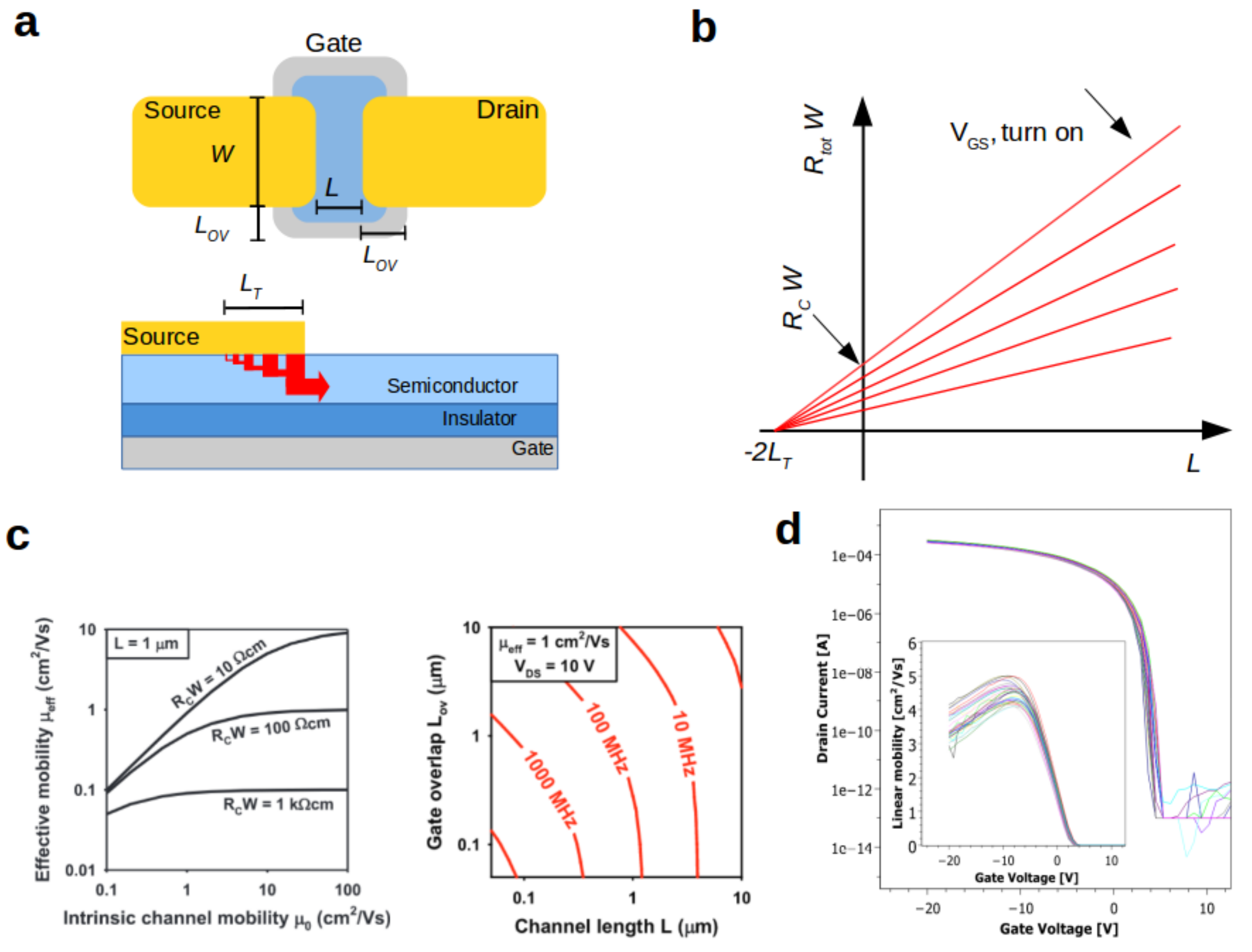}
	\end{center}
	\caption[]{\textbf{a} Top-view of an OTFT structure as used for a switch in, e.g., active-matrix cells. 
		The OSC is patterned to an area smaller than the gate electrode. Below is a cross-section through an OTFT illustrating the current 
		crowding within the transfer length $\mathrm{L}_\mathrm{T}$. \textbf{b} Illustration of transmission-line method 
		(L$_\mathrm{OV}\,\gg\,\mathrm{L}_\mathrm{T}$) for the extraction of the contact resistance. 
		\textbf{c} (left) Relation between intrinsic mobility $\mathrm{\mu}_0$ and effective mobility measured within an OTFT for certain values of contact resistance, and (right) calculated $\mathrm{f}_\mathrm{T}$ for an 
		OTFT with at $\mathrm{V}_{\mathrm{DS}}$=10 V and an effective mobility of 1 cm$^2$/(Vs) (reprinted from \cite{Klauk2018} with permission from John Wiley and Sons). 
		\textbf{d} OTFT with on/off-ratio of $\ge$\,10$^{10}$ in a Corbino architecture (reprinted from \cite{Guo2017}, published under CC BY-NC license for free non-commercial reuse.}
	\label{fig:OTFT_performance}
\end{figure}

Apart from the charge carrier mobility, transistor parameters relevant for circuits are: gain, transconductance, transition frequency, on/off-ratio, device capacitance, and subthreshold slope. In the following, these properties shall be reviewed for state-of-the-art OTFTs.\\
The differential gain is defined by the ratio of drain to gate current, and hence is a measure for the effectiveness of the transistor as an amplifier. For a small signal frequency f and a total device capacitance C$_\mathrm{tot}$, the differential gain is 
\begin{equation}
\mathrm{gain}=\frac{|\mathrm{\partial i_D}|}{|\mathrm{\partial i_G}|}=\frac{\mathrm{g}_{\mathrm{m}}}{2\pi \mathrm{f} \mathrm{C}_\mathrm{tot}}
\end{equation}
where the C$_\mathrm{tot}$ is at least of the size $\tilde{\mathrm{C}}\times(\mathrm{L}+2\mathrm{L}_\mathrm{OV})\times$W (with L$_\mathrm{OV}$ as overlap length, cf. Fig.\,\ref{fig:OTFT_performance}a), and i$_\mathrm{G}$ and i$_\mathrm{D}$ denote the small signal quantities of the drain and gate current, respectively. For non-ideal devices, the transconductance has two contributions - the channel resistance R$_{\mathrm{Ch}} \times$W (which scales with the mobility and the channel length) and the contact resistance R$_{\mathrm{C}} \times$W accounting for injection and extraction of charges. The contact resistance originates from the Schottky-type barrier at the semiconductor-metal interface as well as from the transport through the semiconductor towards the channel \cite{Gruber2013}. Lowest values for R$_{\mathrm{C}} \times$W are in the range of $\le$\,100$\,l\mathrm{\Omega cm}$ \cite{Borchert2020, Yamamura2018, He2017, Choi2016}, and a clear difference between staggered and coplanar transistor designs is not present. Several methods have been proposed to extract the contact resistance \cite{Natali2012, Liu2015}. Among them, the transmission-line method (TLM) is most frequently used. It relies on a systematic variation of the channel resistance R$_{\mathrm{Ch}} \times$W by means of the channel length L in order to obtain R$_{\mathrm{C}} \times$W from extrapolation (L$\rightarrow 0$) of the total device resistance R$_{\mathrm{tot}}\times$W=R$_{\mathrm{Ch}} \times$W+R$_{\mathrm{C}} \times$W (cf. Fig.\,\ref{fig:OTFT_performance}b). The intersection of the total resistance function with the abscissa can be associated with the so-called transfer length L$_\mathrm{T}$, which is the length underneath the source electrode needed to provide the required amount of current \cite{Richard2007} (also known as current crowding effect). Hence, the geometric overlap L$_\mathrm{OV}$ of gate and source electrode should be at least L$_\mathrm{T}$ in order to avoid a bottleneck for charge carrier injection. Values for L$_\mathrm{T}$ typically vary between a few hundreds of nanometer and several tenths of micrometer \cite{Ante2012}.\\
Since the contact resistance constitutes a significant fraction of the total resistance for short channel devices with L$\le\,20\,$\textmu m, the influence of the contact resistance often leads to a significantly underestimated field-effect mobility derived from the slope of the transfer curve for short-channel TFTs \cite{Ante2012, Kleemann2012} (cf. Fig.\,\ref{fig:OTFT_performance}c). In case
of a strongly non-linear contact resistance that depends on the gate-source voltage \cite{Bittle2016a}, however, the field-effect mobility may be overestimated if only a small voltage range is investigated (cf. Figure \ref{fig:OTFT_materials_2}d). Unfortunately, many reports claiming new record values for the intrinsic mobility in OSCs  \cite{Li2012, Giri2011, Yuan2014, Lee2016}* did not account for this effect resulting in significantly overestimated values.\\
The contact resistance has also a strong impact on the transition frequency $\mathrm{f}_\mathrm{T}$ defined as 
\begin{equation}
\mathrm{f}_\mathrm{T}=\frac{\mathrm{g}_{\mathrm{m}}}{2\pi\mathrm{C}_\mathrm{tot}}\le\frac{\mathrm{g}_{\mathrm{m}}}{2\pi \mathrm{W}\times(\mathrm{L}+2\mathrm{L}_\mathrm{OV})^2}.
\label{eq:f_t}
\end{equation}
Since $\mathrm{f}_\mathrm{T}$ is defined by the unity-gain condition, it is a figure of merit of the effectiveness of the TFTs acting as a switch depending on the frequency. A high 
$\mathrm{f}_\mathrm{T}$ requires $\mathrm{L}_\mathrm{OV}$ to be as low as possible. However, in case $\mathrm{L}_\mathrm{OV}$ approaches 
$\mathrm{L}_\mathrm{T}$, f$_T$ cannot be increased further since the transconductance is limited by the contact resistance (cf. Figure \ref{fig:OTFT_performance}c).\\
The best horizontal OTFT reported operates at up to 38 MHz \cite{Yamamura2020}\footnote{Very recently also higher values have been published but this work is not peer-reviewed yet \cite{perinot2020organic}} (higher frequencies have been achieved for vertical organic transistors, cf. \ref{section:VOT}). This record has been achieved for a high-mobility OSC (10 cm$^2$/(Vs)), reducing $\mathrm{L}_\mathrm{OV}$ and minimizing contact resistance by means of surface treatments of the electrodes. Although this $\mathrm{f}_\mathrm{T}$ would be sufficiently high for many applications (e.g., rectifier in an 13.56 MHz RFID-tag or as a switch TFT in an active-matrix display, cf. next section), it should be emphasized that all $\mathrm{f}_\mathrm{T}$ values $\ge$ 20 MHz have been reported for large devices with a channel width of $\ge$ 750 \textmu m \cite{Kitamura2011, Perinot2016, Yamamura2018, perinot2020organic}. However, as discussed in the next section, for most applications, e.g., in active-matrix cells, small transistors (W/L$\simeq$1) are required for effective circuit operation.\\
Apart from the transition frequency, the on/off-ratio and the subthreshold slope are two other important parameters for electric circuit design because they mainly govern the static (leakage) and dynamic (switching) power dissipation in logic components (e.g., inverters) or dynamic memory cells (e.g., the switch TFT in an active-matrix cell). Due to the continuous optimization of low-defect gate insulator materials \cite{Klauk2007, Yamamura2018, Kalb2007}, the interface trap density can be as small as 10$^{11}$\,cm$^{-2}$ \cite{Haeusermann2016} facilitating the fabrication of TFTs with a subthreshold slope very close to $\sim$60\, meV/dec at 25$^\circ$C \cite{Klauk2007, Kalb2007, Yamamura2018}. The on/off-ratio and in particular the off-current dictate the static power consumption. Organic TFTs allow for a very high on/off-ratio due to the large energy gap of most OSCs compared to silicon. However, the intrinsic background doping of many OSCs, e.g., by oxygen \cite{Nayak2013}, which can be as large as 10$^{16}$\,cm$^{-3}$ \cite{Pahner2013}, restricts the on/off-ratio due to parasitic current paths. The influence of such parasitic paths is effectively suppressed for TFTs with a large $\mathrm{W}$ or specific device layouts (Corbino layout) \cite{McCall2014}. For such devices, on/off-ratios as large as 10$^{10}$ \cite{Guo2017, SmartKem2017} (cf. Figure \ref{fig:OTFT_performance}d) and an off-current below 0.1\,fA/\textmu m (current per $\mathrm{W}$) have been obtained. However, for small TFTs which are favorable, e.g., for displays \cite{Gelinck2004}, such methods for leakage current suppression cannot be applied, and it remains a challenge for the future to find semiconductors with very low intrinsic doping or explore alternative integration techniques allowing for the fabrication of small OTFTs with high on/off-ratios.

\subsection{Applications of OTFTs}
\begin{figure}[!htb]
	\begin{center}
		\includegraphics[width=.99\textwidth,clip]{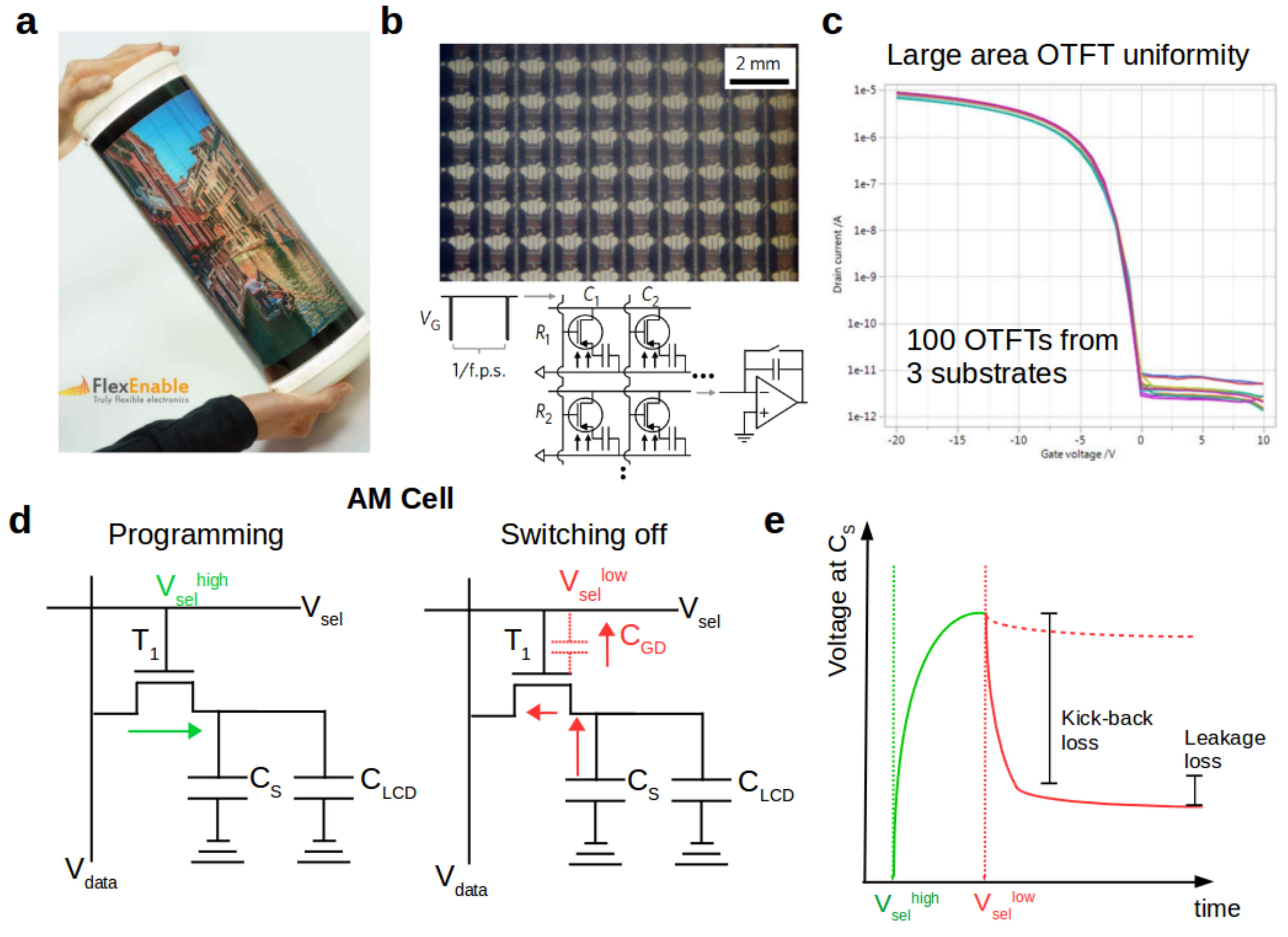}
	\end{center}
	\caption[]{\textbf{a} 12.1" AM LCD display driven by OTFTs with a  curvature radius of 60 mm (reprinted from \cite{Harding2017} with permission by John Wiley and Sons). 
		\textbf{b} Photograph of a printed organic photodetector array and associated readout circuit below (reprinted from \cite{Pierre2017} with permission by Springer Nature). \textbf{c} Uniformity data 
		taken from 100 OTFTs (reprinted from \cite{Harding2017} with permission by John Wiley and Sons). \textbf{d} Scheme of an AM cell for a LCD cell (C$_{LCD}$) shown 
		during programming (left) and switching (right). 
		The direction of current flow is indicated by red and green arrows as well as the discharging of $\mathrm{C}_\mathrm{S}$ 
		via $\mathrm{C}_\mathrm{GD}$. \textbf{e} Voltage at $\mathrm{C}_\mathrm{S}$ vs. time diagram during programming (green) and switching off (red). The voltage loss due to the VKBE and leakage through T$_1$ is indicated (dotted line shows behavior without VKBE).}
	\label{fig:OTFT_applications}
\end{figure}

The promises of organic electronics are low production costs, large area fabrication, and mechanical flexibility at a higher performance level than amorphous silicon. In order to justify these promises, a variety of different electronic circuits have been demonstrated on flexible substrates using low-cost printing techniques, among them active-matrix cells for LCD driving \cite{Harding2017}, OLED driving \cite{Zhou2006, Steudel2012} (cf. Fig.\,\ref{fig:OTFT_applications}a), electrophoretic displays \cite{Gelinck2004}, organic photodetectors \cite{Pierre2017} (cf. Fig.\,\ref{fig:OTFT_applications}b), pressure sensor arrays \cite{Someya2004}, digital and analog components such as inverters \cite{Zhang2009}, shift-registers \cite{Gelinck2004, Myny2012},  amplifiers \cite{Sekitani2016}, DA-converters \cite{Kheradmand-Boroujeni2016}, full RFID-tags \cite{Myny2010, Fiore2015} and many more. This development has been accompanied on the single device level by continuous improvements of performance, stability, and uniformity \cite{Fukuda2014,Ji2018, Zschieschang2017} (cf. Fig.\,\ref{fig:OTFT_applications}c).
However, except for flexible electrophoretic displays \cite{PlasticLogic2018}, OTFTs are currently not being used in commercial applications. One of the most promising applications, however, is the OTFT driven active-matrix (AM) backplane. This technology can be employed in a multitude of applications such as LCD and OLED displays, or photodetector, pressure, and chemical sensor arrays. In the following, developments on OTFTs necessary to allow for improved performance of AM backplanes will be discussed exemplary.\\
An AM array is composed of small subunits denoted as pixels or AM cells. The most basic circuit of any AM cell is depicted in Fig.\,\ref{fig:OTFT_applications}d. It contains one transistor T$_1$ to address and program the pixel, one capacitor $\mathrm{C}_\mathrm{S}$ to maintain the pixel state, and one active element (e.g. LCD cell, OLED etc.). The most demanding component of an AM cell is the transistor T$_1$ since its on-current must be high enough to charge $\mathrm{C}_\mathrm{S}$ during the short addressing time of the pixel ($<$ 1\,\textmu s for VGA resolution), while its off-current should be low enough to prevent discharging of $\mathrm{C}_\mathrm{S}$ during a frame cycle of the display (16.67\,ms for 60\,frames per second). As exemplary shown by Sirringhaus \cite{Sirringhaus2014} for a 200\,ppi display, the transistor T$_1$ should satisfy the condition \textmu$\mathrm{W/L}\ge 3\,\mathrm{cm^2/(Vs)}$ for the on-current and the on/off-ratio should be $\ge$\,$10^6$. Both conditions mentioned above can be fulfilled by flexible OTFTs, which would allow for high-resolution displays. The field-effect mobility in state-of-the-art OTFTs \cite{Harding2017, Cowin2014} would even allow for more complex backplane circuits.
However, two substantial problems yet prevent OTFTs from being used in demanding AM applications - firstly, the high overall OTFT capacitance compared to competing technologies, and secondly, the decreasing on/off-ratio for small OTFTs. The high OTFT capacitance is a direct consequence of the overlap of source/gate and drain/gate (typical overlap is $\ge 3\,$\textmu m due to alignment tolerances). This large capacitance is of great significance for circuit design since it gives rise to the voltage-kick-back-effect (VKBE), a rapid and unintentional discharging of $\mathrm{C}_\mathrm{S}$ (cf. Fig.\,\ref{fig:OTFT_applications}d and e) which ultimately leads to a complete loss of the stored information in the AM pixel. The problem of the VKBE cannot be solved easily since this would require so-called self-aligned TFT architectures having virtually no geometric overlap of source/gate and drain/gate. Self-aligned TFTs are readily available in silicon technology. However, in the case of OTFTs, self-alignment has only been demonstrated using non-scalable fabrication processes \cite{Noh2007, Higgins2015}. Hence, it remains one of the unsolved challenges for OTFTs to demonstrate a truly self-aligned device using processes that can be adopted in a production line. The fabrication of such self-aligned OTFTs would boost the potential of organic transistors for applications enormously. As discussed, AM cells would benefit from the low TFT capacitance, enabling high-resolution screens. Furthermore, self-aligned geometries would help to improve the transition frequency of OTFTs from currently max. 38 MHz to above 100 MHz due to the reduction of overlap capacitances (cf. Figure \ref{fig:OTFT_performance}c)\footnote{Recently, the potential of reduced overlap capacitances has been underlined by Perinot et al. They demonstrated a transition frequency of 160\,MHz at V$_{\mathrm{DS}}$=40\,V. However, the work is currently under peer-review \cite{perinot2020organic}.}. In this case, organic transistors would even become interesting for medium and long-range radio-frequency identification applications - an application which cannot be covered by low-cost amorphous silicon TFTs and currently requires expensive crystalline silicon to be used.\\
In summary, reliable, uniform, and stable operation of OTFTs on flexible substrates has been demonstrated for various applications. Due to their low surface defect density and their high on/off-ratio, state-of-the-art OTFTs can outperform established amorphous silicon transistors in terms of their static performance. However, due to their comparably large capacitance (due to parasitic overlap of electrodes), organic transistors have currently limited potential for medium- and high-level performance applications such as AM displays, large area sensor arrays, or RFID tags. From a business perspective, though, it is essential for OTFTs to enter this medium- and high-level segment, which is due to two aspects. Firstly, there is a fierce competition with the well-established amorphous silicon for low-level applications, and secondly, entering the medium- and high-level segment will open up new fields of applications which cannot be realized by amorphous silicon TFTs.
\subsection{Recent Trends in Organic Transistor Research}
\subsubsection{Vertical Organic Transistors}
\label{section:VOT}
\begin{figure}[!htb]
	\begin{center}
		\includegraphics[width=.99\textwidth,clip]{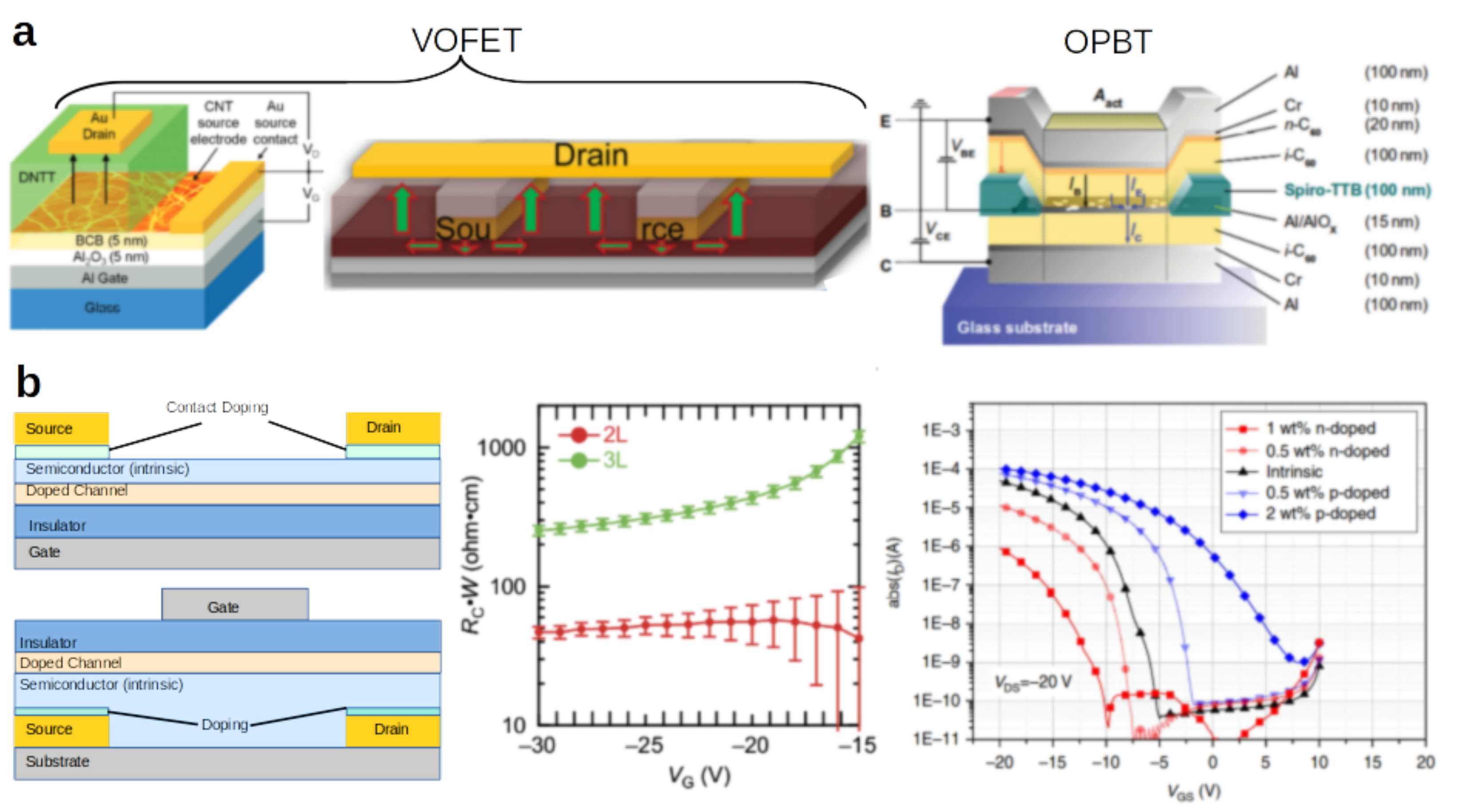}
	\end{center}
	\caption[]{\textbf{a} Selected vertical OTFT structures. From left to right: Carbon nanotube (CNT)-based VOFET by \cite{McCarthy2011} (reprinted with permission by the American Chemical Society), 
		lithographically integrated VOFET by \cite{Kleemann2013}, and the OPBT structure by \cite{Klinger2015}. \textbf{a} 
		Doping in OTFTs. From left to right: layer scheme of a contact and channel doped OTFT, reduction of contact resistance in doped few-layer OTFTs \cite{Yamamura2018} (published under CC BY-NC license for free non-commercial reuse),  and control of threshold voltage and inversion operation by controlled channel doping \cite{Luessem2013}.}
	\label{fig:OTFT_trends_1}
\end{figure}

Applications such as fast oscillator circuits for wireless communication are less demanding than AM cells regarding on/off-ratio and overlap capacitance. However, they require a high transition frequency. Following the definition of the transition frequency, a maximum transconductance per capacitance is needed for high-frequency transistors, which can be achieved using high-mobility semiconductors or short channel lengths. As discussed in the review article
\begin{quote}
   \textit{"A Review on Vertical Organic Transistors" by H. Kleemann, K. Kevin,
A. Fischer, \& K. Leo, Adv. Funct. Mat. 1907113 (2020),}
\end{quote}
vertical organic transistors (VOTs) present a promising alternative to lateral TFT geometries since in ideal vertical devices, the channel length is defined by the thickness of the semiconductor film. Thus, an effective channel length of $\le$\,300\,nm can be achieved without high-resolution patterning techniques, leading to high transconductance. A variety of different VOT concepts have been proposed (see review article by L\"{u}ssem et al. \cite{Luessem2015}), and a selection of structures is shown in Fig.\,\ref{fig:OTFT_trends_1} a.\\
There are two groups of VOTs that can be classified into: 1) organic triodes (with the gate electrode sandwiched between source and drain) \cite{Fischer2012b, Klinger2015, Klinger2017}, and 2) vertical organic field-effect transistors (with the gate electrode structured underneath the source electrode) \cite{McCarthy2011, Ben-Sasson2012, Kleemann2013}.
In vertical organic field-effect transistors (VOFET), the potential at the gate electrode controls the current between the source and drain electrodes by influencing the charge carrier injection underneath the source electrode \cite{Ben-Sasson2012, Sawatzki2018}. Contrary, in organic triodes, the gate is used to tune the transmission of charges by the electric field like in vacuum tubes.\\
Both transistor structures have been continuously improved and, in particular, organic permeable-base transistors (OPBTs), an embodiment of an organic triode (see Fig.\,\ref{fig:OTFT_trends_1}a right), are among the best-performing OTFTs. Key device parameters of OPBTs are: on/off-ratio $\ge$ 10$^8$, driving voltage $\le$\,2V, channel length $\le$\,300 nm, intrinsic gain 35\,dB, maximum current density 1\,kA/cm$^2$, and a transition frequency of 40\,MHz \cite{Klinger2017, Klinger2018, Boroujeni2018}. Thus, OPBTs are currently the organic transistors with the highest frequency of operation. Most remarkably, this record has been reached employing a semiconductor material with an intrinsic mobility of only 0.06\,cm$^2$/(Vs) - more than 200 times less than the record mobility in horizontal TFTs \cite{Xiang2016}. This aspect underlines the fact that the contact resistance limits the performance of such short channel vertical transistors, and methods such as contact doping need to be employed for efficient charge injection \cite{Fischer2012a}.\\
In summary, VOTs are an exciting alternative to conventional OTFTs since they offer an additional strategy for device improvements besides mobility optimization. However, VOTs are still at an early stage of development, and more research is required to understand the principles of device operation. Furthermore, concerning device and circuit integration, VOTs are not at the level of conventional OTFTs yet because the device fabrication requires methods such as soft-lithography or self-assembly \cite{Ben-Sasson2012}, being difficult to adopt in a production process.

\subsubsection{Doped Organic Transistors}
On the one hand, organic semiconductors used in TFTs are typically undoped, which ensures a low off-current and minimal power losses. On the other hand, undoped OTFTs suffer from the fact that small changes either in the background doping or interface trap concentration strongly influence TFT parameters such as V$_\mathrm{th}$, which ultimately limits the device-to-device reliability. Controlled semiconductor doping is a well-established technology in silicon-based electronics, though it has been less popular for OTFTs for a long time. Bigger interest in this topic arose due to the increasing importance of contact resistance and the need to tune the threshold voltage \cite{Luessem2016} for logic circuits.\\
Contact doping is used to improve charge injection, and hence reduce contact resistance. It either requires a thin layer of pure dopant or a doped semiconductor layer between the source/drain electrode and semiconductor (see Fig.\,\ref{fig:OTFT_trends_1}b). Depending on the specific materials (metal/semiconductor) and the device architecture, several mechanisms have been proposed to describe the origin of the improved charge carrier injection. Among them are charge carrier tunneling through a narrow Schottky barrier, image force lowering of the barrier due to the presence of free charges, tuning of a surface dipole, and an increased conductivity of the semiconductor itself (cf. review article by L\"{u}ssem et al. \cite{Luessem2016} and Liu et al. \cite{Liu2015}). Record values of $\mathrm{R}_\mathrm{C}\times \mathrm{W}$ for contact-doped OTFTs are as low as 10\,$\Omega$cm \cite{Borchert2020} while values for OTFTs without contact doping/untreated contacts are larger by at least a factor of 10 \cite{Ante2012}.\\
However, even if contact doping is convenient to use for single devices, it faces limitations concerning device integration. For staggered bottom-gate devices, as shown in Fig.\,\ref{fig:OTFT_trends_1}b, contact doping requires an additional patterning step, which adds complexity and costs to the process. For staggered top-gate TFTs (see Fig.\,\ref{fig:OTFT_trends_1}b), this additional patterning step can be circumvented by using SAMs \cite{Liu2015, Nicht2014}. In this configuration though, often the surface tension of the SAM treated electrodes and the substrate differs substantially, leading to dewetting of the semiconductor \cite{Lee2012}.\\
Besides improving charge carrier injection, molecular doping might also be employed to fine-tune the threshold voltage which is of particular importance for logic components. In comparison to contact-doping, controlled channel doping in OTFTs is an even more challenging task since it requires small and reliable doping ratios in order to maintain the control of the channel conductance by the gate electrode. For vacuum-deposited small molecule OSCs, this adjustability has been demonstrated, allowing to realize the first OTFTs operating in the inversion regime \cite{Luessem2013} (see Fig.\,\ref{fig:OTFT_trends_1}b) as well as organic junction field-effect transistors \cite{Luessem2014}. However, despite first attempts \cite{Lee2016, Kang2016}, precise channel doping for solution-processed OSCs has not been demonstrated. This situation is due to the fact that the formation of uniformly doped films from solution is challenging to control. Furthermore, the molecular arrangement of crystalline organic thin-films is particularly sensitive to small amounts of additives during the film growth, resulting in reduced structural order and hence lower charge carrier mobility \cite{Kleemann2012a}.\\
Despite these challenges, it is a rewarding task to explore the use of doping in TFTs since it is the most promising way to minimize the impact of contact resistance, and threshold-voltage-control adds an important degree of freedom for circuit design.

\subsection{Novel Organic Transistor Concepts - Own Contributions}
In the following, my efforts in the development of novel organic transistor concepts are presented. First, vertical organic transistors offering ultra-short channel length, and hence, high-frequency operation are discussed. Although such devices offer significant benefits over lateral transistor structures in terms of performance, their manufacturing process is more complex. For example, often additional patterning steps or unconventional fabrication methods are required raising significant concerns about their application potential. Hence, in the second part of this section, new concepts for lateral OTFTs are introduced with a special focus on self-aligned devices and new material concepts for high-mobility thin-film crystals.

\subsubsection{Vertical Organic Transistors}
\begin{figure}[!htb]
	\begin{center}
		\includegraphics[width=.99\textwidth,clip]{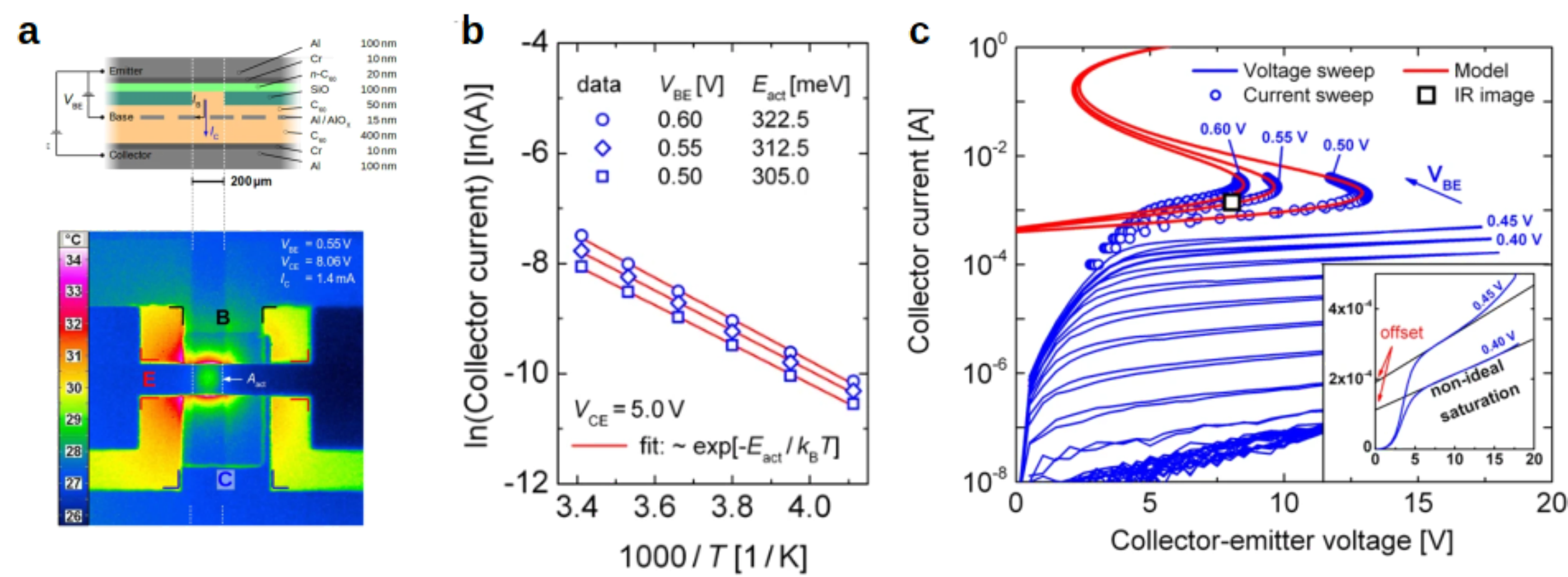}
	\end{center}
	\caption[]{\textbf{a} top: Schematic device cross-section and electrical circuit in common-emitter configuration. Materials: aluminum (Al), chrome (Cr), n-doped C$_{60}$ (n-C$_{60}$), intrinsic (undoped) C$_{60}$ (i-C$_{60}$), native aluminum-oxide (AlO$_X$). Arrows indicate the electron flow. Additional insulating layers (SiO) are inserted for defining and down-scaling the active area. bottom: Thermal imaging during the S-NDR measurement showing the active area A$_{act}$ of the OPBT. \textbf{b} Extraction of the activation energy of conductivity for different base-emitter voltages. \textbf{c} Output characteristic of an OPBT. At low base-emitter voltages, a voltage sweep (blue lines) is used. In order to stabilize the NDR at base-emitter voltages starting from V$_{BE}$= 0.5\,V, a current controlled measurement (blue circles) is used. All curves are measured with forward and backward sweep, demonstrating the repeatability of the self-heating effect. The model (red lines), assuming an Arrhenius-like temperature activation of the conductivity, leads to a reasonable agreement with the experimental data. The black squared point indicates where the thermal image is taken in subfigure b. Inset: The OPBT shows a non-ideal saturation behavior. It can be described by a linear curve with an additional off-set. Images are taken from \cite{Klinger2018}.}
	\label{fig:self_heating}
\end{figure}

Among the multitude of devices out of the class of vertical organic transistors, organic permeable base transistors and vertical organic field-effect transistors stand out due to their simplicity and performance. The focus of my work in this field is to push the electrical performance even further but also to address issues related to device integration. 
\paragraph{Organic Permeable Base Transistors}\vspace{2mm}
The organic permeable base transistor design is inspired by the structure of the vacuum tube triode. Besides charge injecting and extracting electrodes, denoted as emitter and collector, respectively, the OPBT is composed of two semiconducting layers, which are separated by a thin metallic layer named the base electrode (cf. Figures \ref{fig:OTFT_trends_1}a or \ref{fig:self_heating}a). The base electrode is characterized by the special property that charge carriers coming from the emitter can pass through the base electrode and reach the collector. This feature is enabled by nanometer-sized pinholes in the base electrode, which are created by internal strain during the oxidation process of the base electrode which is composed of aluminum. Thus, a thin layer of AlO$_X$ with a thickness of approximately 1-2\,nm passivates the base electrode which leads to transmission of majority charge carriers from the emitter to the collector electrode through the base. Only a small number of charge carriers falls into the base.\\
The operation of the OPBT can be understood comparing it to a lateral field-effect transistor with an ultra-short channel. In fact, the channel of the OPBT is given by the multitude of AlO$_X$-passivated pinholes in the base electrodes, leading to a total channel length of 10-50\,nm. The semiconducting layers with a thickness of approximately 100\,nm on both sides of the base electrode contribute only as a kind of access-line resistance. In particular, the short channel length of the OPBT combined with the high capacitance of the AlO$_X$ ($\sim\,1\,$\textmu\,F/cm$^2$, \cite{Fischer2012a}) allows for high a charge carrier density in the channel of 10$^{20}$\,cm$^{-3}$ \cite{Kaschura2016} and hence, an exceptional high channel conductivity. Consequently, the overall conductivity of the OPBT is not limited by the channel resistance but more by the conductivity of the semiconducting layers and the contact barriers at emitter and collector. As previously shown \cite{Klinger2017}, using optimized electrode materials and strong interface doping in order to reduce the contact resistance, OPBTs may operate in a space-charge-limited-current regime (SCLC) enabling current densities up to 1\,kA/cm$^2$.\\
Despite the excellent performance of OPBTs - small operation voltage of $<$\,10\,V, high on/off-ratio of $>$\,10$^8$, and high current densities - there are still several hurdles to overcome. Firstly, high-frequency operation and stability need to be proven. Secondly, new fabrication methods for a more reliable formation of the base oxide layer need to be developed, and thirdly, OPBTs need to be optimized for specific application scenarios where they can benefit from their inherent advantages. During my research, I focused on all three aspects which led to the following list of publications \cite{Klinger2018, Boroujeni2018, Dollinger2019, Dollinger2019b, Dollinger2020, Wu2020, Guo2020, Guo2020a, Guo2020b, Guo2021}:
\begin{itemize}
    \item \begin{quote}
    \textit{"Non-Linear Self-Heating in Organic Transistors Reaching High Power Densities" by M.P. Klinger, A. Fischer, H. Kleemann, \& K. Leo, Scientific Reports \textbf{8}, 9806 (2018).}
    \end{quote}
\item \begin{quote}
   \textit{"A Pulse-Biasing Small-Signal Measurement Technique Enabling 40 MHz Operation of Vertical Organic Transistors" by B.K. Boroujeni, M.P. Klinger, A. Fischer, H. Kleemann, K. Leo, \& F. Ellinger, Scientific Reports \textbf{8}, 7643 (2018).}
\end{quote}
\item \begin{quote}
   \textit{"Vertical Organic Thin‐Film Transistors with an Anodized Permeable Base for Very Low Leakage Current" by F. Dollinger, K.G. Lim, Y. Li, E. Guo, P. Formánek, R. H\"ubner, A. Fischer, H. Kleemann, K. Leo, \& F. Ellinger, Advanced Materials \textbf{31} (19), 1900917 (2019).}
 \end{quote}
 \item \begin{quote}
   \textit{"Electrically Stable Organic Permeable Base Transistors for Display Applications" by F. Dollinger, H. Iseke, E. Guo, A. Fischer, H. Kleemann, K. Leo, \& F. Ellinger, Advanced Electronic Materials \textbf{5} (12), 1900576 (2019).}
\end{quote}
\item \begin{quote}
   \textit{"Unraveling Structure and Device Operation of Organic Permeable Base Transistors" by G. Darbandy, F. Dollinger, P. Formanék, R. H\"ubner, S. Resch, C. Roemer, A. Fischer, K. Leo, A. Kloes, \& H. Kleemann, Advanced Electronic Materials, \textbf{6}, 2000230 (2020).}
\end{quote}
\item \begin{quote}
   \textit{"Vertical Organic Permeable Dual-Base Transistors for Logic Circuits" by E. Guo, Z. Wu, G. Darbandy, S. Xing, S.J. Wang, A. Tahn, M. G\"obel, A. Kloes, K. Leo, \& H. Kleemann, Nature Communications \textbf{11}, 4725 (2020).}
\end{quote}
\item \begin{quote}
   \textit{"Efficient and low-voltage vertical organic permeable base light-emitting transistors" by Z. Wu, Y. Liu, E. Guo, G. Dharbandy, R. H\"ubner, S.J. Wang, A. Kloes, H. Kleemann, \& K. Leo, Nature Materials \textbf{20},   1007–1014 (2021).}
\end{quote}
\item \begin{quote}
   \textit{"High-Performance Static Induction Transistors Based on Small-Molecule Organic Semiconductors" by E. Guo, K. Leo, \& H. Kleemann, Advanced Materials Technologies \textbf{5}, 2000361 (2020).}
\end{quote}
\item \begin{quote}
   \textit{"Integrated complementary inverters and ring oscillators based on vertical-channel dual-base organic thin-film transistors" by E. Guo, S. Xing, F. Dollinger, R. H\"ubner, S.-J. Wang, Z. Wu, K. Leo, \& H. Kleemann, Nature Electronics \textbf{4}, 588–594 (2021).}
\end{quote}
\item \begin{quote}
   \textit{"Organic Permeable Base Transistors–Insights and Perspectives" by E. Guo, F. Dollinger, B. Amaya, A. Fischer, \& H. Kleemann, Advanced Optical Materials \textbf{9}, 2002058 (2021).}
\end{quote}
\end{itemize}

\subparagraph{Self-Heating in Organic Transistors} \vspace{2mm} 
Efficient contact doping and optimization of the layer structure enable OPBTs work at high current densities at comparably low voltage, which is important if aiming for operation at high frequency. However, as shown by Klinger et al. \cite{Klinger2017}, significant device heating caused by the electrical power occurs at power densities of $>$\,10\,W/cm$^2$. Consequently, the device does not operate in a steady-state anymore, but the current continuously increases over time if a voltage is applied. At higher power densities of $>$\,100\,W/cm$^2$, only pulsed current-voltage measurements can be carried out due to thermal destruction of devices. Thus, it is important to discuss the effects of thermal heating before going towards high-frequency operation.\\
As shown in \cite{Klinger2018}, OPBTs (as most other organic semiconductor devices) show a positive thermal feedback effect. As an electrical current causes Joule heating, the conductivity as well as the mobility of the organic
semiconductor increase over time due to the hopping-type charge carrier transport. In turn, this effect causes the current to increase further and, hence, accelerates the heating of the devices closing the circle of the so-called positive thermal feedback loop.\\ 
While organic light-emitting diodes or solar cells operate at low power densities, thermal feedback is of utmost importance for organic transistors, which are designed to operate at high frequency and hence high power density. For this reason, a model to describe and parameterize the thermal feedback behavior is developed. As commonly observed in experiments, this model assumes an Arrhenius-like thermal activation of the conductivity $\sigma (T)=\sigma_0F(T)$ with 
\begin{equation}
    F(T)=exp(-\frac{E_{act}}{k_B}(T^{-1}-T_a^{-1}))
\end{equation}
where E$_{act}$ is the thermal activation energy and T$_a$ the ambient temperature. The current-voltage curve of the OPBT is described in this model with a general power law according to
\begin{equation}
    I(V,T)=I_{ref}(\frac{V}{V_{ref}})^\alpha F_1(T)+I_{off}F_2(T),
\end{equation}
where F$_1$(T) and F$_2$(T) account for the different thermal activation in the on- and off-state of the transistor. Furthermore, three parameters describing the steady-state current-voltage curve in the on-state at ambient temperature are introduced: an exponent $\alpha$, a reference point of the curve with the current I$_{ref}$, and the voltage V$_{ref}$. Thermal effects are included, assuming that the power dissipation due to the electrical current has to equal the heat Q which can be transported by the substrate (considering a temperature-independent thermal resistance of the substrate).\\
As shown in Figure \ref{fig:self_heating}, depending on the selected base-emitter voltage (referring to V$_{ref}$), the OPBTs enter the self-heating regime at a current above 1\,mA. Most interestingly, the current-voltage curve even exhibits an S-shaped negative differential resistance (S-NDR), which causes rapid destruction of the device if not driven in with short voltage pulses (pulsed regime). In this case, the temperature in the device might quickly exceed 100$^\circ$C. The model fits the experimental behavior well, and in principle, is suited to predict the steady-state temperature of the device for given pulse duration and duty cycle during the electrical measurement.\\
This kind of thermal feedback loop is usually not observed in lateral organic transistors since the power densities do not reach these high levels. However, for short-channel transistors, the effects of thermal heating need to be taken into account if transistor parameters such as the transition frequency are determined. The most critical parameters which determine the strength and on-set of the thermal feedback are the thermal activation energy of the conductivity E$_{act}$ and the thermal resistance of the substrate $\Theta_{th}$. In particular, in order to be able to operate devices at high power densities, small E$_{act}$ and $\Theta_{th}$ are required. Unfortunately, the semiconductor material used for OPBTs - the buckyball molecule C$_{60}$ shows a very strong thermal activation of transport with E$_{act}$=300\,meV \cite{Klinger2018, Fischer2015} in the investigated range (see Fig.\,\ref{fig:self_heating}b). Presumably, this high value originates from the comparably low charge carrier density in the bulk of the semiconductor (10$^{16}$-10$^{17}$\,cm$^{-3}$) which is shown to significantly decrease for a higher density of charge carriers \cite{Schwarze2019}. Other semiconductor materials such as oligobenzenes or oligobenzothiophenes could offer much lower values \cite{Dong2007} for E$_{act}$ if they could be employed for OPBTs. A lower value of E$_{act}$ would also be beneficial to reduce the requirements concerning the thermal resistance of the substrate $\Theta_{th}$. In particular, aiming for truly flexible devices where in general, the thermal resistance of the substrate is higher than, e.g., on silicon substrates, heat dissipation becomes a problem. Hence, the reduction of the thermal activation of transport in the semiconductor seems to be the only possible path to reach higher current densities and higher frequencies of operation. Unfortunately,  oligobenzenes or oligobenzothiophenes such as pentacene perform poorly in OPBTs due to a different morphology of the permeable base layer. Thus, it remains an open task for the future to find semiconductor materials with low E$_{act}$ and performance in OPBTs comparable to C$_{60}$.
\subparagraph{High-Frequency Operation}\vspace{2mm} With this knowledge about the severe impact of self-heating on the device performance, the devices are optimized in order to evaluate their high-frequency operation, i.e., the transition frequency. Although the name suggests being a dynamic characterization, the hypothesis of the conventional transition frequency measurement is that the device operates in a steady-state. In particular, a DC-voltage is
\begin{figure}[!htb]
	\begin{center}
		\includegraphics[width=.99\textwidth,clip]{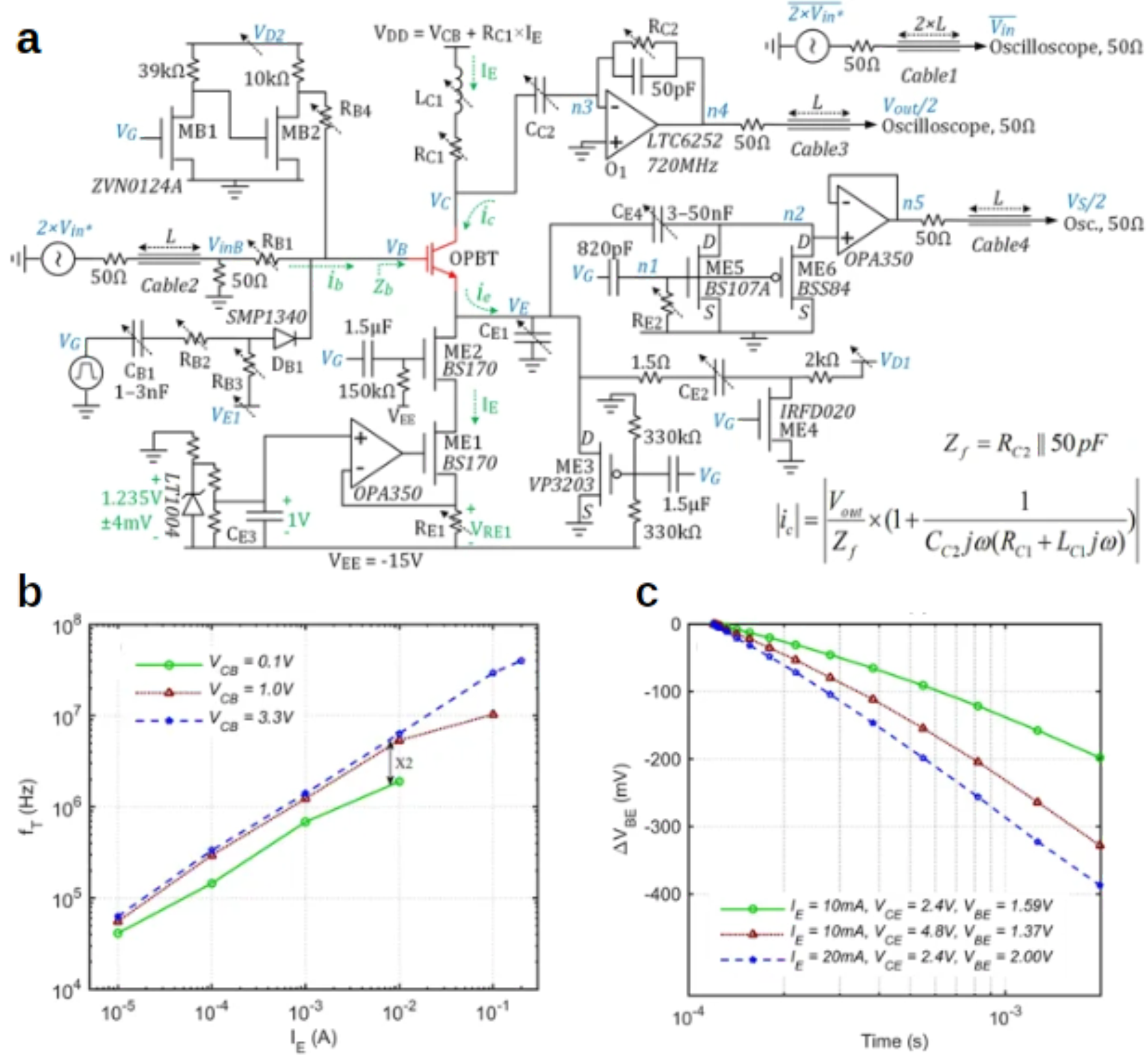}
	\end{center}
	\caption[]{\textbf{a} Simplified schematic of the new pulse-biasing small-signal measurement setup. \textbf{b} Transition frequency versus pulse-biased emitter current for a pulse duration of 10\,\textmu s. \textbf{c} The OPBT at high power density showing a self-heating introduced shift of V$_{BE}$. Figures reprinted from \cite{Boroujeni2018}.}
	\label{fig:OPBT_ft}
\end{figure}
applied between emitter and collector in order to measure the collector current, while a sinusoidal signal is sent to the base to record the base displacement current. If the collector current equals the base displacement current, the unity-gain-condition is reached, and the device operates at its transition frequency.\\
However, as discussed in the previous paragraph, the collector current constantly increases over time due to self-heating, which in turn means that a static transition frequency measurement in the self-heating regime cannot be carried out. For this reason, the group of Prof. Frank Ellinger developed a pulse-biasing characterization circuit which enables the characterization of the device performance at a constant device temperature or even to study the temporal evolution of the performance as self-heating progresses \cite{Boroujeni2018}.\\
The circuit shown in Figure \ref{fig:OPBT_ft}a, developed by B.K. Boroujeni, is capable of applying an accurate current or voltage to the device within less than 10\,\textmu s and measure all important small-signal parameters such as the gain h$_{21}$, the transconductance g$_m$, the intrinsic gain A$_{v0}$, and the transition frequency f$_T$. The devices used for self-heating experiments are optimized to reach high frequency operation. More specifically, the layer thickness of the semiconductor is reduced to 100\,nm, and strong contact doping is employed. Thus, it is possible to drive an emitter current up to 200\,mA (or 500\,A/cm$^2$) through the device during the small-signal characterization. This current is only limited by the maximum power that can be supplied by the measurement circuit.\\
As shown in Figure \ref{fig:OPBT_ft}b, the transition frequency f$_T$ increases linearly with the emitter current, and reaches a value of 40\,MHz at the highest possible current of 200\,mA. This value of 40\,MHz represents a new record value for the transition frequency of organic thin-film transistors. Moreover, it is worth mentioning that this f$_T$ measurement is carried out for the shortest possible pulse duration of 10\,\textmu s and hence the device operates at room temperature without the influence of self-heating.\\
The best way to visualize the self-heating effect on the device performance is to study the drift of the base-emitter voltage $\Delta$V$_{BE}$ over time (cf. Fig.\,\ref{fig:OPBT_ft}c). Here, negative $\Delta$V$_{BE}$ refers to the situation that the same current can be applied at a lower voltage, while positive $\Delta$V$_{BE}$ means that more voltage would be required. As shown in Fig.\,\ref{fig:OPBT_ft}c, there is a negative $\Delta$V$_{BE}$ of several hundreds of meV once the device is turned on. Most interestingly, the magnitude of $\Delta$V$_{BE}$ varies depending on the overall power going through the device, which clearly confirms that self-heating is dominant in this regime and causes the device to operate even more effectively due to the increased conductivity. In principle, this means that an f$_T$ measurement carried out with a pulse duration of 10\,ms would deliver an even larger value than 40\,MHz. A positive $\Delta$V$_{BE}$ is only obtained for low power densities and a pulse duration $\ge$1000\,s which presumably caused by long-term degradation or bias-stress of the device \cite{Dollinger2019b}.\\
Although it might appear as an academic trick to measure the small-signal parameters in a pulsed regime excluding self-heating and bias-stress, there is a multitude of applications where transistors only operate in a pulsed mode. For example, the switching transistors in active-matrix displays are turned on only for several hundreds of nanoseconds while remaining in the off-state for several milliseconds, or the logic gates in a ring oscillator. For these kinds of applications, the pulse-biasing method provides small-signal parameters that describe the operation of the devices more accurately, and hence could be used to improve circuit models.      
\subparagraph{Stability and Device Reliability}\vspace{2mm} The suitability for applications does not only depend on the highest performance a transistor technology can possibly reach but also on the long-term stability during operation. In this regard, organic transistors are often suspected to perform worse \cite{Bobbert2012} than established inorganic thin-film transistor technologies such as low-temperature poly-crystalline silicon (LTPS) or transparent conductive oxides. In particular, n-type organic semiconductors are prone to degradation if they are
\begin{figure}[!htb]
	\begin{center}
		\includegraphics[width=.99\textwidth,clip]{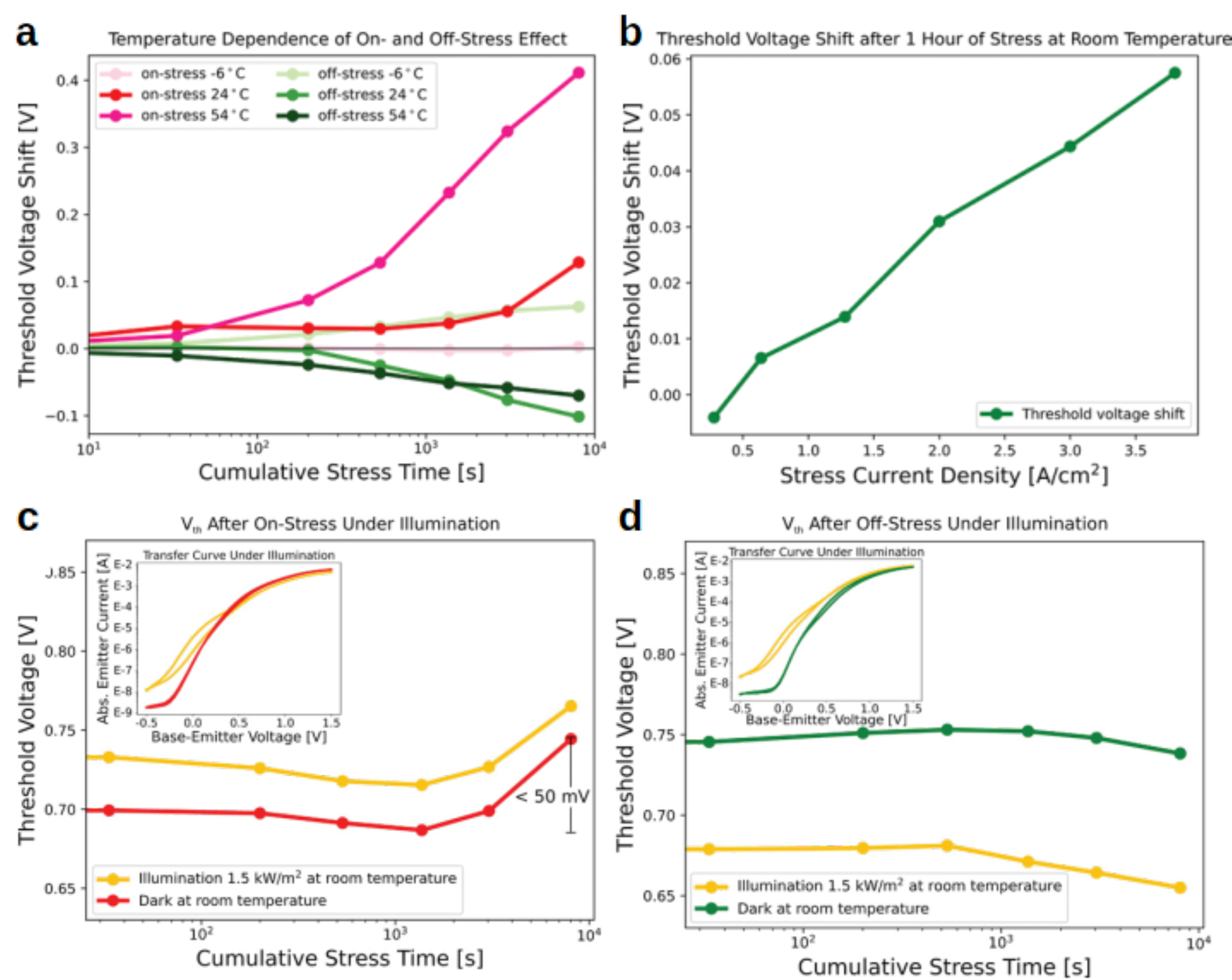}
	\end{center}
	\caption[]{\textbf{a} Threshold voltage shift under on- and off-stress at temperatures varied from -6 to 54\,$^\circ$C. At low temperatures, V$_{Th}$ is stable with marginal changes below 50\,mV. Elevated temperatures activate a larger on-stress \textbf{b} The threshold voltage shift after one hour of stress depending on the stress current. The stress effect scales linearly with the applied current. \textbf{c} and \textbf{d} Development of the absolute value of the threshold voltage V$_Th$ during extended stress with and without strong illumination. \textbf{c} On-stress at V$_{BE}$=1\,V and \textbf{d} off-stress at V$_{BE}$=-1\,V. Insets show changes in the initial transfer curve under illumination. Figures reprinted from \cite{Dollinger2019b}.}
	\label{fig:stability}
\end{figure}
exposed even to the smallest concentrations of oxygen and water. Additionally, due to the low temperature of the substrate during processing (room temperature), a layer such as the native AlO$_X$ or in general metal-semiconductor interfaces are expected to have a high density of defect states \cite{Haeusermann2016, Kalb2007, Geiger2018}. An acceptable degree of operational stability can only be achieved with low defect density interfaces, e.g., prepared using self-assembled monolayers or Teflon-like dielectrics \cite{Kalb2010}. Additionally, proper device encapsulation is required for the stable operation of n-type organic transistors. For p-type OTFTs, excellent stability of operation can be achieved even without encapsulation if semiconductor materials with an ionization energy of $>$ 5.2e\,eV are used \cite{Jia2018}. In this case, the stability of organic TFTs can be even on-par with other commercial TFT technologies.\\
In order to understand whether the exceptional electrical performance of OPBTs is only reached on the expenses of reduced operational stability, a detailed stability analysis under varying operation conditions is performed \cite{Dollinger2019b}. This analysis includes the stability in the on-state and off-state, at elevated temperature, as well as under illumination.\\
The results of this stability analysis are summarized in Figure \ref{fig:stability}. As shown in Figure \ref{fig:stability}a, stressing the device by setting it permanently to either the on- or off-state causes a shift of the threshold voltage with the opposite sign. Thus, the device can be recovered after long on-state stress by setting it to the off-state. This recovery also occurs without biasing the devices but at a lower recovery rate. The strength of the threshold voltage shift is only weakly affected by the biasing conditions and current density (cf. Figure \ref{fig:stability}b). Instead, it mainly depends on the external temperature, and a strong increase in the threshold voltage shift is observed for elevated temperature. Of course, operating the device at current densities where
\begin{figure}[htb]
	\begin{center}
		\includegraphics[width=.99\textwidth,clip]{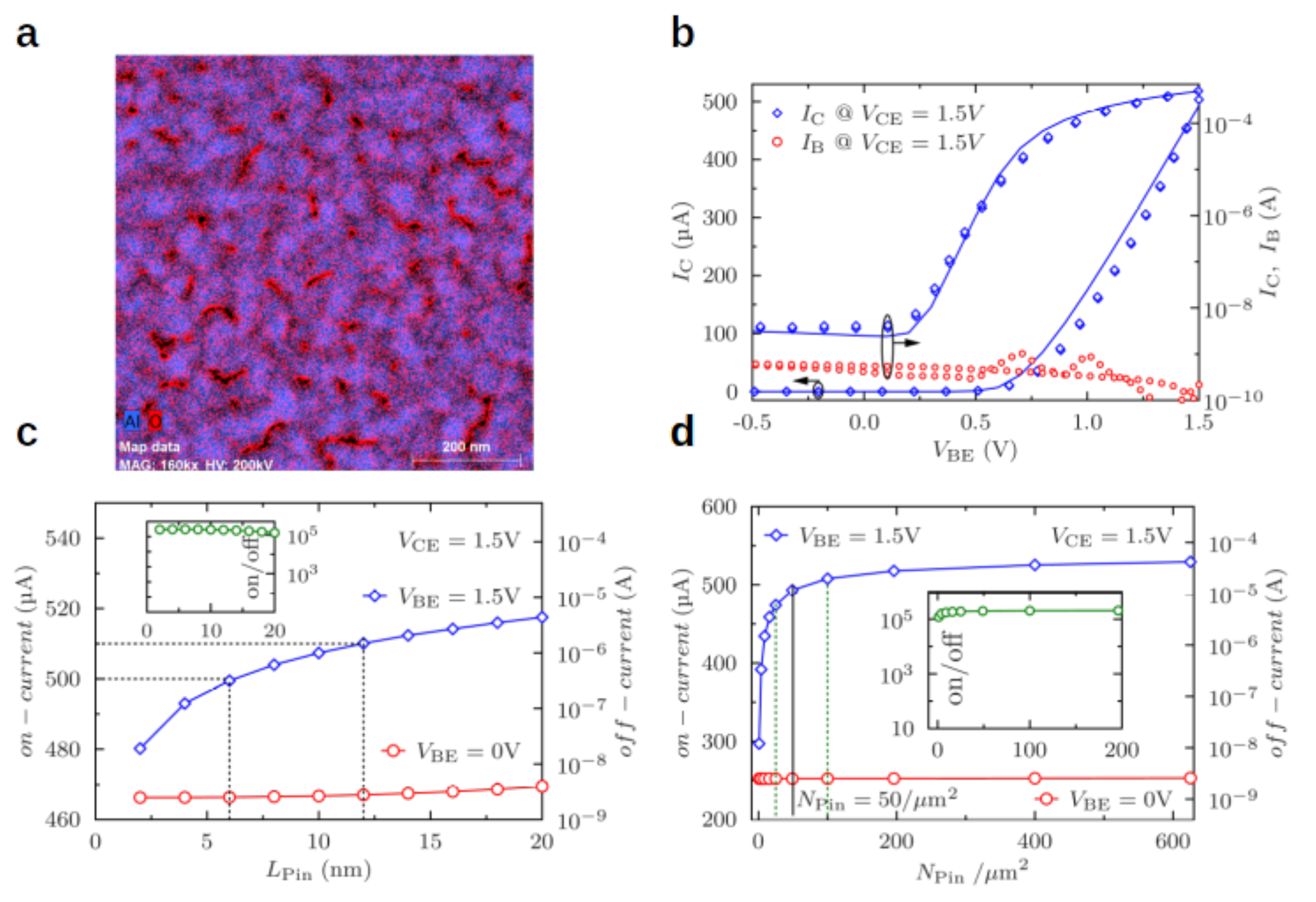}
	\end{center}
	\caption[]{\textbf{a} Energy-dispersive X-ray spectroscopy based element maps of the base layer showing well passivated pinholes in the metallic film. Superimposed Al and O distributions, proving that the sample is uniformly oxidized and
that the pinholes are free of both elements. \textbf{b} Collector current I$_C$ vs. the base-emitter voltage V$_{BE}$ as obtained by experiment (symbols) compared with data from TCAD simulation (solid lines) including
the base leakage current I$_B$ (circles) for a collector-emitter voltage of V$_{CE}$=1.5\,V. \textbf{c} and \textbf{b} On- and Off-state current obtained from simulation for varying pinhole diameter L$_{Pin}$ and pinhole density N$_{Pin}$ (inset shows the I$_{on}$/I$_{off}$). Images reprinted from \cite{Dollinger2020}.}
	\label{fig:Reliability}
\end{figure}
self-heating occurs would further accelerate the performance degradation. Most interestingly, a much stronger thermal activation of the bias-stress is obtained in the on-state. From this finding and the fact that the bias-stress is reversible, it can be concluded that the threshold voltage shift mainly originates from electromigration effects (e.g., defect migration in the AlO$_X$ layer due to high electric fields) rather than trapping or de-trapping of charge carriers. This conclusion is further supported by the fact that bias-stress is not influenced by illumination either in the on- nor off-state (cf. Fig.\,\ref{fig:stability}c and d), and hence, it can be ruled out that recombination and trapping significantly contribute to the bias-stress.\\
Overall, the strength of the threshold voltage shift at room temperature, which is observed even at high current densities and strong illumination of 1.5\,kW/cm$^2$, remains below 100\,mV even for several hours of stress. This excellent stability is presumably caused by two effects: 1) the device encapsulation using a rigid glass lid and 2) the required post-fabrication annealing step at 180\,$^\circ$C which is necessary in order to obtain the desired morphology of the base electrode. In particular, the annealing step might lead to a reduced density of defects in the AlO$_X$ which is essential for operational stability.\\
\\
Another possible problem that is often discussed with regard to short-channel transistors or devices that rely on
self-assembly of layers such as the base electrode in OPBTs, is the device reliability and scaling. In particular, for OPBTs the question arises whether the size and distribution of the pinholes in the base electrode have a substantial influence on the electrical characteristics of the devices. For this reason, the morphology of the base electrode is analyzed in order to determine the density, size, and distribution of pinholes. Furthermore, together with Dr. Darbandy, advanced 3D TCAD simulations are used (Technology Computer Aided Design) to study whether variations in pinhole size and distribution have a significant influence on the on- or off-state of the OPBT.\\
In Fig.\,\ref{fig:Reliability}a an elemental map for aluminum and oxygen obtained by energy-dispersive X-ray spectroscopy (EDXS) in a transmission-electron-microscope (TEM) is shown. The image clearly shows that the aluminum base electrode is not a closed thin film but rather porous and covered with pinholes. Close to the pinholes, a stronger signal for the oxygen channel is observed, indicating the formation of AlO$_X$. Using the EDXS-TEM image, the density of pinholes is determined to be 54\,\textmu m$^2$ and the majority of pinholes has a size between 25 and 65\,nm$^2$. Only very few pinholes with a size $\ge$\,100\,nm$^2$ are obtained \cite{Dollinger2020}.\\
Using this information about the pinhole geometry and adding further parameters describing the charge carrier transport in the semiconductor \cite{Fischer2012b, Klinger2017, Kaschura2016}, the field and charge carrier distribution in the device as well as the current-voltage curve are simulated using the sentaurus TCAD solver \cite{Synposes2018} (cf. Figure \ref{fig:Reliability}b). The model is fully able to quantitatively describe the operation of the device in the off-state, subthreshold regime, and on-state.\\
In order to evaluate the influence of the base morphology on the device performance, a systematic variation of the pinhole parameters (density N$_\mathrm{Pin}$ and size L$_\mathrm{Pin}$) is carried out. Hereby, for the sake of simplicity, cylindrical pinholes with a diameter of L$_{\mathrm{Pin}}$ are assumed. As it can be seen in Figure \ref{fig:Reliability}c and d, neither the on-state nor off-state current are strongly affected by the size or density of pinholes as obtained by the experiment. The pinhole density only becomes relevant for the on-state current for N$_{Pin}\,\le$\,10\,/\textmu m$^2$, because the overall current is limited by the charge carrier density in the pinholes rather then the injection of charge carriers and resistance of the bulk semiconductor material \cite{Kaschura2016}. The off-state current is expected to increase for L$_{Pin}\ge$\,30\,nm since the base-potential and the capacitance of the AlO$_X$ are not large enough to fully deplete the semiconductor by means of the field-effect. However, comparing this finding to the experimentally obtained size distribution where only very few pinholes with a diameter $\ge$\,10\,nm have been observed, it can be concluded that the distribution of pinhole diameter has no impact on the off-state current by practical means.\\
Overall, this study shows that although OPBTs rely on a self-assembly process, the device performance is robust in a large range against parameter variations of the base electrode.

\begin{figure}[htb]
	\begin{center}
		\includegraphics[width=.99\textwidth,clip]{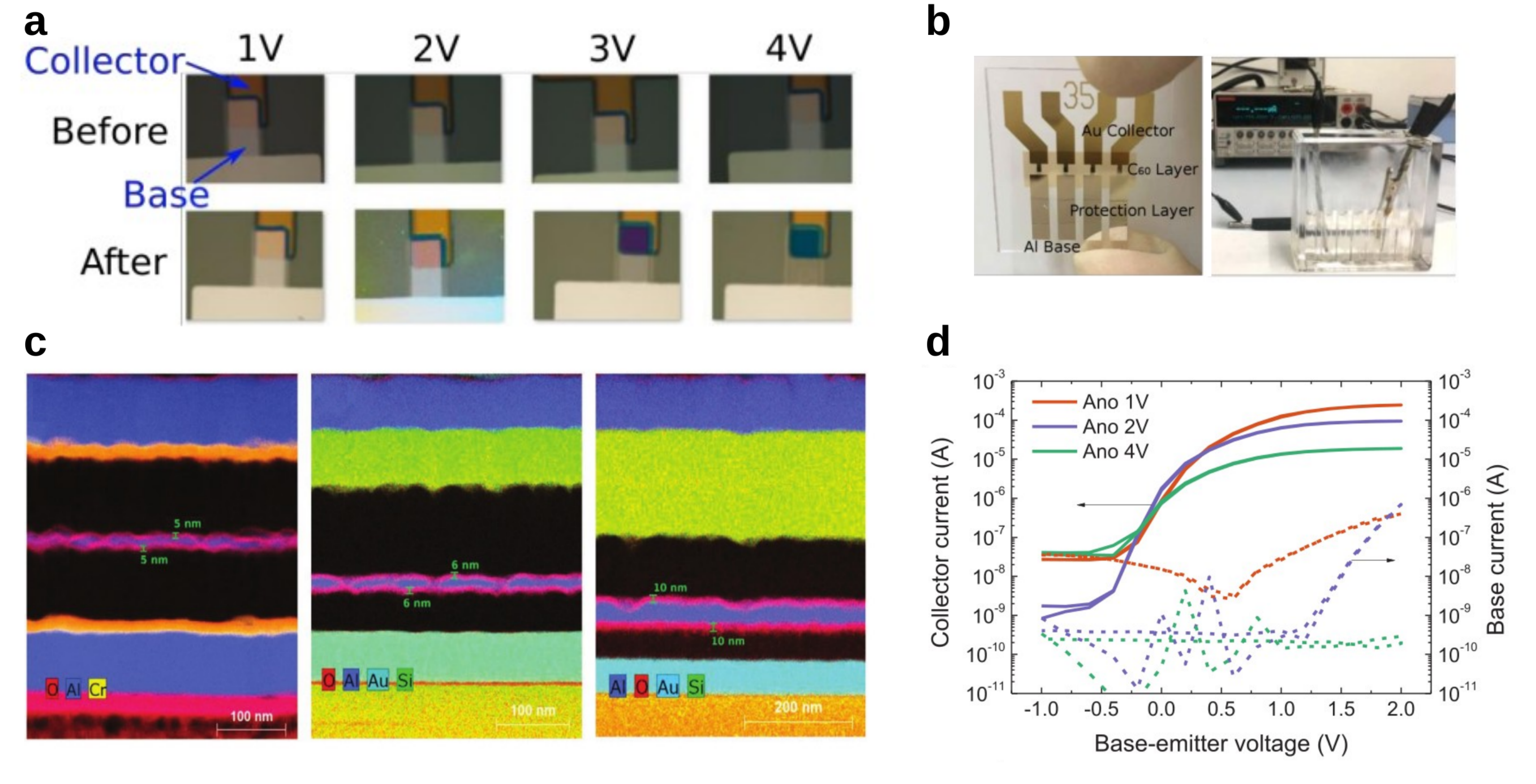}
	\end{center}
	\caption[]{\textbf{a} Photograph of the base electrode before and after anodization at given anodization voltage. With increasing voltage, the base electrode becomes increasingly transparent proving the formation of AlO$_X$. \textbf{b} Experimental setup for the wet-chemical anodization. \textbf{c} Energy-dispersive X-ray spectroscopy on a TEM cross-section showing the formation of the AlO$_X$ layer for different anodization conditions (from left to right): oxidation by air, anodization at 2\,V (50\,nm base thickness). \textbf{d} Collector current I$_C$ and base current I$_B$ vs. the base-emitter voltage V$_{BE}$ for different anodization conditions (ANO) from 1\,V to 4\,V. Images taken from \cite{Dollinger2019}.}
	\label{fig:anodization}
\end{figure}

\subparagraph{Wet-Chemical Anodization for Base Layer Formation}\vspace{2mm} Although the base layer morphology does not critically influence the OPBT reliability, the fabrication process based on self-assembly is still unsatisfying. For technical reasons it is desirable to be able to control the thickness of the AlO$_X$ layer as well as the overall thickness of the base layer. In particular, relying on the self-assembly process, the base layer thickness is limited to $\le$\,20\,nm due to the suppression of pinhole formation for thicker films. Furthermore, the thin AlO$_X$ restricts the transistor gain due to undesirable base leakage currents.\\
In order to overcome these restrictions, new fabrication techniques for the base electrode are explored in this work.\\
Since the formation of pinholes is caused by strain during the oxidation process, additive techniques to form AlO$_X$, such as atomic layer deposition, cannot be employed. Stronger oxidation techniques such as plasma-assisted oxidation are also not suitable due to the self-limiting growth of AlO$_X$ and the destructive character of the plasma environment to the organic semiconductor material. Wet-chemical anodization though, allows for precise thickness control of the oxide layer via the applied voltage, and has been shown to create high-quality dielectric interfaces for organic thin-film transistors \cite{Kaltenbrunner2011}. However, 
wet-chemical anodization has not been carried out yet atop of an organic semiconductor material because of concerns with regards to degradation caused by the harsh oxidation environment.\\
Despite these concerns, wet-chemical anodization is employed in order to control the formation and thickness of the AlO$_X$ layer at the base electrode. For this purpose, the OPBT fabrication under ultra-high-vacuum conditions (UHV) is interrupted after the base layer, and the structure is dipped into a beaker containing citric acid (1\,mM/L in water) and two electrodes to apply the anodization potential (cf. Figure \ref{fig:anodization}a and b). If a potential between collector and base is applied for 60s, the base electrode becomes increasingly transparent with
increasing anodization potential, indicating the formation of AlO$_X$ (Figure \ref{fig:anodization}a). As shown by energy-dispersive X-ray spectroscopy in a TEM cross-section (Figure \ref{fig:anodization}c), the thickness of the AlO$_X$ is increased upon anodization from approximately 5\,nm to more than 10\,nm for an anodization voltage of 4\,V. Due to this increased thickness of the AlO$_X$ passivation, the base leakage current in anodized OPBTs is substantially reduced to values $\le$\,0.1\,nA (limited by the resolution of the current measurement, see Figure \ref{fig:anodization}d). In consequence, the anodization technique leads to exceptionally high transmission factors (collector current divided by emitter current) of 99.9996\% corresponding to a current gain of 2.5$\times$10$^5$. This value is 2-3 orders of magnitude better than the state-of-the-art for OPBTs oxidized by air.\\
Most strikingly though, wet-chemical anodization also enables thicker base electrode layers to be used in OPBTs. Due to the increased strain during wet-chemical anodization compared to anodization by air, the layer thickness can be increased up to 50\,nm without a significant drop of the current density \cite{Dollinger2019}.\\
However, the question remains why wet-chemically anodization can be seamlessly employed for OPBTs without severe degradation?\\
The answer to this question contains two important aspects. Firstly, the semiconductor material used for OPBTs (C$_{60}$) possesses a comparably high ionization potential of approximately 7\,eV, which leads to only the aluminium being anodized but not the organic semiconductor if the anodization potential is kept below 4\,V. Secondly, although C$_{60}$ is a material whose electrical performance severely degrades upon oxygen or water exposure, it can be recovered by an annealing step under inert conditions which allows oxygen and water to desorb from the layer. In particular, oxygen and water are only physisorbed by C$_{60}$ and a chemisorption is not obtained. Thus, an annealing step at 150\,$^\circ$C for two hours is sufficient to fully recover the inert charge carrier transport properties of C$_{60}$.\\
Overall, the wet-chemical anodization technique is a powerful method that enables full process control on the AlO$_X$ thickness and renders the possibility to employ base electrode thicknesses of up to 50\,nm without compromising the performance of the transistors. 

\begin{figure}[htb]
	\begin{center}
		\includegraphics[width=.99\textwidth,clip]{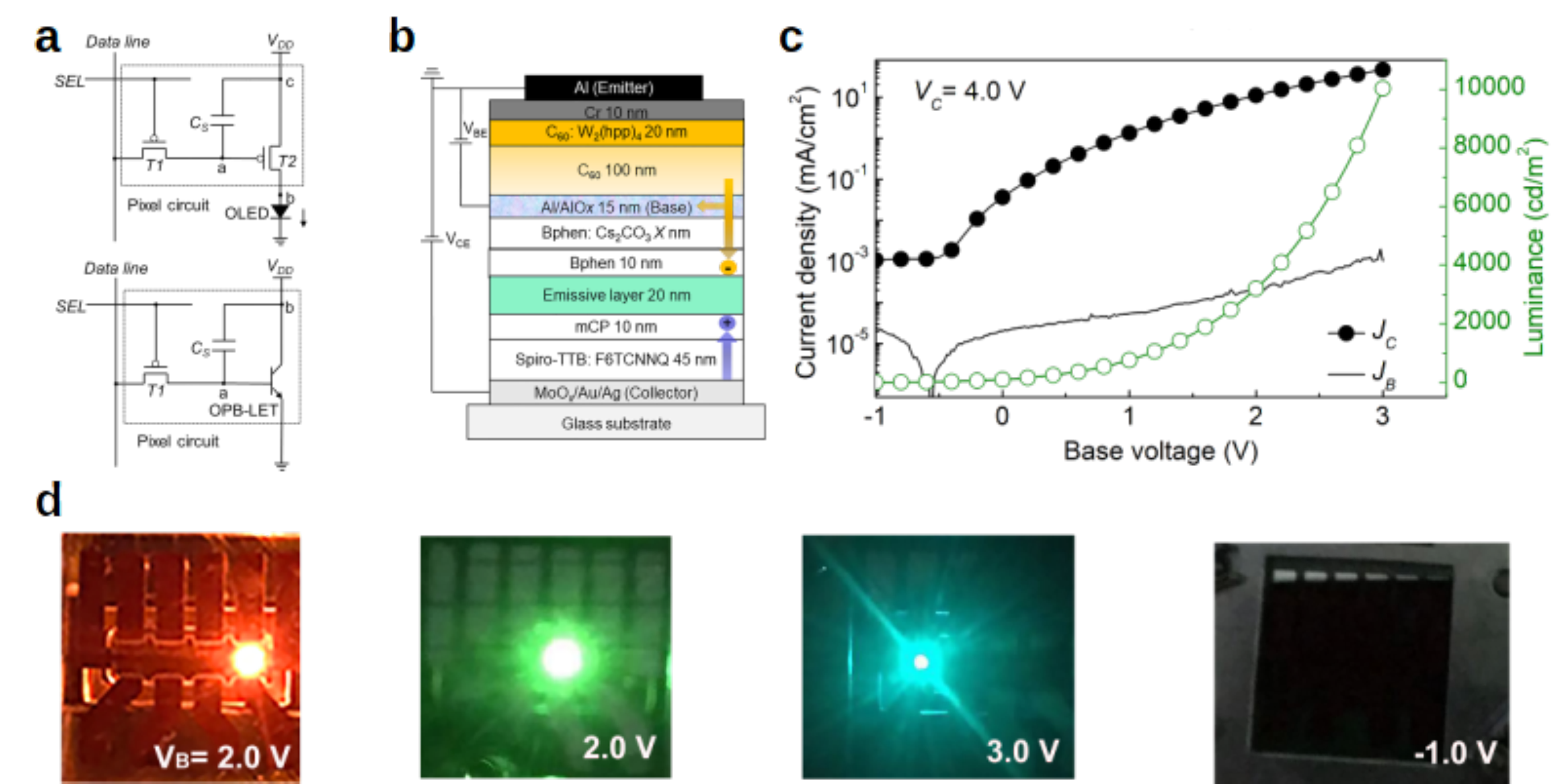}
	\end{center}
	\caption[]{\textbf{a} Pixel circuit of traditional 2T1C digital drive scheme for active-matrix OLED (AMOLED) and circuit of organic permeable base light-emitting transistors (OPB-LET) used for AMOLED. \textbf{b} Schematic device structure and hole/electron injection paths of an OPB-LET. The basement of the device (i.e., the substrate, emitter, collector, base and channel layers) is the same for each color device. The chemical structure and full names of all molecules are given in \cite{Wu2020}. \textbf{c} Transfer characteristics of an OPB-LET with a green emitter molecule. \textbf{d} Photographs of a red, green, and blue OPB-LET at the indicated base-emitter voltage. Additionally, the green OPB-LET is shown in its off-state. Figures taken from \cite{Wu2020}.}
	\label{fig:LOPBT}
\end{figure}

\begin{figure}[htb]
	\begin{center}
		\includegraphics[width=.99\textwidth,clip]{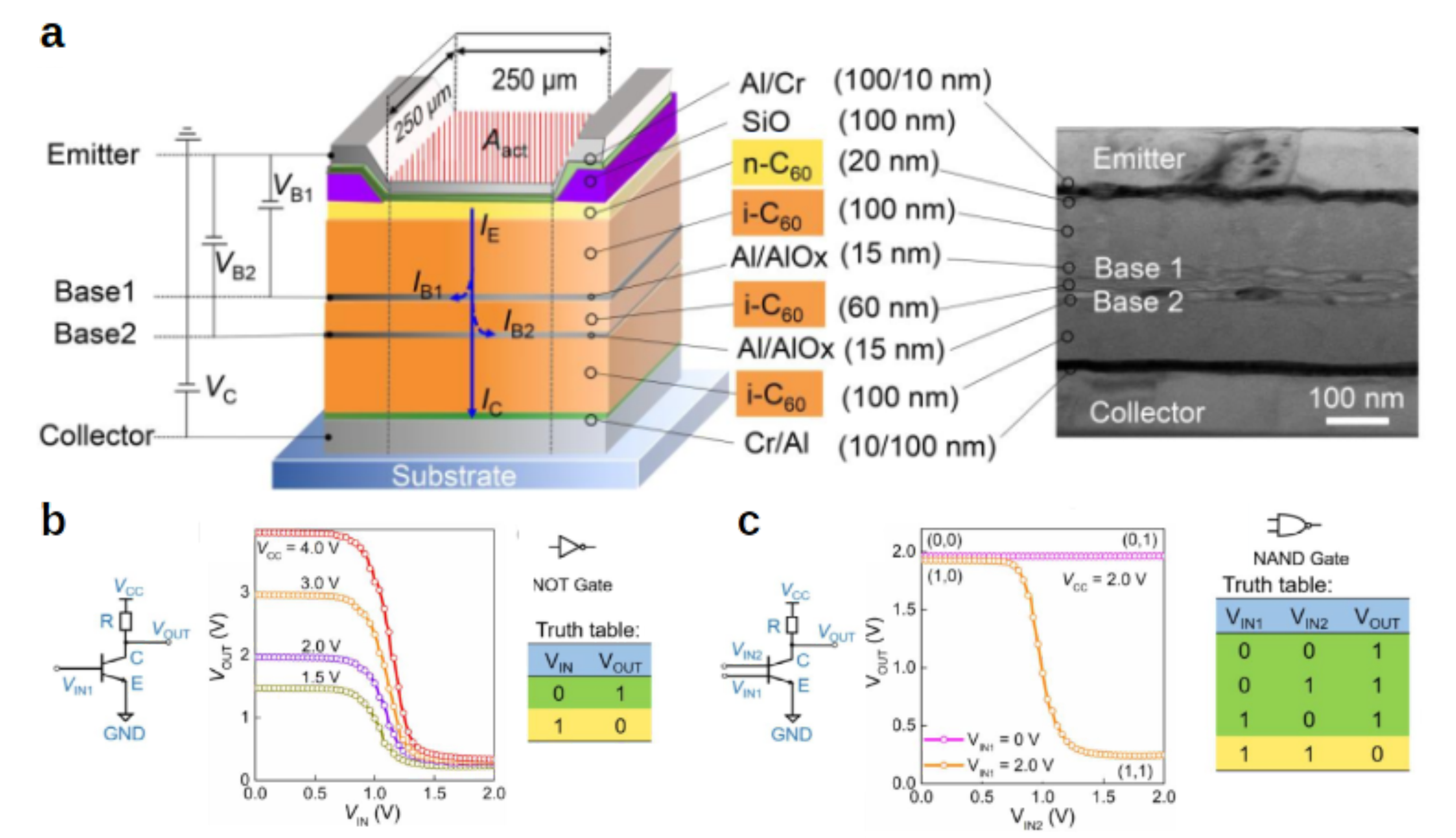}
	\end{center}
	\caption[]{\textbf{a} Scheme of an DB-OPBT measured using common emitter configuration, and cross-section TEM image of an DB-OPBT.  \textbf{b} and \textbf{c} Circuit diagram, transfer characteristics, and truth table for a NOT and NAND realized by a DB-OPBT. Figures taken from \cite{Guo2020}.}
	\label{fig:DB-OPBT}
\end{figure}

\begin{figure}[htb]
	\begin{center}
		\includegraphics[width=.99\textwidth,clip]{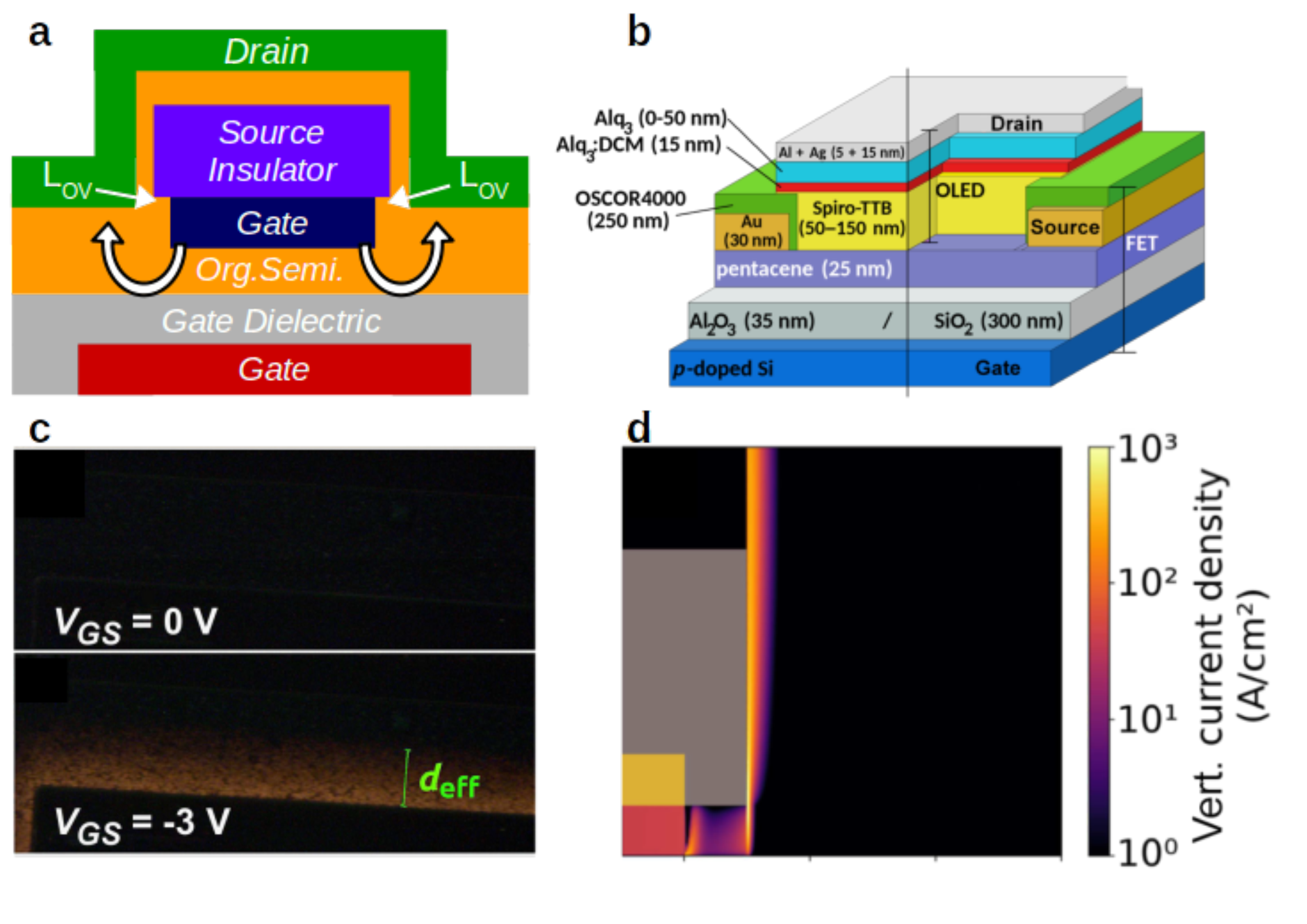}
	\end{center}
	\caption[]{\textbf{a} Scheme of a VOFET. The direction of current flow is indicated by white arrows. \textbf{b} Sample   structure and stack of experimental VOLET  devices. The full names of the chemical compounds are given in \cite{Sawatzki2018} \textbf{c}  Light   distribution measurement   of   a VOLET  (V$_{DS}$\,=\,-10\,V). Upper panel: V$_{GS}$\,=\,0\,V (off-state). Lower panel: V$_{GS}$\,=\,-3\,V (on-state). \textbf{d} Vertical component of the current density as obtained from 2-dimensional drift-diffusion simulations. The current distribution is shown for the on-state (V$_{GS}$\,=\,-2.5\,V). Figures reprinted from \cite{Sawatzki2018}.}
	\label{fig:VOFET_1}
\end{figure}

\subparagraph{Applications of OPBTs}\vspace{2mm} Having the performance and fabrication process of OPBTs optimized, this paragraph focuses on the application of OPBTs. In particular, novel embodiments of the device are developed in order to make use of the unique vertical architecture of the OPBT. The first structure proposed employs OPBTs as an efficient light-emitting transistor, while the second structure utilizes the vertical configuration for multi-state logic switches.\\
\\
\textbf{Organic Permeable Base Light-Emitting Transistors:} In an active-matrix light-emitting diode display, the light-emitting component is controlled by at least two transistors denoted as driving and switching transistor. While the switching transistor is used to program the brightness information on the pixel, the driving transistor acts as a constant current source to the light-emitting element. If the driving transistor could be combined with the light-emitting component in one element, the active area of light-emission could be vastly increased, improving the overall power-efficiency of the display. For this reason, light-emitting organic transistors have obtained much attention \cite{Muccini2006}. In particular, vertical organic transistors are interesting as a light-emitting device due to the optical cavity character which is inherent to these devices \cite{Mccarthy2010}.\\
For the organic permeable base light-emitting transistor (OPBT-LET) a pin-OLED is incorporated into the collector side of the OPBT structure (cf. Figure \ref{fig:LOPBT}a). Hence, the current passed from the emitter to the collector can be controlled via the base potential, which is used as a tuning knob to control the light emission. On the collector side a hole injection layer is used in order to supply holes to the pin-OLED for radiative recombination. The thick and fully reflective emitter electrode together with the thin collector electrode (total thickness of 10\,nm) is used to define the optical cavity of the system. Additionally, the thin base electrode is utilized to define a second cavity, which adds an additional degree of freedom for the device design.
Using this stack configuration, all-vacuum-processed, vertical organic light-emitting transistors operating at low driving voltages, with unprecedentedly high efficiencies and luminance in three primary colors are demonstrated (cf. Figure \ref{fig:LOPBT}b). The fabricated red, green and blue devices, operating at driving voltages of below 5.0\,V, reach peak external quantum efficiencies of 19.6\,\%, 24.6\,\% and 11.8\,\%, and current efficiencies of 20.6\,cd/A, 90.1\,cd/A, and 27.1\,cd/A along with maximum luminance of 9,833\,cd/m$^2$, 12,513\,cd/m$^2$ and 4,753\,cd/m$^2$, respectively \cite{Wu2020}. These values clearly stand out in all regards. In particular, the current efficiencies, highest luminance as well as the external quantum efficiencies surpass previously published light-emitting transistors by more than a factor of 2. These high efficiencies became possible due to the tunable microcavity formed between collector and base, and due to the highly efficient injection of electrons into the pin-OLED. Furthermore, the base electrode can effectively regulate the charge injection and exciton formation, allowing for a high optical contrast of $\ge$10$^5$. The most striking difference though is the exceptionally small driving voltage, which is comparable to the individual reference OLEDs. These small voltages are achieved because, as previously shown, the OPBT can carry currents in the order of A/cm$^2$, while the point of operation of the pin-OLED is in the range of tenths of mA/cm$^2$. Hence, there is no additional voltage drop across the emitter side of the OPBT.\\
\\
\textbf{Dual-Base Organic Permeable Base Transistors:} Another possible application, which is costly to realize with lateral TFTs, but straight-forward with OPBTs, are multi-state logic switches. For example, the implementation of a NAND-element requires at least two lateral TFTs with separated gate electrodes. However, this implementation consumes a significant amount of area which makes it expensive for wafer integration.\\
In this regard, the OPBT architecture facilitates the use of dual- or even multi-base arrangements. Figure \ref{fig:DB-OPBT}a depicts such a dual-base organic permeable base transistor (DB-OPBT) where emitter and collector are separated by two independently addressable base electrodes. Thus, apart from the access electrodes, the DB-OPBT has the same areal footprint as a common OPBT but offers the possibility to act as a multi-state logic switch \cite{Guo2020}.\\
Due to the high gain of the OPBTs, either base electrode in the DB-OPBT can be employed to switch the OPBT from the on- to the off-state independently. Furthermore, the implementation of multiple base electrodes does not affect the overall current density, which can be driven by the OPBT, and gain factors of DB-OPBTs reach similar values as for single-base OPBTs. Thus, implementing two base electrodes in one OPBT, several logic functions are realized, among them the NOT, NAND, and AND gate function (cf. Figure \ref{fig:DB-OPBT}b and c). Furthermore, connecting two DB-OPBT enables an OR or XOR function.
\paragraph{Vertical Organic Field-Effect Transistors}\vspace{2mm}
An alternative vertical transistor configuration is the vertical organic field-effect transistor (VOFET). In contrast to the OPBT, the VOFET does not resemble a vacuum triode, but its design is closer to the conventional lateral thin-film transistor. However, in contrast to lateral TFTs, the source and drain electrodes in the VOFET are vertically stacked and separated by the semiconductor and an additional insulating spacer (source-insulator, cf. Figure \ref{fig:VOFET_1}a). Due to this vertical arrangement, an ultra-short channel with a length of $\le$\,300\,nm is formed, which facilitates high on-state current densities and possibly also high-frequency operation. One notable advantage of VOFETs over OPBTs is their structural similarity to lateral TFTs which renders the possibility to apply well-established TFT fabrication methods such as lithography, wet-etching, printing, etc.. Consequently, VOFETs are well-suited for high-density integration into more complex circuitry. Due to this fact, VOFETs are a very active field of research, and open questions are mainly linked to the device fabrication (i.e., the definition of the overlap of source and source-insulator), the basics of device operation, and the optimization for high-frequency operation.
During my research, I focused on all three aspects, which led to the following list of publications \cite{Sawatzki2018, Kneppe2021, Kleemann2020a, Lim2020, Fan2019, Hoppner2019}:
\begin{itemize}
     \item \begin{quote}
    \textit{"Organic Transistors Approaching Operation at 100MHz" by M. H\"oppner, D. Kneppe, B.K. Boroujeni, F. Ellinger, K. Leo, \& H. Kleemann, \textbf{in preparation}, (2021).}
    \end{quote}
      \item \begin{quote}
    \textit{"Solution-processed pseudo-vertical organic transistors based on TIPS-pentacene" by D. Kneppe, F. Talnack, B.K. Boroujeni, C. da Rocha, M. H\"oppner, A. Tahn, S. Mannsfeld, F. Ellinger, K. Leo, \& H. Kleemann, Materials Today Energy \textbf{21}, 100697 (2021).}
    \end{quote}
    \item \begin{quote}
    \textit{"Oxide-Semiconductor based Vertical Field-Effect Transistors" by J. Kim, B.K. Boroujeni, F. Ellinger, K. Leo, \& H. Kleemann, \textbf{in preparation}, (2021).}
    \end{quote}
    \item \begin{quote}
    \textit{"Anodization for Simplified Processing and Efficient Charge Transport in Vertical Organic Field‐Effect Transistors" by K.G. Lim, E. Guo, A. Fischer, M. Qiao, K. Leo, \& H. Kleemann, Advanced Functional Materials \textbf{30}, 2001703 (2020).}
    \end{quote}
    \item \begin{quote}
    \textit{"Megahertz operation of vertical organic transistors for ultra-high resolution active-matrix display" by H. Kleemann, G. Schwartz, S. Zott, M. Baumann, \& M. Furno, Flexible and Printed Electronics \textbf{5}, 014009 (2020).}
    \end{quote}
     \item \begin{quote}
    \textit{"Precise patterning of organic semiconductors by reactive ion etching" by M. H\"oppner, D. Kneppe, H. Kleemann, \& K. Leo, Organic Electronics \textbf{76}, 105357 (2020).}
    \end{quote}
      \item \begin{quote}
    \textit{"Method for producing an organic transistor and organic transistor" by H. Kleemann, G. Schwartz, \& J. Blochwitz-Nimoth, US Patent, number 10497888 (2019).}
    \end{quote}
    \item \begin{quote}
     \textit{"Method for producing a vertical organic field-effect transistor, and vertical organic field-effect transistor by H. Kleemann, \& G. Schwartz, US Patent, number 10170715 (2019).}
    \end{quote}
     \item \begin{quote}
    \textit{"High-performance ultra-short channel field-effect transistor using solution-processable colloidal nanocrystals" by X. Fan, D. Kneppe, V. Sayevich, H. Kleemann, A. Tahn, K. Leo, V. Lesnyak, \& A. Eychm\"uller, Physical Chemistry Letters \textbf{10}, 4025 (2019).}
    \end{quote}
    \item \begin{quote}
   \textit{"Balance of Horizontal and Vertical Charge Transport in Organic Field-Effect Transistors" by F.M. Sawatzki, D.H. Doan, H. Kleemann, M. Liero, A. Glitzky, T. Koprucki, \& Leo, Physical Review Applied \textbf{10}, 034069 (2018). }
\end{quote}
\end{itemize}
\subparagraph{Understanding the Operation of Vertical Organic Field-Effect Transistors}\vspace{2mm}
The device scheme of a VOFET is shown in Figure \ref{fig:VOFET_1}a. A severe conceptional and technological hurdle in VOFET fabrication is the definition of the overlap between the source electrode and the source-insulator L$_{OV}$ (cf. Fig.\,\ref{fig:VOFET_1}a). Ideally, this overlap length should be as small as possible since it acts as an additional lateral channel, limiting the overall current density. However, in practice a non-vanishing overlap is always given, e.g., due to alignment tolerances of masks \cite{Sawatzki2018, Kleemann2013}, under-etching effects \cite{Kleemann2020a}, or shadowing effects due to an unfavorable mask-substrate distance \cite{Gunther2016}. As a matter of fact, this undesirable overlap has been shown to vary between a few hundreds of nanometer to several micrometers \cite{Kleemann2020}. Consequently, speaking more strictly, the channel is not truly vertical but rather a complex 2-dimensional object with lateral and vertical contributions. Thus, in order to be able to predict further improvements, it is important to understand the distribution of charge carriers and the electric field within the device.\\
In order to investigate and literally visualize the charge carrier distribution, a pin-OLED is integrated into the VOFET architecture, resulting in a vertical light-emitting transistor (VOLET) (cf. Figure \ref{fig:VOFET_1}b). Using an optical microscope, the lateral spread of the current underneath the drain electrode is recorded as a function of drain-source V$_{DS}$ and gate-source voltage V$_{GS}$. These dependencies are used to develop a quantitative model describing the 2-dimensional charge carrier distribution \cite{Sawatzki2018}.\\
Experimentally, an exponentially decaying light-emission starting from the edge of the source electrode is observed. This finding allows for the definition of an effective lateral channel depth d$_{eff}$, which is a measure for the current spreading underneath the drain electrode (cf. Figure \ref{fig:VOFET_1}c). Furthermore, experimentally a linear dependence of d$_{eff}$ on V$_{GS}$ and V$_{DS}$ is obtained, which is used for the validation of 2-dimensional drift-diffusion simulations which are carried out by the collaborators at WIAS Berlin. It should be noted that the lateral current spreading is as large as several tenths of micrometers in the VOLET devices. This unacceptably large current spreading is mainly caused by the low charge carrier mobility of the electron-transport layer (approx. 10$^{-4}$\,cm$^2$/(Vs)) of the light-emitting diodes compared to the hole mobility in the underlying layer of pentacene (0.45\,cm$^2$/(Vs)). As it will be shown, using a uniform charge carrier mobility, the current spreading is reduced to less than 100\,nm.\\
To demonstrate this effect, two-dimensional drift-diffusion simulations are employed to obtain the spatial charge carrier distribution. For the simulation, a uniform but anisotropic charge carrier mobility for the benchmark material pentacene is taken into account (by choosing the lateral mobility a hundred times larger than the vertical mobility \cite{Gunther2015}). Exemplary, a cross-section through the device, visualizing the spatial distribution of the vertical current density in the on-state is depicted in Figure \ref{fig:VOFET_1}d. While in lateral TFTs, charge carriers predominantly move along the gate dielectric interface (except the areas underneath the source and drain electrode), charge carriers are no longer restricted to the gate dielectric interface in VOFETs. In fact, in VOFETs the vertical electric field pulls up charge carriers to accumulate in a second channel underneath the source-insulator as indicated by the vertical fin in the current density next to the source electrode. The density of charge carriers depends on the thickness of the source-insulator and the strength of the drain-source field. Apparently, this second channel's appearance with an increased charge accumulation for higher drain-source voltage might explain the loss of saturation, which is often observed in VOFETs \cite{Kleemann2013}. The most striking finding from the simulations though, is the formation of the vertical channel adjacent to the source-insulator with a current spreading in the lateral direction of less than 100\,nm. This finding is even more surprising, knowing that the vertical charge carrier mobility in the semiconducting layer is a hundred times smaller than in lateral direction. The reason for the formation of such a narrow vertical channel lies in the field distribution. In particular, based on the simulation, it can be predicted that the electric field's vertical component is several orders of magnitude larger than the lateral component. Hence, there is a minimal driving force for free charge carriers to move in the lateral direction, which ultimately causes the strong vertical confinement of the channel (vertical electric field $\sim$0.5\,MV/cm).
Overall, simulations and experiments provide four important insights into the operation of VOFETs:
\begin{enumerate}
    \item despite the anisotropy of charge carrier mobility, the spreading of the vertical channel in lateral direction is in the range of 100\,nm,
    \item the charge carrier transport in the lateral channel underneath the source-insulator is not restricted to the gate dielectric interface but rather a second channel at the source-insulator is formed due to the drain-source voltage,
    \item due to the vertical drain-source field, the lateral channel underneath the source-insulator can carry more charges than in an equivalent lateral TFT,
    \item and finally, due to the strong vertical drain-source field/ low lateral field, the lateral channel might limit the average current density in the device although the lateral channel is shorter than the vertical channel.
\end{enumerate}
\subparagraph{New Fabrication Processes for Vertical Organic Field-Effect Transistors}\vspace{2mm}
As mentioned above, the precise and reliable definition of the lateral overlap of the source-insulator over the source electrode represents a substantial technological hurdle. While methods such as double-layer lithography \cite{Sawatzki2018} and shadow mask patterning \cite{Gunther2016} enable precise and reliable control over L$_{OV}$, the practically achievable overlap length is larger than 2-3\textmu m which is undoubtedly too long to justify the name vertical transistor. Contrary to that, methods such as lift-off \cite{Kleemann2013} or chemical wet-etching \citep{Kleemann2020a} facilitate VOFET structures with an overlap in the nanometer range. However, these methods are difficult to control in experiments since, for example, the under-etching strongly depends on parameters such as layer adhesion, etchant distribution, heat dissipation during etching, feature size, and many more. Furthermore, etching methods and lift-off bear the problem that thick layers of photoresist (typically $\ge$\,400\,nm) are used as a source-insulator. Such high step-edges though, might cause difficulties during the following process steps, e.g., non-conformal coverage during spin-coating or metal deposition. Thus, there is an obvious need to develop alternative fabrication methods which allow for a precise and reliable definition of the source-insulator overlap in the sub-micrometer range and layer thicknesses of the source-insulator of less than 100\,nm.\\
Since the wet-chemical anodization turned out to be very useful improving OPBTs, this technique is adapted to VOFETs, too \cite{Lim2020}. Furthermore, in the case of VOFETs, the anodization technique can be applied for the fabrication of the gate dielectrics as well as for the source-insulator offering a substantial added value in terms of process simplification.\\
\begin{figure}[htb]
	\begin{center}
		\includegraphics[width=.99\textwidth,clip]{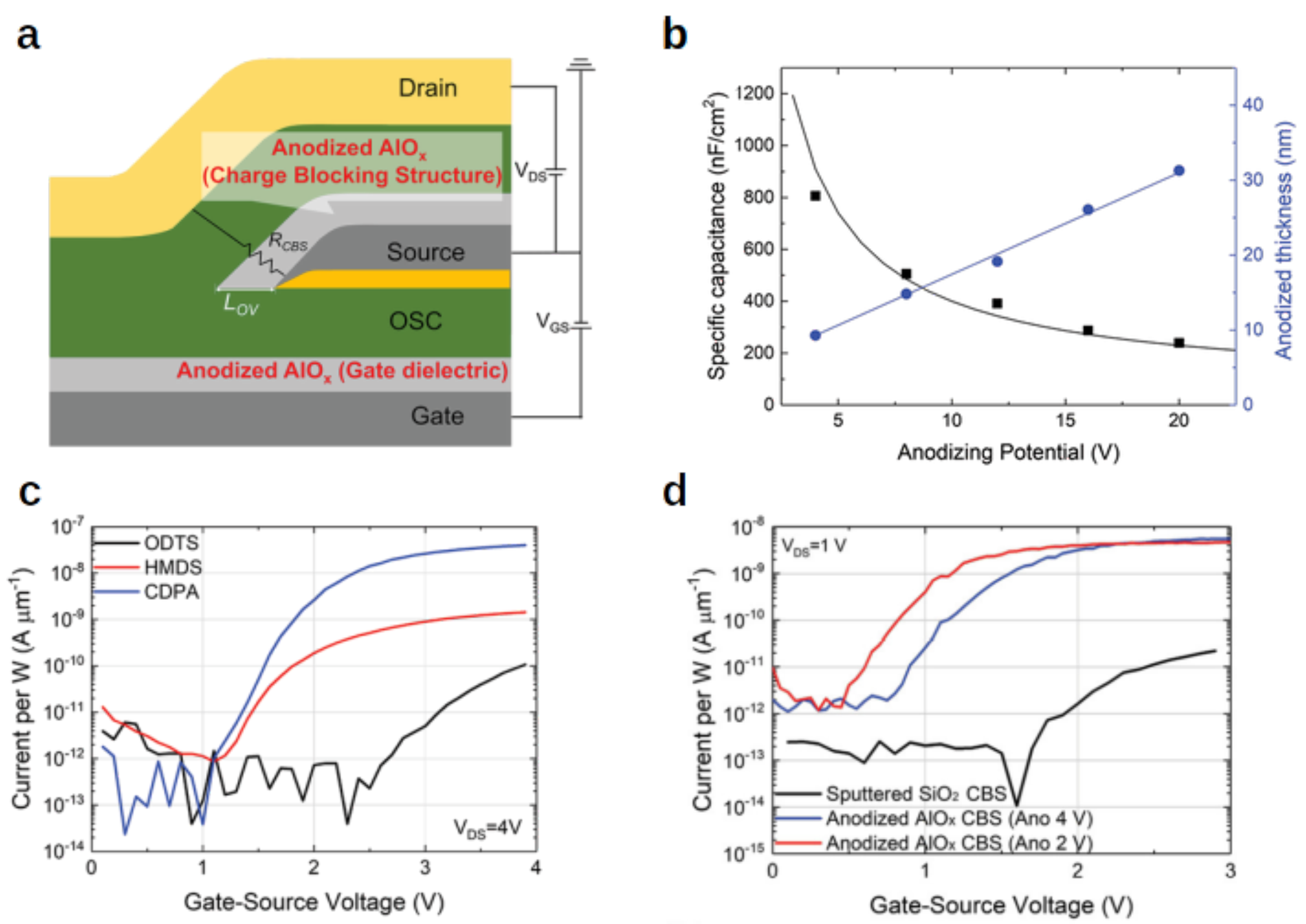}
	\end{center}
	\caption[]{\textbf{a} Scheme of a VOFET with anodized gate dielectrics and source insulator. \textbf{b} Thickness and dielectric capacitance vs. anodization voltage for Al$_2$O$_3$ films formed by wet-chemical anodization of aluminum in citric acid. \textbf{c} Reference OTFTs with anodized gate dielectrics and different types of self-assembled monolayers (channel length 100\,nm). \textbf{d} Transfer curve of VOFETs with anodized gate dielectrics and source insulator for different anodization conditions of the source-insulator (C$_{60}$ as semiconductor material). For reference, the transfer curve of a VOFET with sputtered SiO$_2$ is shown. Figures reprinted from \cite{Lim2020}.}
	\label{fig:VOFET_2}
\end{figure}
The VOFET structure with anodized dielectrics is schematically shown in Figure \ref{fig:VOFET_2}a. In a first step, the gate dielectric is anodized in a solution of citric acid, enabling precise control of the oxide thickness and capacitance (see Figure \ref{fig:VOFET_2}b). Thus, varying the anodization voltage between reference and gate electrode between 4 and 20\,V leads to the formation of a tight Al$_2$O$_3$ layer with a thickness between 9 and 32\,nm, respectively. Hence, high capacitance dielectric layers can be grown allowing for operation voltages of the devices lower than 10\,V.\\
To further improve the quality of the oxide layer, i.e., reducing the number of interface defect states, various types of self-assembled monolayers (ODTS: Octadecyltrichlorosilane, HMDS: hexamethyldisilazane, CDPA: 12-cyclohexyldodecyl-phosphonic acid) are applied to the oxide surface. In particular, as shown in Figure \ref{fig:VOFET_2}c, CDPA allows for a significant reduction of the number of surface defect states, resulting in a charge carrier mobility of 1.2\,cm$^2$/(Vs) for the semiconductor material C$_{60}$ measured in reference OTFTs. This superior behavior of CDPA-functionalized oxide surfaces is presumably caused by the phosphonic acid forming an almost perfect monolayer, while, e.g., HMDS is only weakly ordered or even forms multilayer films.\\
An almost similar fabrication protocol is used for the fabrication of the source-insulator (denoted here as charge-blocking-structure, CBS). A thin gold layer (10\,nm) is deposited onto the semiconductor material through a shadow mask followed by a 50\,nm thick aluminum layer. While the aluminum layer can be anodized as described above, the gold layer ensures good charge carrier injection into the semiconductor since it is not affected by the anodization process. Thus, fully functional VOFETs with exceptionally small off-currents and stability up to V$_{DS}$\,=\,10\,V are built ($\sim$\,7\,MV/cm)). In particular, the tight Al$_2$O$_3$ layer ensures much better robustness against high electric fields then, e.g., a layer of photoresist or a sputtered oxide film (cf. \cite{Lim2020} and Figure 6c). Furthermore, in concern of the on-state current, the VOFET processed with anodized source-insulator can compete with devices processed, e.g., by lift-off \cite{Kleemann2013}.\\
However, it should also be noted that there are two problems related to the anodization processes which are not solved yet.
Firstly, a noble metal such as gold is needed at the source electrode in order to impede the formation of the oxide layer at the charge injecting interface. The use of noble metals though, might cause the formation of large contact barriers in case of electron-transporting semiconductor materials due to the mismatch of work functions. Secondly, the anodization technique is currently only applicable for semiconducting materials such as C$_{60}$ due to its high ionization potential ($\sim$\,7\,eV). For typical hole-transporting materials such as pentacene, the wet-chemical oxidation process causes the formation of pentacene-quinone molecules, which show severely degraded charge carrier transport properties \cite{Zschieschang2010}. In order to solve this problem, hole-conducting semiconductor materials with an ionization potential higher than 6.8eV need to be used. In practice though, this solution is difficult to realize since hole injection will be severely deteriorated due to the presence of a high hole-injection barrier even if noble metals are used as electrodes. 
\subparagraph{High-Frequency Operation of Vertical Organic Field-Effect Transistors}\vspace{2mm}
Along with the improved understanding of device operation and the advancements in transistor fabrication, the vertical organic field-effect transistor structures are also optimized towards high-frequency operation in this work. In particular, due to the unique vertical electrode configuration, new scaling rules concerning device capacitance need to be considered if the device geometry is to be optimized for high frequencies. Furthermore, a special focus is put on the realization of small transistor devices. This aspect is of particular importance for application because all reports on organic transistors operating above 20\,MHz are employing devices with a channel width of at least 100\,\textmu m \cite{Borchert2020} and often larger than 500\,\textmu m \cite{Kitamura2011, Kitamura2011a, Borchert2019, Yamamura2018}. For active-matrix displays though, small devices with a channel width in access of 10\,\textmu m are required due to the availability of space in the circuit.\\
The VOFET fabrication procedure developed in this work is fully compatible with state-of-the-art display backplane technologies. In this regard, this work differs substantially from previous reports about vertical organic transistors where sophisticated fabrication methods have been employed, which cannot seamlessly be adopted in
\begin{figure}[htp]
	\begin{center}
		\includegraphics[width=.85\textwidth,clip]{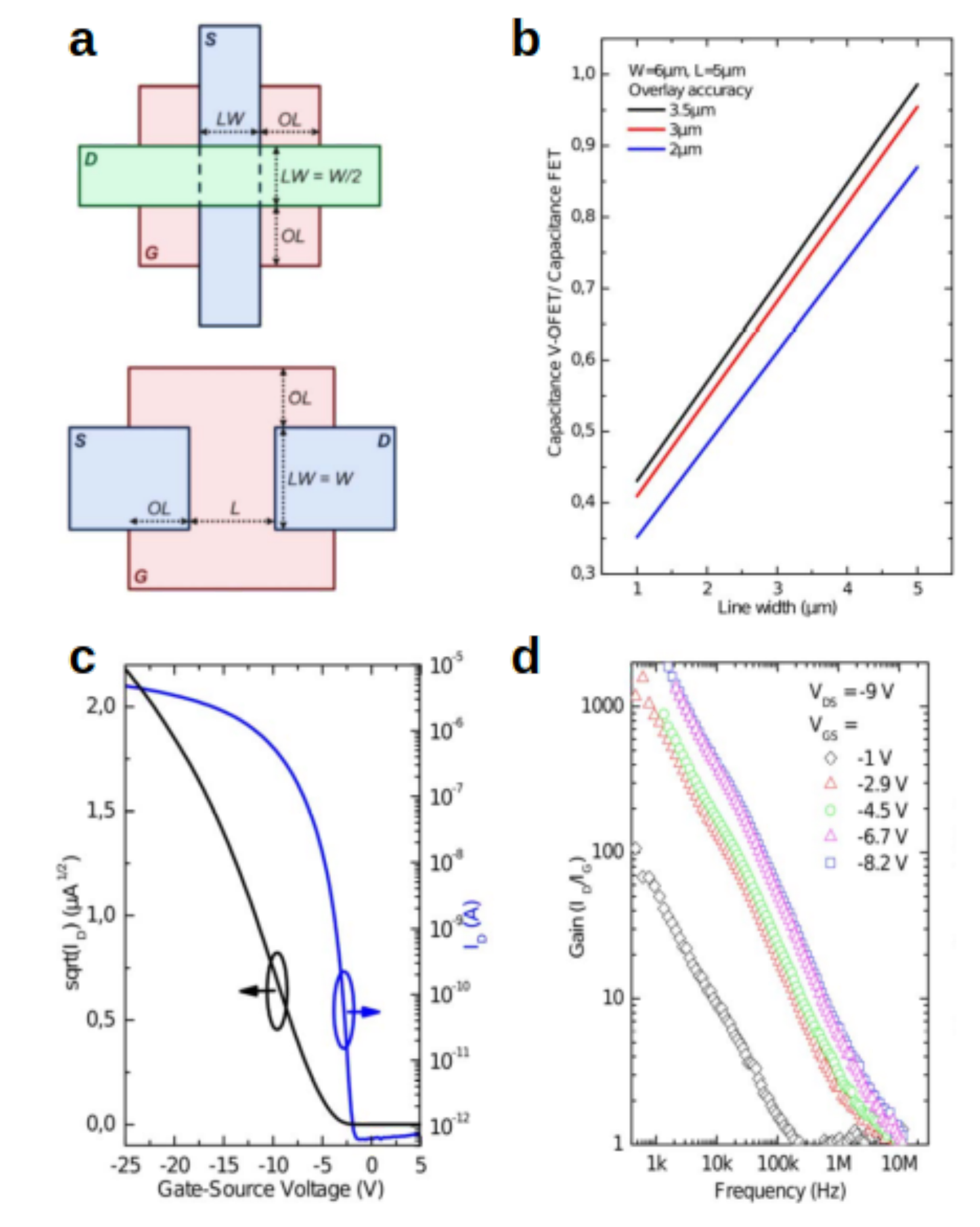}
	\end{center}
	\caption[]{\textbf{a} Comparison of VOFET (upper panel) and OTFT layouts (lower panel). \textbf{b} VOFET over OTFT capacitance versus metal line width for various overlay tolerances. \textbf{c} Transfer curve for a VOFET (W\,=\,20\,\textmu m) at V${_{DS}}$\,=\,-10\,V approaching the saturation. \textbf{d} Current gain |I$_D$|/|I$_G$| obtained from AC-measurements and extraction of frequency f$_T$\,=\,f(|I$_D$|/|I$_G$|\,=\,1). Figures reprinted from \cite{Kleemann2020a}.}
	\label{fig:VOFET_3}
\end{figure}
production lines for high-resolution displays. Such processes are: shadow mask patterning \cite{Fischer2012a}, lift-off \cite{Kleemann2013}, nm-scale lithography \cite{Ben-sasson2012c, Stutzmann2003}, and unconventional electrode or gate insulator materials \cite{Mccarthy2010, Ma2004}. Here, only well-established fabrication techniques for high throughput industrial processes, such as conventional photolithography, wet and dry etching are adapted \cite{Furno2015, Kleemann2012, Hoppner2019}. More specifically, a gate dielectric with a specific capacitance of 50\,nF/cm$^2$ (hybrid dielectrics composed of 25\,nm of Al$_2$O$_3$ and 30\,nm of the fluoropolymer CYTOP) is deposited by atomic layer deposition and spin-coating, respectively. Source and drain electrodes are composed of Au and are patterned on the organic semiconductor by photolithography followed by a wet-etching in standard-etchant for gold (potassium monoiodide). The gate electrode is also made of Al, and is patterned by photolithograhy and wet etching. The source-insulator is composed of a highly cross-linked photoresist (NLOF2020) with a thickness of 400\,nm. Finally, the semiconductor film is patterned by reactive ion etching using oxygen \cite{Hoppner2019}. All processing steps are performed under ambient conditions (except the vacuum deposition steps) and at process temperatures $\le$\,120\,$^\circ$C.\\
Vertical organic diode structures are able to operate up to 1\,GHz and beyond \cite{Steudel2005, Kleemann2012c, Kleemann2012d, Sawatzki2021}. Although similar organic semiconductor materials are used, the frequency of operation of lateral organic transistors is still in the range of tenths of megahertz. This large discrepancy is caused by the large parasitic capacitance of lateral devices and hence low current density (if referred to the overall areal footprint of the devices). Going from a lateral OTFT configuration to a vertical transistor configuration, the ratio of channel capacitance to parasitic capacitance improves substantially. In Figure \ref{fig:VOFET_3}a, the electrode layout for an OTFT and a VOFET are shown. Both devices nominally have the same channel width W. While in the OTFT, the channel width is given by the line width LW of the metal electrode, the VOFET has a channel width which is twice the line width of the electrode. Assuming the same design rules for the overlay of layers OL and the line width of electrodes (e.g., limited by the photolithography), the VOFET requires a significantly smaller device area in order to realize the same W for both transistors. In order to illustrate this big advantage of VOFETs over OTFTs, Figure \ref{fig:VOFET_3}b shows the ratio of the VOFET over OTFT capacitance for a transistor with a channel length of 5\,\textmu m, a channel width of 6\,\textmu m, and various values for the alignment accuracy. For a line width larger than 5-6\,\textmu m, the VOFET architecture has no advantage over OTFTs in terms of device capacitance. However, for smaller line widths (e.g., 2.5-3\,\textmu m), which are relevant for applications in displays, the VOFET has a capacitance of only 50 to 60\% compared to the lateral OTFTs which represents a significant improvement. Furthermore, one should keep in mind that the VOFET with an ultra-short vertical channel can provide much more current than an OTFT with a channel length of 5\,\textmu m. Thus, the product of device capacitance and resistance is expected to be much smaller for VOFETs which directly translates into high operation frequency.\\
In Figure \ref{fig:VOFET_3}, the transfer curve of a VOFET is shown. This VOFET is fabricated with a line width of 3\,\textmu m and an electrode overlay of 2\,\textmu m. As can be seen, the VOFET transfer curve is characterized by a steep subthreshold slope of $\sim$60\,mV/dec and an on/off-ratio of 5\,$\cdot$10$^6$ which is excellent for such an ultra-short channel device. Overall, the current density (taking the full areal footprint of the device) is in the range of 10-20\,A/cm$^2$. These values though are still more than 10x smaller than for OPBTs which is due to unavoidable parasitic areas in the 2-dimensional electrode configuration. However, compared to OPBTs which are fabricated using shadow-masks, these VOFETs are fully integrated.\\
In order to investigate the operation frequency of these devices, a unity-gain transition frequency measurement is performed. For moderate voltages up to V$_{GS}$\,=\,8.2\,V, these devices can be operated up to a frequency of 10\,MHz. A prediction based on the DC transfer curve up to V$_{GS}$\,=\,25\,V gives values up to 40\,MHz. Unfortunately, the signal generator used for these experiments does not allow to apply voltages larger than 8.2\,V at frequencies about 20\,MHz, which currently restricts the evaluation of the transition frequency. Although the 10\,MHz reached by VOFETs is not a new record value for organic transistors, the performance is remarkable since these devices are the first small organic transistors (W\,=\,20$\,$\textmu m) which can operate at these frequencies. For large scale devices (LW\,$\gg$\,OL, e.g., W\,$\ge$\,500$\,$\textmu m), as used by other research groups \cite{Kitamura2011, Yamamura2018}, the VOFET is predicted to work at frequencies well above 80\,MHz due to the negligible influence of unavoidable overlay capacitances \cite{Hoppner2020}. However, such large devices are no relevant for applications such as active-matrix displays.\\
Overall, VOFETs can reach a DC performance (on/off-ratio, subthreshold slope) similar to the best OTFTs with the additional advantage of having a smaller device capacitance. In the future, further improvements, e.g., by using interface doping for improved injection or using organic semiconductor materials with higher intrinsic mobility, will help to bring the transition frequency to values close to 100\,MHz. The substantial problem which remains at this point is the increased heat generation while operating at high frequencies, which puts a big question-mark on whether organic transistors could possibly operate at 1\,GHz.
\subsection{New Approaches for Lateral Organic Transistors}
Besides the advancements in the development of vertical organic transistors, there are still many possibilities to further improve lateral OTFTs. Concerning high-frequency operation, two specific challenges need to be addressed, which could also be advantageous for vertical organic transistors. The first challenge is the fabrication of so-called self-aligned transistors, which have been successfully employed in silicon-technology to achieve low-capacitance devices and high-frequency operation. In such devices, the gate electrode is used as a hard-mask for the patterning of source and drain. In this way, there is virtually no source-gate or drain-gate overlap. However, in organic semiconductors, truly self-aligned transistors are challenging to realize since it would require an etching process of the gate dielectrics, which does not severely damage the organic semiconductor underneath. The second challenge refers to the vision of using highly ordered organic thin-film crystals rather than poly-crystalline films in order to advance charge carrier mobility further and reduce contact resistance. In this regard, it remains a significant hurdle to grow large-area and uniform organic thin-film crystals due to the weak van der Waals interactions.\\
Both problems are investigated in this work, which led to the development of pseudo-self-aligned organic transistors and dopable organic thin-film crystals based on rubrene. These studies resulted in the following publications \cite{Vahland2021, Sawatzki2021, Wang2021, Lashkov2021}:
\begin{itemize}
    \item J. Vahland, K. Leo, and H. Kleemann, "Pseudo-Self-Aligned Organic Thin-Film Transistors", Submitted to Advanced Electronic Materials (2021).
    \item S.J. Wang, M. Sawatzki, H. Kleemann, I. Lashkov, O. Janson, D. Wolf, A. Lubk, E. Hieckmann, F. Talnack, S. Mannsfeld, Y. Krupskaya, J. van den Brink, B. B\"uchner and K. Leo, "Vacuum processed large area doped thin-film crystals: A new approach for high-performance organic electronics", Materials Physics Today \textbf{17}, 100352  (2021).
    \item I. Lashkov, K. Krechan, K. Ortstein, F. Talnack, S.J. Wang, S.C.B. Mannsfeld, H. Kleemann, \& K. Leo, "Modulation Doping for Threshold Voltage Control in Organic Field-Effect Transistors", ACS Applied Materials \& Interfaces \textbf{13}, 8664-8671 (2021).
    \item M. Sawatzki, H. Kleemann, B. K.-Boroujeni, F. Ellinger, and K. Leo, "Doped Highly Crystalline Organic Films: Toward High‐Performance Organic Electronics", Advanced Sciences \textbf{8}, 2003519 (2021).
\end{itemize}
\subsubsection{Pseudo-Self-Aligned Coplanar Organic Transistors in Top-Gate Configuration}
Self-aligned transistor structures are most common in silicon-based integrated circuit technology. Using a coplanar top-gate transistor architecture, the gate insulator and gate electrode are employed as a hard-mask during the
\begin{figure}[htb]
	\begin{center}
		\includegraphics[width=.90\textwidth,clip]{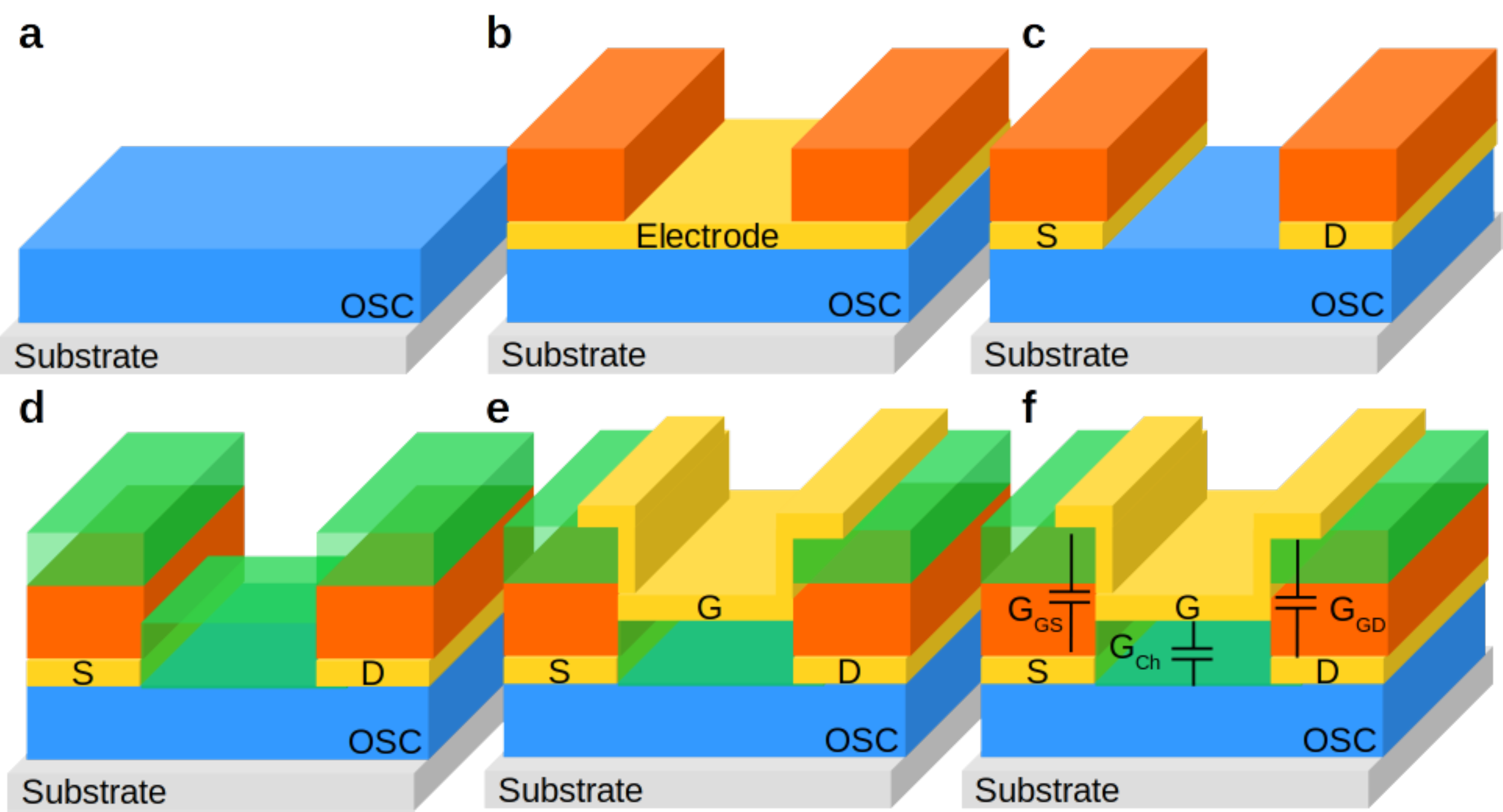}
	\end{center}
	\caption[]{Fabrication procedure and structure of the pseudo-self-aligned OTFT. \textbf{a} DNTT as organic semiconductor (OSC) on the substrate, \textbf{b} deposition of metallic electrode (Au) and patterning of photoresist (orange), \textbf{c} wet-chemical etching to define source and drain electrodes, \textbf{d} deposition of the gate dielectric material, and \textbf{e} deposition and patterning of the gate electrode. \textbf{f} shows a cross-section of the pseudo-self-aligned OTFT with the gate-source C$_\mathrm{GS}$, gate-drain C$_\mathrm{GD}$, and channel capacitance C$_\mathrm{Ch}$.}
	\label{fig:OTFT_1}
\end{figure}
etching process followed by subsequent doping (e.g., by ion-implantation). In this way, highly doped contact regions for source and drain next to the gate insulator are established, while the transistor channel remains protected. The doping technology and the compatibility of the semiconductor and the gate insulator material are the enabling factors for the self-aligned configuration. For organic semiconductor-based transistors, though, both processes are not readily available, and hence, the self-aligned structure as employed for silicon-based TFTs cannot be adapted. In particular, the main challenge for OTFTs lies in finding combinations of semiconductor material and gate insulator, which offer sufficient compatibility during the etching process of the gate insulator.\\
In order to fabricate self-aligned OTFTs and circumvent the above-described problems, researchers developed alternative processes  \cite{Noh2007, Higgins2015}. Unfortunately, these alternative processes might involve lift-off steps or so-called through-substrate lithography (exposing from the back of the substrate), which cannot be adapted in mass-production processes due to device yield concerns or other limitations. Thus, it is still an open challenge to develop self-aligned OTFTs without employing non-conventional, non-scalable processes.\\
In order to tackle this problem, a pseudo-self-aligned organic transistor structure in a coplanar top-gate architecture is envisioned (Figure \ref{fig:OTFT_1}). In this structure, rather than using the gate as a hard-mask to define the source and drain electrode, self-alignment is achieved using source and drain electrodes with an additional spacer layer to minimize the overlap capacitance. The spacer layer is an approximately 2\,\textmu m thick photo-resist, which stays on the source and drain electrode after the wet-chemical etching process. Hence, if the gate electrode overlaps with the source and drain electrodes, the thick spacer layer will significantly reduce the undesired overlap capacitance. The most crucial fabrication step though, is the
wet-chemical etching since it is carried out directly on the most sensitive interface, namely the transistor channel. The entire metal film needs to be removed from the channel interface during the etching process without damaging the organic semiconductor material. Furthermore, the etching process should be optimized so that under-etching underneath the spacer layer does not occur. We evaluated two different etchants (aqua regia and KI/I$_2$) concerning their ability to remove the metallic layer (Au) from the underlying organic semiconductor (here, the oligobenzothiophene DNTT). Due to the crystallinity of
\begin{figure}[htb]
	\begin{center}
		\includegraphics[width=.99\textwidth,clip]{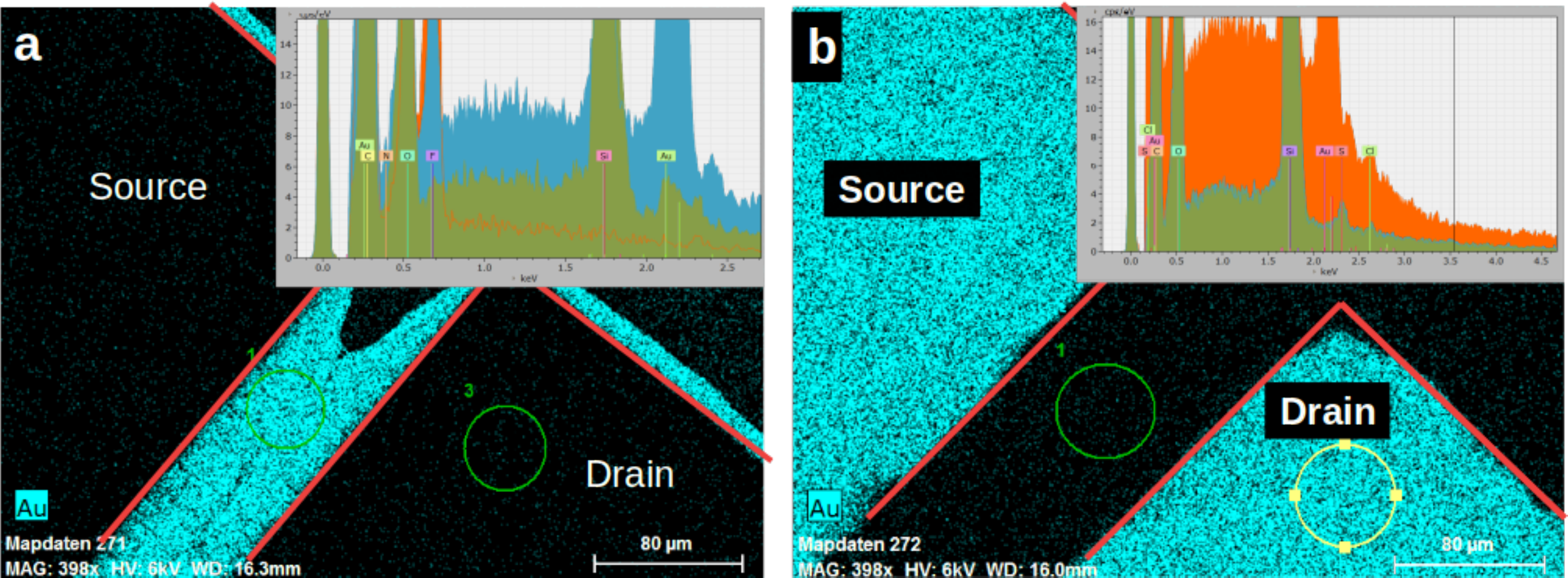}
	\end{center}
	\caption[]{Energy-dispersive X-ray (EDX) spectroscopy mapping (Au channel) on a pseudo-self-aligned OTFT fabricated according to the recipe shown in Figure \ref{fig:OTFT_1}. The Au layer is etched with \textbf{a} KI/I$_2$ or \textbf{b} aqua regia. Apparently, even after an extended duration of etching with KI/I$_2$, Au residuals are still visible in the OTFT channel. The insets so the EDX spectra comparing different positions on the sample (channel area vs. electrode area). In \textbf{a} the electrodes do not give a strong signal on the Au channel since they are covered with photoresist. Figures adapted from \cite{Vahland2021}.}
	\label{fig:OTFT_2}
\end{figure}
this semiconductor material and the related surface roughness, the etching time for the 30\,nm thin gold layer is significantly longer compared to the etching of a gold layer in direct contact with the underlying substrate. Presumably, the poor wetting of the etchant on the rough surface reduces the effectiveness of the etchant solution. This effect is particularly strong for KI/I$_2$ where even after several minutes of etching, small particles of Au remain on the surface and cannot be etched anymore (Figure \ref{fig:OTFT_2}). Aqua regia though, allows for faster and more uniform etching of the Au layer, and in consequence, the layer can be removed from the
\begin{figure}[htb]
	\begin{center}
		\includegraphics[width=.99\textwidth,clip]{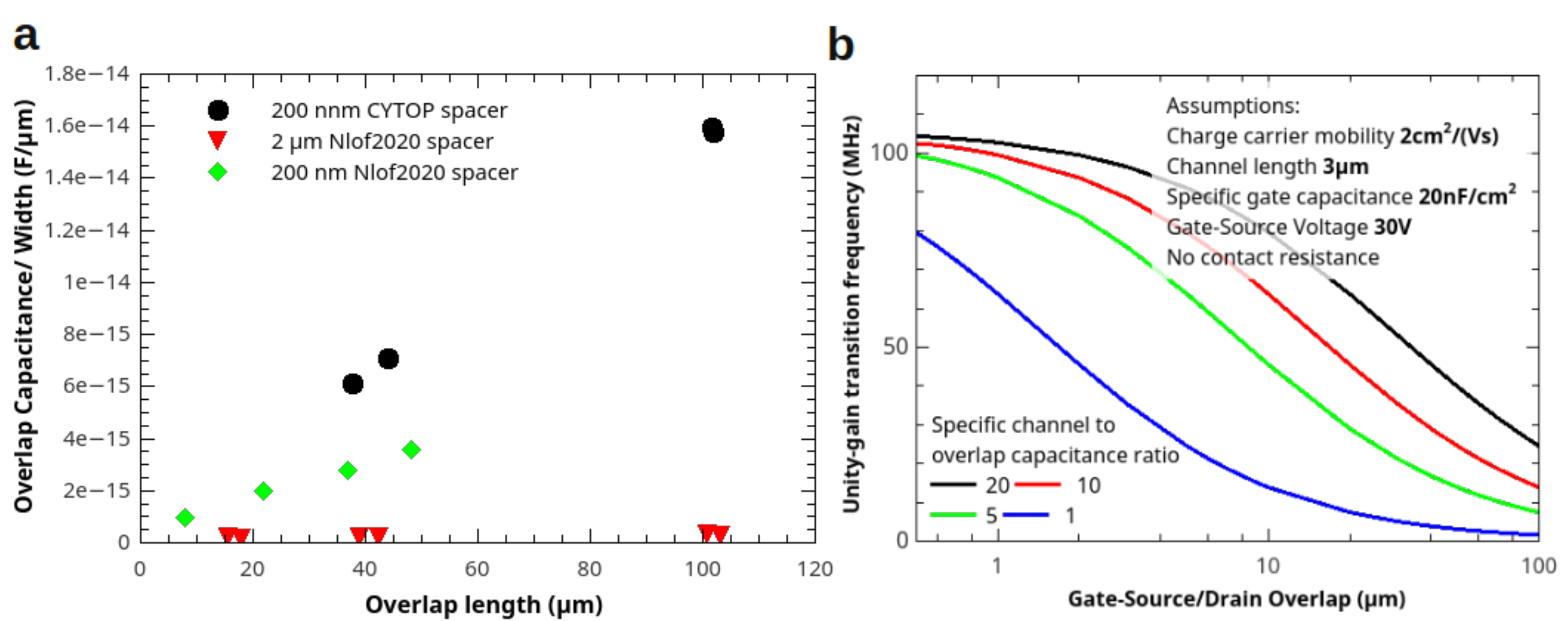}
	\end{center}
	\caption[]{\textbf{a} Measurement of the total overlap capacitance (normalized by the channel width) of a pseud-self-aligned OTFT as a function of the overlap length for different spacer layers. From the slope, the ratio of the specific channel to the overlap capacitance is determined to be $\sim$\,10.  \textbf{b} Calculation of the theoretical unity-gain transition frequency vs. the electrode overlap according to Equation \ref{eq:f_t} for different ratios of the specific channel capacitance over the specific gate-source/gate-drain capacitance. Further assumption for the calculation are given in the inset. Figure \textbf{a} adapted from \cite{Vahland2021}.}
	\label{fig:OTFT_3}
\end{figure}
organic semiconductor completely without causing a significant under-etching underneath the spacer layer (Figure \ref{fig:OTFT_2}).\\
Figure \ref{fig:OTFT_3}a shows an evaluation of the capacitance scaling in the pseudo-self-aligned OTFTs for different thicknesses and materials of the spacer layer. The dependence for CYTOP (200\,nm) is shown as a reference, representing the capacitance of the channel. Since direct measurement of the overlap capacitance is difficult for an overlap length of 2\,\textmu m, the variation of the overlap capacitance with the overlap length is shown. When using a spacer layers atop of the drain and source electrode, e.g., 200\,nm or 2\,\textmu m of Nlof2020, the specific overlap capacitance is significantly reduced compared to the pure CYTOP layer (maximum ratio of ten, roughly corresponding to the ratio of thicknesses, 2\,\textmu m/200\,nm \,=\,10). Thus, compared to the non-self-aligned case, the parasitic overlap capacitance is reduced by a factor of ten for overlap length <\,10\,\textmu m.\\
The advantage of the reduction of the overlap capacitance is highlighted in Figure \ref{fig:OTFT_3}b. Here, the calculated unity-gain transition frequency according to Equation \ref{eq:f_t} is plotted versus the gate-source/gate-drain overlap length for an OTFT with a channel length of 3\,\textmu m and a charge carrier mobility of 2\,cm$^2$/(Vs). For this transistor configuration, the pseudo-self-alignment provides a
\begin{figure}[htb!]
	\begin{center}
		\includegraphics[width=.99\textwidth,clip]{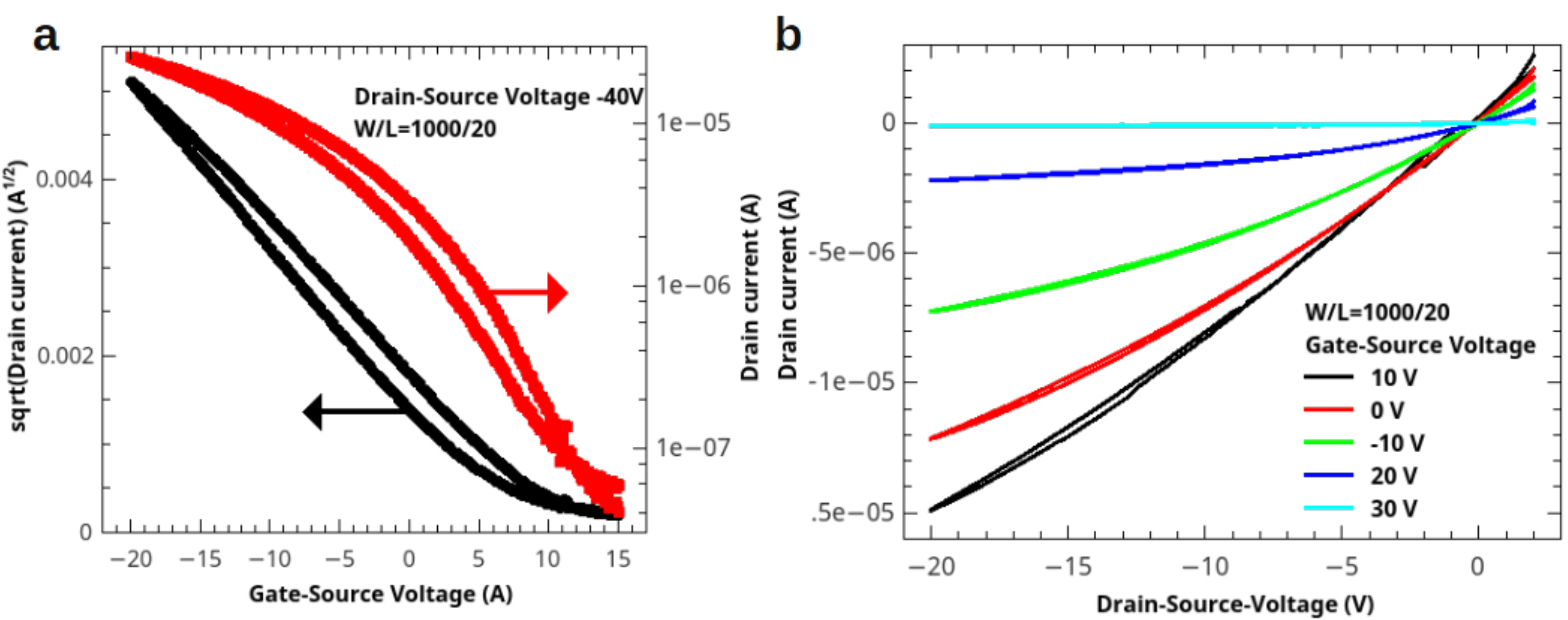}
	\end{center}
	\caption[]{\textbf{a} Transfer and \textbf{b} output curve of a pseudo-self-aligned OTFT based on DNTT. The gate insulator is a hybrid composed of 200\,nm of CYTOP and 25\,nm of Al$_2$O$_3$. The spacer layer on source and drain is a 400\,$\mathrm{nm}$ thick layer of Nlof2020. The channel length and width of the device is 1000\,$\mathrm{mu}$m and 20\,\textmu m, respectively. The charge carrier mobility is determined to be 2\,cm$^2$/(Vs). Adapted from \cite{Vahland2021}.}
	\label{fig:OTFT_4}
\end{figure}
substantial advantage over non-self-alignment technologies. In the technologically relevant overlap region between 2\,$\mathrm{\,mu}$m and 5\,\textmu m, the pseudo-self-alignment configuration would provide a higher unity-gain transition frequency f$_\mathrm{T}$ and the increase of f$_\mathrm{T}$ is directly proportional to the ratio between the specific channel and overlap capacitance. Hence, for the 2\,\textmu m thick layer of Nlof (spacer) and the 200\,nm thick layer of CYTOP (channel dielectrics), the transition frequency is increased by a factor of 3-8.\\
In Figure \ref{fig:OTFT_4}, the transfer and output curve of a pseudo-self-aligned OTFT based on DNTT as organic semiconductor are shown. A charge carrier mobility of up to 2\,cm$^2$/(Vs) is obtained in these devices, proving that the transistor channel is intact and not severely damaged during the etching process. Assuming the experimental values for the charge carrier mobility as well as the channel and overlap capacitance, it is predicted that the unity-gain transition frequency would improve from 12.5\,MHz for a non-self-aligned devices to 84\,MHz for a pseudo-self-aligned OTFT, as proposed here (overlap length of 10\,\textmu m, channel length of 3\,\textmu m). The scaling and dynamic characterization of the pseudo-self-aligned devices is still ongoing, and hence, experimental values cannot be reported yet. However, the prediction based on the experimental input is very encouraging and underlines the advantage of the proposed device structure. 
\subsubsection{New Materials for Vertical Organic Thin-Film Transistors}
Structural order and charge carrier mobility are inherent linked to each other. In amorphous or nano-crystalline organic semiconductor materials, which are widely used for organic thin-film electronics, hopping-like motion between localized states is the predominant transport mechanism. Consequently, charge carrier mobility in these material systems is usually restricted to the range $\le$\,1\,cm$^2$/(Vs). In highly ordered organic crystals though, even band-like transistor might occur allowing for charge carrier mobility of >\,1\,cm$^2$/(Vs). In recent years, substantial progress has been achieved in growing large-area organic thin-film crystals from solution \cite{Yamamura2020}. However, such techniques require specific surfaces and complex fabrication methods which prevent adopting these approaches to arbitrary substrate materials.\\
In the last two years, an alternative technique to grow large-area organic thin-film crystals composed of rubrene is developed at IAPP. This techniques allows for rubrene crystal growth from the vacuum phase, which may be used on almost arbitrary substrates. Rubrene is a famous organic semiconductor material, offering charge carrier mobility values above 10\,cm$^2$/(Vs) \cite{Podzorov2004}. However, it is predominantly grown in a tube-furnace process that might not allow for the preparation of thin-film crystals. Instead, a thermal annealing step of vacuum processed amorphous rubrene thin-films is explored with the use of an underlayer material to promote the crystal growth and surface coverage of the crystals. We further investigated the effect of epitaxial molecular doping on thin-film rubrene crystal’s structural and charge transport properties which has not been proven to be possible with the conventional tube-furnace method.\\
Amorphous rubrene thin films deposited on oxide surfaces might be transformed into highly crystalline rubrene thin films by thermal annealing \cite{Fusella2017, Fielitz2016}. 
\begin{figure}[htb!]
	\begin{center}
		\includegraphics[width=.99\textwidth,clip]{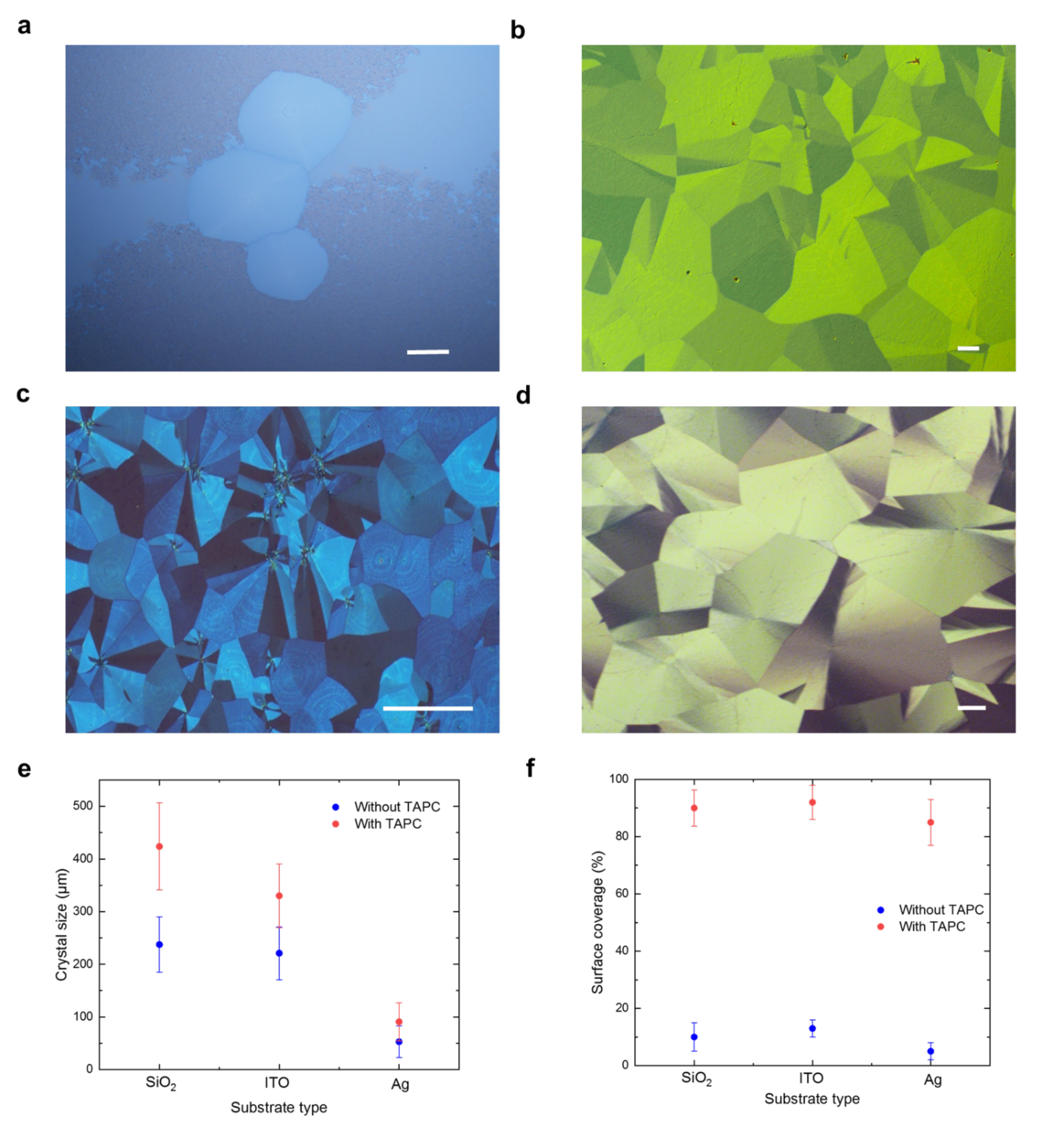}
	\end{center}
	\caption[]{Crystallized rubrene thin films (40\,nm) on \textbf{a} SiO$_2$ (300\,nm) coated Si substrate annealed for 3 minutes at 160$^\circ$C. \textbf{b} TAPC (5\,nm)/ SiO$_2$ (300\,nm) coated Si substrate annealed for 4 minutes at 160$^\circ$C. \textbf{c} TAPC (5\,nm)/ Ag (20\,nm)/ Glass annealed for 4 minutes at 160$^\circ$C. \textbf{d} TAPC (5\,nm)/ Indium-Tin-Oxide (100\,nm)/ Glass annealed for 4 minutes at 160$^\circ$C. The scale bars correspond to 100\,\textmu m. Crystal size \textbf{e} and surface coverage \textbf{f} of orthorhombic rubrene crystals grown on different substrates. The error bars are the standard deviation from three different films. Figure adapted from \cite{Wang2021}.}
	\label{fig:OTFT_5}
\end{figure}
Using a thin amorphous underlayer, this approach can be generalized to grow crystalline rubrene thin films on a variety of surfaces, from oxides to metals (Figure \ref{fig:OTFT_5}). Furthermore, the crystallization of thin films deposited directly on silicon dioxide yielded only a few domains of orthorhombic crystals (confirmed by X-ray diffraction) while most of the surface is covered by an amorphous layer of rubrene (Figure \ref{fig:OTFT_5}a). The use of underlayer materials with a low glass transition temperature T$_\mathrm{G}$ in the range of 75-83$^\circ$C (e.g., amorphous organic semiconductor materials) though, has been reported to promote the rubrene orthorhombic crystal growth \cite{Fusella2017}. We chose a thin, 5 nm layer of di-[4-(N,N-ditolyl-amino)-phenyl]cyclohexane (TAPC) (hole transporting material, T$_\mathrm{G}$\,=\,82$^\circ$C) or the insulating polymer Poly(methyl methacrylate) (PMMA, T$_\mathrm{G}$\,=\,78$^\circ$C) beneath rubrene, which allows for almost complete surface coverage with domains of orthorhombic rubrene crystal with a few hundred micrometers diameter on silicon dioxide, silver, and ITO surfaces (Figure \ref{fig:OTFT_5}b-d). The most surprising finding though, is the growth of orthorhombic rubrene crystals on metallic surfaces since the interaction between rubrene molecules and the substrate underneath has a large influence on the crystal growth kinetics and the growth of orthorhombic rubrene crystals on metal surfaces is generally not possible. Therefore, a thin layer of TAPC is sufficient to decouple the interaction between rubrene and the metal
\begin{figure}[htb!]
	\begin{center}
		\includegraphics[width=.99\textwidth,clip]{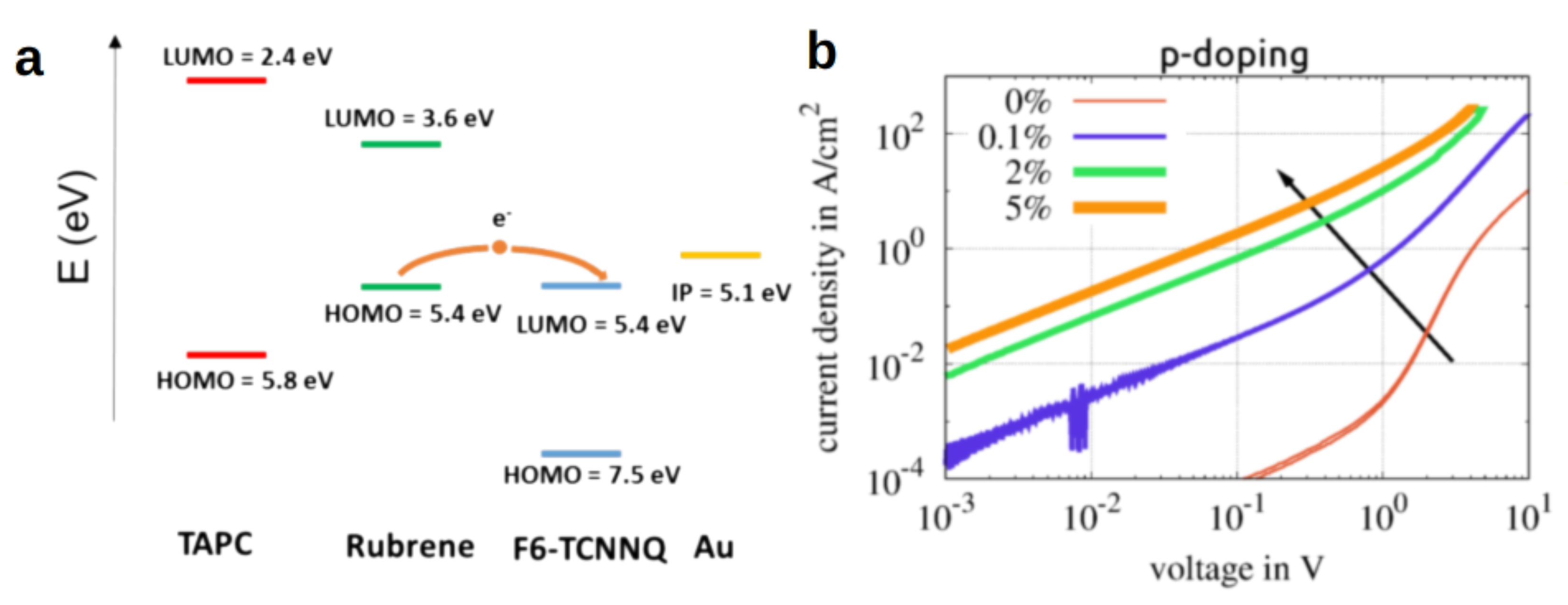}
	\end{center}
	\caption[]{\textbf{a} Energy level diagram of TAPC, rubrene, F6-TCNNQ, gold. Due to the position of the lowest unoccupied molecular orbital (LUMO) of the dopant and the highest occupied molecular orbital (HOMO) of the host charge carrier transfer can take place creating an ionized acceptor and a mobile polaron on the host. \textbf{b} Current-voltage measurement of a p-type doped rubrene film (film thickness 400\,nm sandwiched between two gold electrodes). The concentration of the dopant F6-TCNNQ is given in weight-percent in the inset of the figure. Figure adapted from \cite{Wang2021} and \cite{Sawatzki2021}}
	\label{fig:OTFT_6}
\end{figure}
surface and promote orthorhombic rubrene crystals growth even on the metal surface. In Figure \ref{fig:OTFT_5}e and f, the crystal domain size and surface coverage of the orthorhombic rubrene crystals on different substrate surfaces with and without the thin TAPC underlayer is summarized.\\
Another interesting aspect about this growth mechanism for rubrene thin-film crystals is, that it is possible to intentionally dope the rubrene during the deposition. Up to a certain concentration of the dopant molecule, the growth mode and the integrity of the crystals is maintained. In the experiments, small-molecule-based donors or acceptor molecules are added during the rubrene deposition in order to electrically dope the films n- or p-type, respectively. For example, in the case of the strong electron acceptor 1,3,4,5,7,8-hexafluorotetracyanonaphthoquinodimethane (F6-TCNNQ), up to 5\,mol$\%$ of dopants can be added to the rubrene crystals without affecting the perfection of the crystals, suggesting that the charge carrier mobility might not be affected as well. Using such a high dopant concentration, conductivity values in the range of 0.1 to 1\,S/cm are attained (Figure \ref{fig:OTFT_6}). More detailed investigations about the doping efficiency and the possible incorporation of even stronger (but larger) dopants is currently ongoing.\\
In any case, the combination of highly crystalline organic thin-films with the ability to electrically dope them offers the possibility to build organic thin-film transistors with high charge carrier mobility and low contact resistance. In Figure \ref{fig:OTFT_7}, the structure of a rubrene-based single-crystal thin-film transistor in a Hall-bar geometry is shown (channel length is 20\,\textmu m). Using a doped film at the electrode for improved charge carrier injection and extraction, an average charge carrier mobility of 4.35$\pm$0.76cm$^2$/(Vs) is obtained on TAPC. Unfortunately, due to the semiconducting property of the TAPC underlayer, a second channel is opened up in the TAPC with increasing gate-source voltage, causing an undesirable loss of the drain current vs. gate-source voltage dependence (Figure \ref{fig:OTFT_7}b). As shown in Figure \ref{fig:OTFT_7}c, using the insulating PMMA instead of TAPC as an underlayer, this unfavorable effect can be avoided yielding equally high charge carrier mobility. The contact resistance in these devices has been determined to be $\le$\,1\,k$\mathrm{\Omega}$cm which proves the very efficient injection of charge carriers into the devices.\\
Currently, the crystal growth is further improved. Moreover, fabrication methods are developed to integrate the rubrene thin-film crystal OTFTs into more complex electronic circuits such as oscillators. 
\begin{figure}[htb]
	\begin{center}
		\includegraphics[width=.90\textwidth,clip]{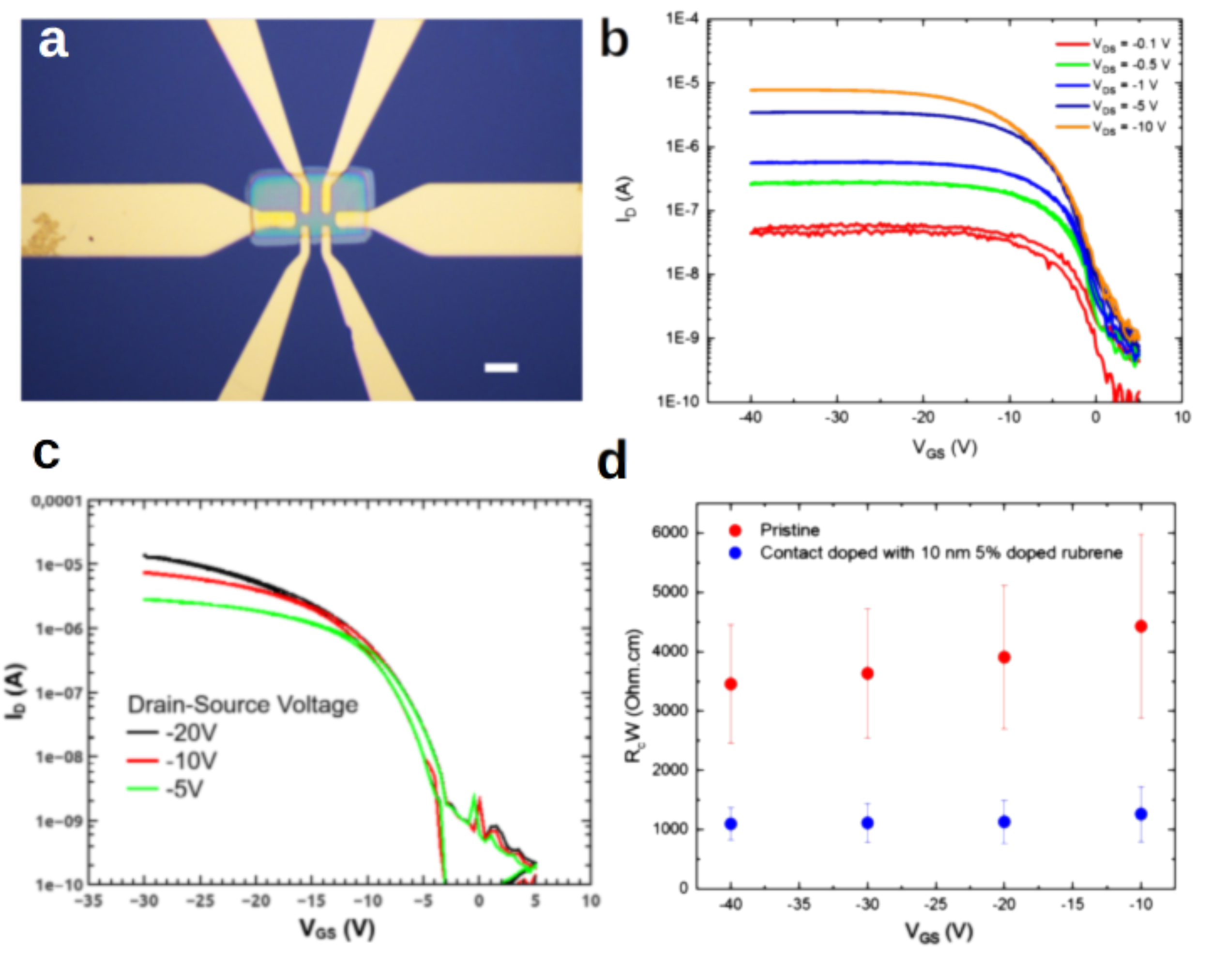}
	\end{center}
	\caption[]{\textbf{a} Photograph of a rubrene thin-film crystal OTFT in a Hall-bar geometry. The channel length is 20\,\textmu m. \textbf{b} Transfer curve of such a rubrene OTFT with TAPC as an underlayer on the SiO$_2$ dielectrics. At V$_\mathrm{GS}\,\le$\,-10\,V, the drain current I$\mathrm{D}$ does not increase anymore which is due to the formation of a second channel in the TAPC underlayer. \textbf{c} Transfer curve of such a rubrene OTFT with TAPC as an underlayer on the SiO$_2$ dielectrics. In this case, the drain current increases with the gate-source voltage as predicted by the gradual channel approximation. Figure adapted from \cite{Wang2021}.}
	\label{fig:OTFT_7}
\end{figure}

\pagestyle{fancy}
\fancyhead{}
\fancyhead[LE]{\scshape \thepage} 
\fancyhead[RE]{\scshape \rightmark}
\fancyhead[LO]{\scshape \rightmark} 
\fancyhead[RO]{\scshape \thepage}
\fancyfoot{}
\renewcommand{\headrulewidth}{1pt}
\cleardoublepage
\setcounter{section}{3}
\setcounter{subsection}{0}
\section*{Organic Electro-Chemical Transistors}
\addcontentsline{toc}{section}{3 \hspace{0.1cm} Organic Electro-Chemical Transistors}
\thispagestyle{empty}
\vspace{2cm}
\setlength{\epigraphwidth}{.5\textwidth} 
\setlength{\epigraphrule}{1pt} 
\epigraph{\large{\textit{'Simplicity is the ultimate sophistication.'}}}{\footnotesize{Aristotle, Philosopher}}
\vspace{2cm}
OTFTs discussed in the previous chapter closely resemble conventional silicon-based TFTs except from their flexibility and their potential for low-cost, large-area integration. However, there is an alternative OTFT concept, the organic electro-chemical transistor (OECT) \cite{Bernards2007, Rivnay2018}, holding great promise for applications in chemical sensors \cite{Lin2010, Gualandi2016}, bio-electronics  \cite{Leleux2015}, and artificial neural networks \cite{Burgt2017, Gkoupidenis2017, Burgt2018}.\\
In particular, OECTs are based on mixed ionic-electronic conductors, and due to the huge dielectric double-layer capacitance, OECTs inherently operate at voltages below 1\,V ensuring low power consumption and compatibilty with biological systems such as nerve cells. Furthermore, due to innovative deposition methods, OECTs render the possibility to build reconfigurable electronic circuits which might be used to mimic the behavior of biological synaptic networks.\\
The focus of my research in the field of OECTs is to develop and explore new fabrication methods and OECT architectures in order to build brain-inspired neural networks (neuromorphic networks) which might be used, e.g., for hardware-based classification of bio-signals (e.g., electroencephalography (EEG) or electrocardiography (ECG)). Additionally, I research on OECTs as ion-sensors which might be combined with the neuromorphic networks to be utilised as intelligent sensor tags. Moreover, the investigations on the ion-sensing mechanism help us to develop a detailed understanding on the ion-signal propagation and electronic response of the mixed ionic-electronic OECT system.\\
In the following, the working mechanism of OECTs is summarized, including an overview on semiconductor materials for OECTs. Moreover, a brief introduction to the concepts of neuromorphic computing is given, focusing on terms such as synaptic plasticity and artificial neurons. In the last section of this chapter, my contributions to the field are presented. 

\subsection{Organic Electro-Chemical Transistors}
\subsubsection{Operation of Organic Electro-Chemical Transistors}
\begin{figure}[htb]
	\begin{center}
		\includegraphics[width=.99\textwidth,clip]{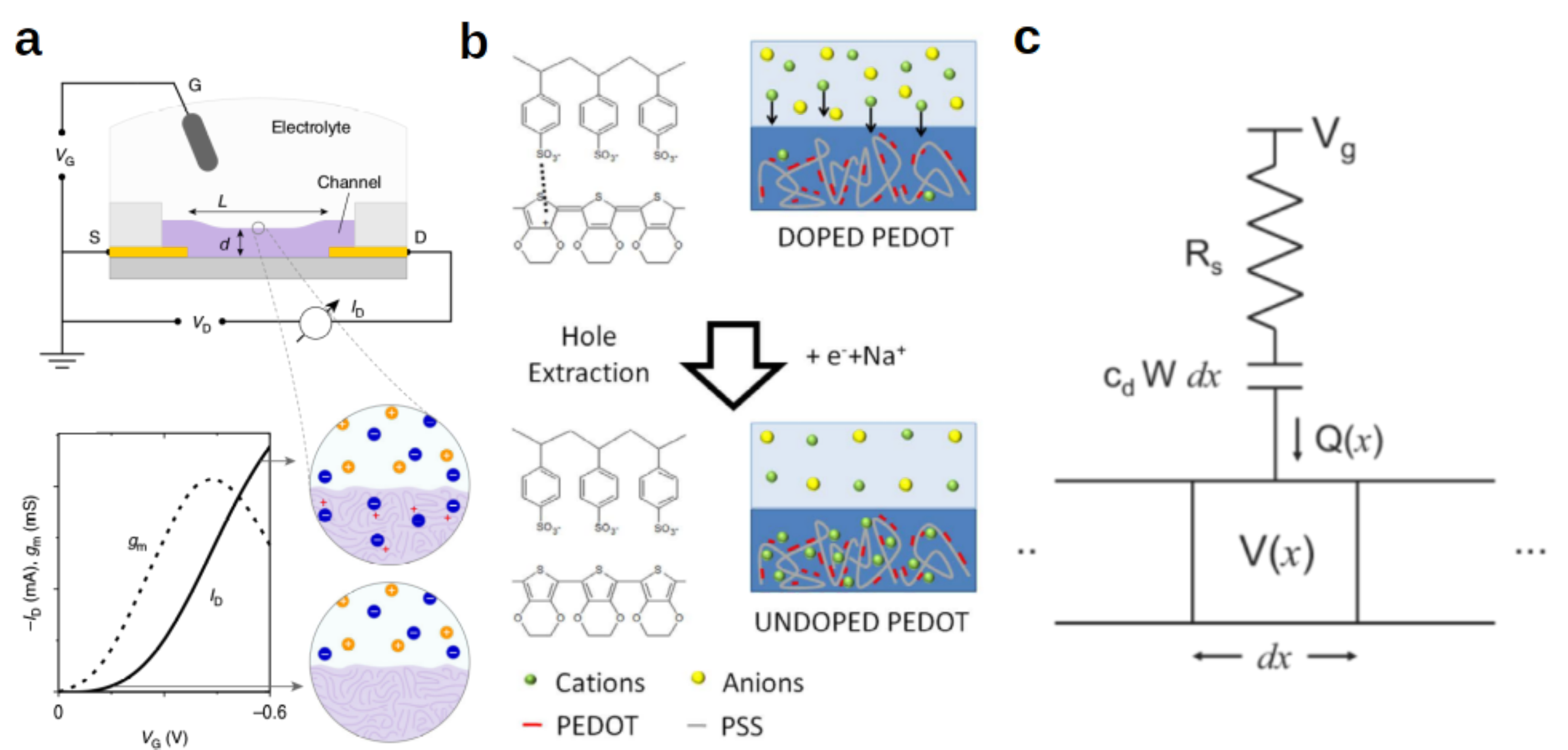}
	\end{center}
	\caption[]{\textbf{a} Scheme of an OECT operating in a saline solution (top) and corresponding transfer curve (bottom). The migration of ions into the OSC is illustrated (Figure reprinted from \cite{Inal2017}, published under CC BY license for free non-commercial reuse). \textbf{b} Doping and dedoping mechanism exemplary shown for PEDOT:PSS (Figure reprinted from \cite{Gualandi2016a}, published under CC BY license for free non-commercial reuse). \textbf{c} Equivalent circuit model for the ionic sub-circuit in Bernards' model. The charge from the ionic circuit is coupled to the voltage in the electronic circuit at a position x along the organic semiconductor (Figure \textbf{c} reprinted from \cite{Bernards2007} with permission by John Wiley and Sons).}
	\label{fig:OTFT_OECT}
\end{figure}

Alike conventional OTFTs, OECTs are three-terminal devices (source, drain, and gate) where source and drain are connected through an organic semiconductor material. Contrary to conventional OTFTs, OECTs employ a gated electrolyte as the dielectric material rather than an insulating polymer or oxide layer. In consequence, OECTs do not possess defined two-dimensional channel interfaces since ions are able to penetrate through the entire semiconductor film. Hence, OECTs are volumetric devices, and the entire semiconductor layer acts as the transistor channel (cf. Fig.\,\ref{fig:OTFT_OECT}a). In OECTs, the ions in the electrolyte are employed to intentionally dope or dedope the semiconductor by a redox-reaction. In this way, polarons are added or removed from the conducting conjugated backbone of the polymer. The probably best-studied material system for OECTs is poly(3,4-ethylenedioxythiophene) polystyrene sulfonate (PEDOT:PSS) operated in a saline solution as shown in Figure \ref{fig:OTFT_OECT}b. Within the OECT, the balance of the redox-reaction 
\begin{table}[h]
	\centering
	\ce{PEDOT$^+$PSS$^-$ + M$^+$ + e$^-$ <=>[reduction][{oxidation}] PEDOT$^0$M$^+$PSS$^-$} \par\ 
\end{table}
can be manipulated by the gate potential where M$^+$ refers the concentration of metal cations (e.g., Na$^+$). In particular, the density of metal cations in the polymer film is governed by the electro-chemical potential E$_\mathrm{F}$ in the PEDOT:PSS film, which by itself is defined by the PSS-dopant concentration, the temperature, and the external gate potential in the framework of Boltzmann-statistics. In the field of electro-chemistry, this dependency is condensed in the Nernst equation written as 
\begin{align}
\mathrm{E_F=E_{F,0}+\frac{k_B T}{ze}ln\frac{[M^+:PSS^-][PEDOT^0]}{[M^+][PEDOT^+:PSS^-]}}
\label{eq:Nernst}
\end{align}
where E$_\mathrm{F,0}$ is the standard electrode potential of the electro-chemical cell (composed of PEDOT:PSS and the electrode) and z is the valence of the electron transfer. In this way, PEDOT can be switched from its insulating state (undoped/reduced state, PEDOT$^0$) to its conductive state (doped/ oxidized state, PEDOT$^+$PSS$^-$), depending on the concentration of metal cations M$^+$ compensating the charged sulfonate. Because the small ions (e.g., Na$^+$) can penetrate through the PEDOT film, every polymer chain can be seen as an independent nanoscale TFT channel. Thus, all these channels are connected in parallel which leads to a large transconductance in excess of 400\,\textmu S/\textmu m (normalized by channel width) \cite{Khodagholy2013} which is more than 400 times
larger than for solid-state OTFTs \cite{Klauk2007, Kitamura2011}. This large transconductance is caused by the high double-layer capacitance in excess of 1...100\,F/cm$^3$ \cite{Inal2017} (volumetric capacitance) which also enables millivolt operation of OECTs. These two aspects, operation in a saline solution and switching at millivolts, make OECTs ideal for application in bio-electronics.\\
In a first approximation, the operation of OECTs is quantitatively described by Bernards' model \cite{Bernards2007}. This model is based on the assumption that the OECT operation can be separated into an ionic and electronic sub-circuit, which are only coupled via the double-layer capacitance. The ion movement is described by an RC element, where the resistance is a measure for the conductivity of the electrolyte, and the capacitance accounts for the interfacial polarization capacitance at the semiconductor to electrolyte and gate electrode to electrolyte interface (Fig.\,\ref{fig:OTFT_OECT}c). Obviously, this approach must be seen as a small-signal approximation, losing its validity if Faradaic processes, such as electrolysis, occur. However, usually the Faradaic regime is avoided for OECTs. The electronic sub-circuit is composed of a series connection of resistances where the conductivity of each resistor is governed by the local polaron concentration (assumption of constant mobility), which is defined by the gate potential. This separation of the sub-circuits causes the system's response to be determined by two independent time-constants, namely the electronic $\mathrm{\tau_e}$ and ionic $\mathrm{\tau_i}$ time-constant, respectively. The electronic time-constant is given by the transit time of electrons/holes from source to drain $\tau_e$=L/(\textmu V$_{\mathrm{DS}}$). The ionic time-constant accounts for the RC-time of the ionic sub-circuit, which breaks down to $\tau_i\sim$l/c$^{1/2}$ (series resistance multiplied by capacitance) with l being the distance between the gate electrode and the semiconductor-electrolyte interface and c being the concentration of the electrolyte (c$^{1/2}$ is proportional to the double-layer capacitance according to the Gouy-Chapmann theory). Overall, the transient current-voltage behavior of an OECT in this model is derived as:
\begin{align}
\mathrm{I(t,\,V_{GS})=I_{SS}(V_{GS})+\Delta I_{SS}\left(1-f\frac{\tau_e}{\tau_i}\right)exp(-t/\tau_i)}
\label{eq:Transient}
\end{align}
where I$_{SS}$(V$_{GS}$) is the steady-state drain current, $\Delta$I$_{SS}$=I$_{SS}$(V$_{GS}$=0)-I$_{SS}$(V$_{GS}$), and f is a constant accounting for the spatial non-uniformity of the doping/de-doping process (f$\in[0,1]$). Within the same model, also the steady-state current-voltage behavior of an OECT can be derived, which obeys
\begin{align}
\mathrm{I}_\mathrm{D}&=\text{\textmu}\mathrm{e}\mathrm{p_0}\mathrm{t_\mathrm{OSC}}\frac{\mathrm{W}}{\mathrm{L}}\left[1-\frac{\mathrm{V_{GS}}-1/2\mathrm{V_{DS}}}{\mathrm{V_{P}}}\right]\mathrm{V_{DS}}&;&\quad \mathrm{V}_\mathrm{DS}\le \mathrm{V}_\mathrm{GS}
\label{eq:OECT_IV_1}\\
\mathrm{I}_\mathrm{D}&=\text{\textmu}\mathrm{e}\mathrm{p_0}\mathrm{t_\mathrm{OSC}}\frac{\mathrm{W}}{\mathrm{L}}\left[\mathrm{V_{DS}}-\frac{\mathrm{V_{GS}}^2}{2\mathrm{V_{P}}}\right]&;&\quad \mathrm{V}_\mathrm{DS}> \mathrm{V}_\mathrm{GS}
\label{eq:OECT_IV_2}
\end{align}
with t$_\mathrm{OSC}$ as the thickness of the semiconductor film, p$_0$ as the initial hole density in the semiconductor, and V$_P$ as the pinch-off voltage, defined as ep$_0$t/c$_d$ (c$_d$ is the specific capacitance per area of the double-layer capacitance).

\subsubsection{Materials for Organic Electro-Chemical Transistors}
\begin{figure}[htp]
	\begin{center}
		\includegraphics[width=.99\textwidth,clip]{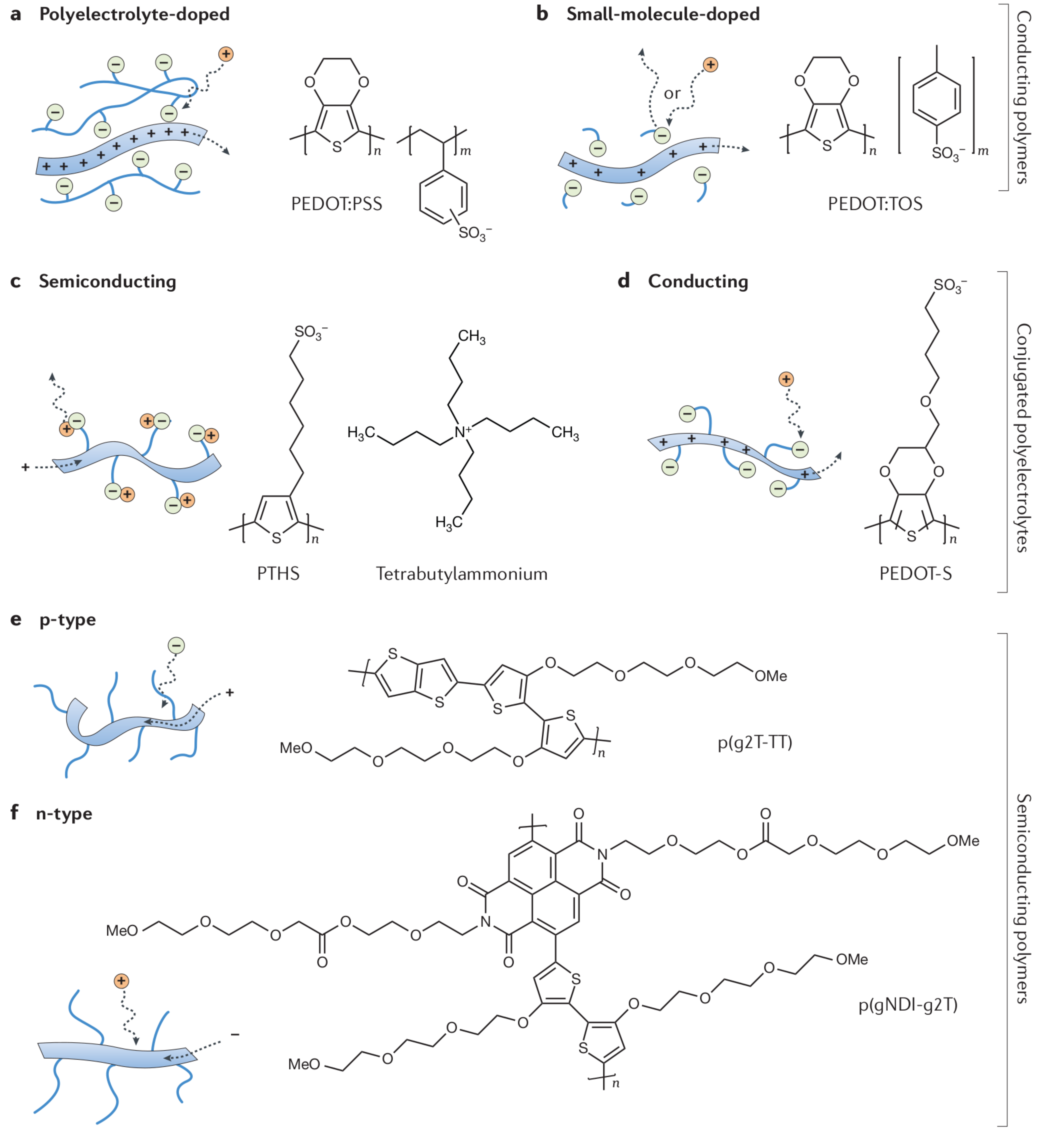}
	\end{center}
	\caption[]{Conducting polymers: \textbf{a} poly(3,4‐ethylenedioxythiophene) doped with poly(styrene sulfonate) (PEDOT:PSS) and \textbf{b} poly(3,4‐ethylene dioxythiophene) doped with tosylate (PEDOT:TOS). Polarons (represented by plus signs) on the PEDOT backbone are compensated by anions (green circles). Dedoping via injection of cations (orange circles) or extraction of anions, accompanied by extraction of holes. Conjugated polyelectrolytes: \textbf{c} poly(6-(thiophene‐3‐yl)hexane‐1‐sulfonate) tetrabutylammonium (PTHS) and \textbf{d}  poly(4-(2,3‐dihydrothieno[3,4‐b][1,4]dioxin‐2‐yl‐methoxy)-1‐butanesulfonic acid) (PEDOT‐S).
Semiconducting polymers: \textbf{e} p-type poly(2-(3,3'-bis(2-(2-(2-methoxyethoxy)ethoxy)ethoxy)-[2,2'-bithiophen]-5-yl)thieno[3,2-b]thiophene) (p(g2T-TT))and \textbf{f} n-type poly((ethoxy)ethyl 2-(2-(2-methoxyethoxy)ethoxy)acetate)-naphthalene-1,4,5,8-tetracarboxylic-diimide-
co-3,3'-bis(2-(2-(2-methoxyethoxy)ethoxy)ethoxy)-(bithiophene)) (p(gNDI-g2T)). These polymers can be doped by the injection of ions. Reprinted from \cite{Rivnay2018}, with permission from Springer Nature.}
	\label{fig:Materials}
\end{figure}

Although it would be possible to use small molecule semiconductors for OECTs, in practice, conductive polymers dominate this field of research. The reason for that lies in the OECT's inherent compatibility with solution-based processing and porous nature which allows ions to deeply penetrate into the semiconductor film yielding high values for the volumetric device capacitance. As mentioned above, PEDOT is the most frequently used conductive polymer utilized in OECTs due its high charge carrier mobility of $\sim$1\,cm$^2$/(Vs) and the solubility of its doped form in aqueous solutions. Furthermore, it can be deposited by an electro-polymerization process of its monomer EDOT which enables seamless integration into micro-electrode arrays, e.g., for sensing bio-signals. As shown in Figure \ref{fig:Materials}a and b, doping of PEDOT is achieved by adding p-type dopants such as small anions (e.g., tosylate, TOS) or polyanions (e.g., poly(styrene sulfonate), PSS). Additionally, the core of PEDOT might be doped more directly by attaching ionic side-chains to the PEDOT chain (Figure \ref{fig:Materials}c and d). In any case, the dopant creates a free polaron on the PEDOT chain resulting in the electrical conductivity of PEDOT being as large as 1000\,S/cm \cite{Wange2017}. When operated in an OECT, the anion causing the doping of PEDOT is neutralized by cations coming from the electrolyte. Thereby, the PEDOT becomes dedoped and hence losses its conductivity.\\
Although PEDOT is widely used in OECTs, it has a couple of disadvantages. In particular, PEDOT:PSS has a complex microstructure (composed of PSS-rich and PSS-poor domains), which reduces the volumetric capacitance. Moreover, PEDOT has a high Young's module which makes it partially incompatible with biological tissue. Furthermore, the acidity of PSS might lead to corrosion of metallic electrodes and undesirable interactions with living tissues.\\
Due to these disadvantages, other conjugated polymers such as polythiophenes or polypyrrole have been investigated as alternative hole-conducting polymers. These materials offer higher charge carrier mobility and a more simple microstructure than PEDOT (Figure \ref{fig:Materials}c). Moreover, even in their undoped state, they can be soluble in a large variety of different solvents, which facilitates the fabrication of accumulation-mode OECTs which cannot easily be achieved for PEDOT (Figure \ref{fig:Materials}e).\\
Aside from hole-conducting polymers, also more complex co-polymers composed of donor- and acceptor-units have been investigated for OECT. The particular advantage of these materials is their inherent ability to conduct holes as well as electrons. Hence, materials such as poly((ethoxy)ethyl 2-(2-(2-methoxyethoxy)ethoxy)acetate)-naphthalene-1,4,5,8‐tetracarboxylic-diimide-co‐3,3'-bis(2-(2-(2-methoxyethoxy)ethoxy)ethoxy)-(bithiophene)) (p(gNDI-g2T)) (cf.\~Figure \ref{fig:Materials}f) enable the fabrication of n-type or ambipolar OECTs \cite{Giovannitti2016}.\\
Another important constituent of OECTs, is the electrolyte used for gating. The electrolyte
might be a liquid such as a saline solution, an ion-conductive solid-state electrolyte \cite{Nilsson2002}, or gels \cite{Khodagholy2012}. Solid-state electrolytes might be ionomers such as Nafion \cite{Burgt2017} or ion-conductive polymer-metal composites such as Li-doped poly(ethylene carbonate) \cite{Tominaga2017} or poly(vinyl alcohol) \cite{Lee2010}. Also bio-compatible ion-conductive polymers such as chitosan or chitin might be used
\cite{Spyropouloseaau2019}. Similar, gels can be composed of ionic liquids added to a polymer matrix \cite{Khodagholy2012}. In any case, the conductivity of gels or solid-state electrolytes remains below the conductivity of liquid electrolytes. However, the advantage of these systems is the potential integration of multiple devices \cite{Ersman2019}. Furthermore, they enable OECT architectures where the gate electrode might be as close as several tenths of nanometers to the conductive polymer, enabling high frequency of operation \cite{Spyropouloseaau2019}, which would impossible using liquid electrolytes. 
\subsection{Applications for Organic Electro-Chemical Transistors}
OECTs act as transducers from ionic to electronic conductivity and vice versa. Obviously, this functions renders the possibility to employ OECTs as ion-sensors. In particular, as ions are used to directly dope or dedope the organic semiconductor in a high-capacitance volumetric structure, OECTs offer a superior sensitivity over ion-sensor devices based on ion-gated silicon transistors. \\
The potential of OECTs as ion-sensors goes clearly beyond the simple detection and quantification of ion concentration as needed e.g., for environmental monitoring. In particular, due to the softness and possible bio-compatibility of many organic semiconductor materials, OECTs hold great potential to be used as bio-sensor devices. For example, organs, certain tissues, or even single cells in the human body show electrical activity, e.g., electroencephalographic or electrocardiographic signals \cite{Cea2020}, and OECTs may act as an electronic interface to connect to electrically active tissues which helps to understand the function/ dysfunction \cite{Leleux2015} of these cells. For example, arrays of OECTs on flexible and bio-compatible substrates have been used to map electrical signals within the body, e.g., electrocorticograms \cite{Lee2017}), which can be used for monitoring, e.g., during brain surgery.\\
However, OECTs may not only revolutionize the field of bio-signal sensors but might even enable a new kind of hydrid-electronics connecting living neurons with artificial neural networks \cite{Keene2020}. In particular, since OECTs resemble in their function the synapses of a neural network, they might become the key-element for efficient brain-inspired neural networks (neuromorphic computing) \cite{Burgt2017a, Burgt2017, Burgt2018, Gkoupidenis2017}. Basic functions of a biological synaptic network such as homeostatic plasticity \cite{Gkoupidenis2017}, paired-pulse depression, dynamic filtering \cite{Gkoupidenis2015}, and short-term and long-term memory \cite{Gkoupidenis2017, Burgt2017} can be simulated. Neuromorphic systems composed of OECTs are a rapidly developing field of research with the aim to use these networks for advanced computational tasks such as pattern recognition. However, it remains an substantial challenge to increase the integration density of OECTs, improve the functionality of artificial synapses, and develop strategies for efficient learning of these systems.
\subsubsection{OECTs as Ion-Sensors}
Due to the doping and dedoping process caused by ions, OECTs are an obvious choice for ion-sensors \cite{Gualandi2019}. In particular, due to the large volumetric capacitance and the fact that the ions directly affect the channel interface of the transistor, OECTs offer excellent ion-sensitivity outperforming other sensor
\begin{figure}[htb]
	\begin{center}
		\includegraphics[width=.90\textwidth,clip]{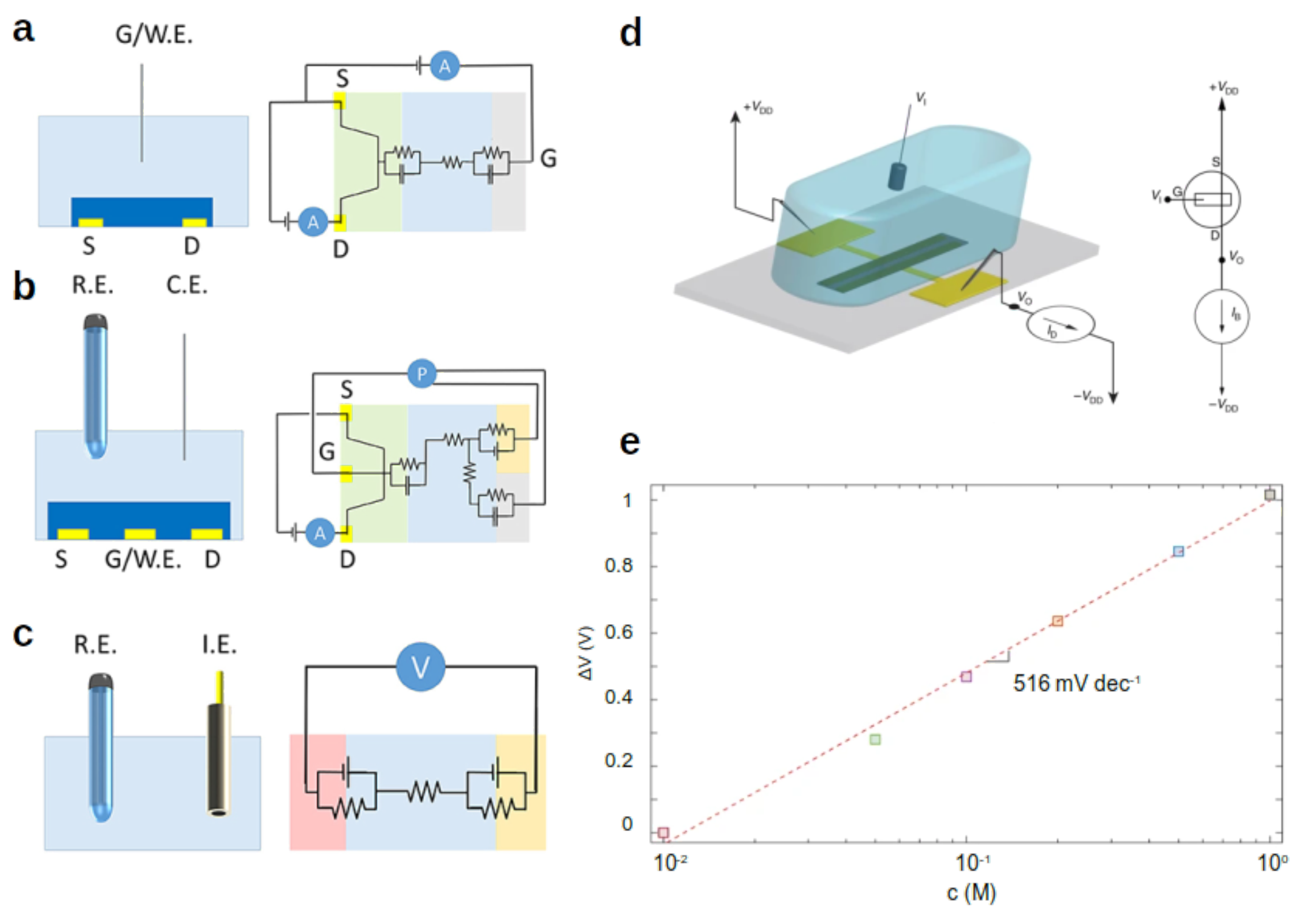}
	\end{center}
	\caption[]{\textbf{a} Typically OECT setup for ion-sensing and corresponding equivalent circuit with source (S), drain (D) and gate (G) or working electrode (W.E.). \textbf{b} OECT setup for ion-sensing and equivalent circuit including a potentiostat with reference electrode (R.E.) and counter-electrode (C.E.). \textbf{c} Conventional setup of a potentiometric measurement and corresponding electrical circuit with a R.E. and indicator electrode (I.E.). \textbf{d} Measurement setup for ion-sensing using an OECT in a current-driven inverter-like configuration, and \textbf{e} Sensitivity curve (voltage shift $\Delta$V vs. concentration c) for the structure shown in \textbf{d} (NaCl as electrolyte). \textbf{a}, \textbf{b}, and \textbf{c} reprinted from \cite{Gualandi2019}, with permission from Frontiers under Creative Commons Attribution License (CC BY). \textbf{d} and \textbf{e} reprinted from \cite{Ghittorelli2018}, with permission from Springer Nature under CC BY license.}
	\label{fig:Sensor}
\end{figure}
technologies based on inorganic semiconductor materials \cite{Ghittorelli2018, Wu2018}. Furthermore, OECTs and polymer-covered electrodes can be conveniently miniaturized, enabling the fabrication of compact, highly integrated sensor devices which is difficult to achieve with conventional electrode materials used in potentiostats.\\
Two functions are essential for ion-sensors. Firstly, the sensor must provide a measure for the ion concentration in the electrolyte. Secondly, it must be able to resolve the different kinds of ions from which it can ideally reconstruct the composition of the electrolyte. For OECT-based ion-sensors, the ion concentration is conveniently accessed through the channel resistance or the turn-off voltage V$_\mathrm{TO}$ of the OECT if the gate potential is varied (Figure \ref{fig:Sensor}\,a). As previously described (Equation \ref{eq:Nernst}), assuming Boltzmann-statistics, V$_\mathrm{TO}$ is linked to the ion concentration c through
\begin{align}
\mathrm{V_{TO}\sim -const.~ln(c).}
\label{eq:Nernst_2}
\end{align}
This simple approach though has a couple of shortcomings. Firstly, with this equation it is not possible to determine the concentration independently since the constant in Equation \ref{eq:Nernst_2} depends on the electrode materials, solvent, and many more. Secondly, this equation does not distinguish between cations and anions, and hence, it is a priori not clear which type of ions is actually detected. The first problem may be solved by recording a calibration curve for defined electrolyte concentrations. The second problem is related to the choice of the electrode materials. For example, PEDOT:PSS is only sensitive to the cation concentration (due to the high density of PSS$^-$-anions), and hence, if PEDOT:PSS is used as the channel and gate electrode material, only cations are being sensed. In contrast, using so-called electrodes of the second kind as the gate electrode (e.g., Ag, AgCl), the electro-chemical potential of the gate electrode is ruled by the activity of anions in the solution. Hence, since the electro-chemical potential of the gate electrode governs the transconductance of the OECT, AgCl electrodes might be used for selective sensing of Cl-anions. This approach of reactive electrode materials might be generalized (e.g., Ag/Ag$_2$S, Ag/AgI, Ag/AgBr, etc.) in order to achieve broad ion selectivity. Even polymer-based electrodes might be employed as second-order-electrodes in order to achieve selective cation-sensing \cite{Wu2018}. Furthermore, if in addition to the OECT, the electro-chemical cell is equipped with a potentiostat (see Figure \ref{fig:Sensor}b), even the absolute electro-chemical potential of the gate electrode may be determined which allows for a quantitative evaluation of the ion concentration, even without a calibration curve \cite{Gualandi2019}.\\
Despite the above-described electro-chemical cell configurations containing OECTs for ion-sensing, the most simple configuration for ion detection is given by a two-electrode structure with a reference electrode (R.E.) and indicator electrode (I.E.) as shown in Figure \ref{fig:Sensor}c. In this configuration, ion-sensing is realized by a high impedance analysis, measuring the voltage between R.E. and I.E. in the absence of an electrical current. Although this approach is rather simple and convenient to use, it bears the problem that a high impedance analysis is required (rather than a simple threshold-voltage measurement as for OECTs) and miniaturization of the cell is challenging due to an increasing complexity of the equivalent circuit caused by e.g., a non-uniform electric field.\\
All in all, OECT-based ion-sensors bring a clear advantage over existing electrode materials and cell configurations. As shown by Ghitorelli et al. \cite{Ghittorelli2018} (compare Figure \ref{fig:Sensor}d), OECTs can be integrated into an current-driven inverter-like configuration enabling precise determination of the ion concentration due to the differential readout of the inverter structure. In this way, the authors have proven that an unprecedented sensitivity of $\ge$500\,mV/dec  (Figure \ref{fig:Sensor}e) is reached which is approximately 20 times better than inorganic transistor-based ion-sensors. Furthermore, the authors highlighted that ion-selectivity might be achieved for this high-quality sensor if ion-selective membranes are used atop of the OECT stack. More specifically, they demonstrated selective ion-sensing of Na$^+$ and K$^+$ with a sensitivity as high as 400\,mV/dec.
\subsubsection{Neuromorphic Structures}
Today's computers are the masters of calculation. Up to 10$^{18}$ floating-point operations can be carried out during a single second (exaflop) and the performance is going to increase even further. Additionally, with the advent of advanced machine learning algorithms (artificial intelligence), computers are nowadays able to carry out complex tasks such as handwritten number recognition, speech recognition, etc.. However, despite all this progress, a severe problem became evident, which is the significantly higher energy consumption of computers for certain tasks (e.g., speech recognition) compared to the human brain. The origin of this problem lies in the so-called von-Neumann architecture, which is the basis of virtually all personal computers today. In this von-Neumann architecture, data storage and data processing are physically separated, and the data transfer between the storage unit and the processing unit adds a big share to the overall power consumption of the system. Thus, although it is possible to use the massive computational capacitance of today's computers to implement functions of
\begin{figure}[htb]
	\begin{center}
		\includegraphics[width=.80\textwidth,clip]{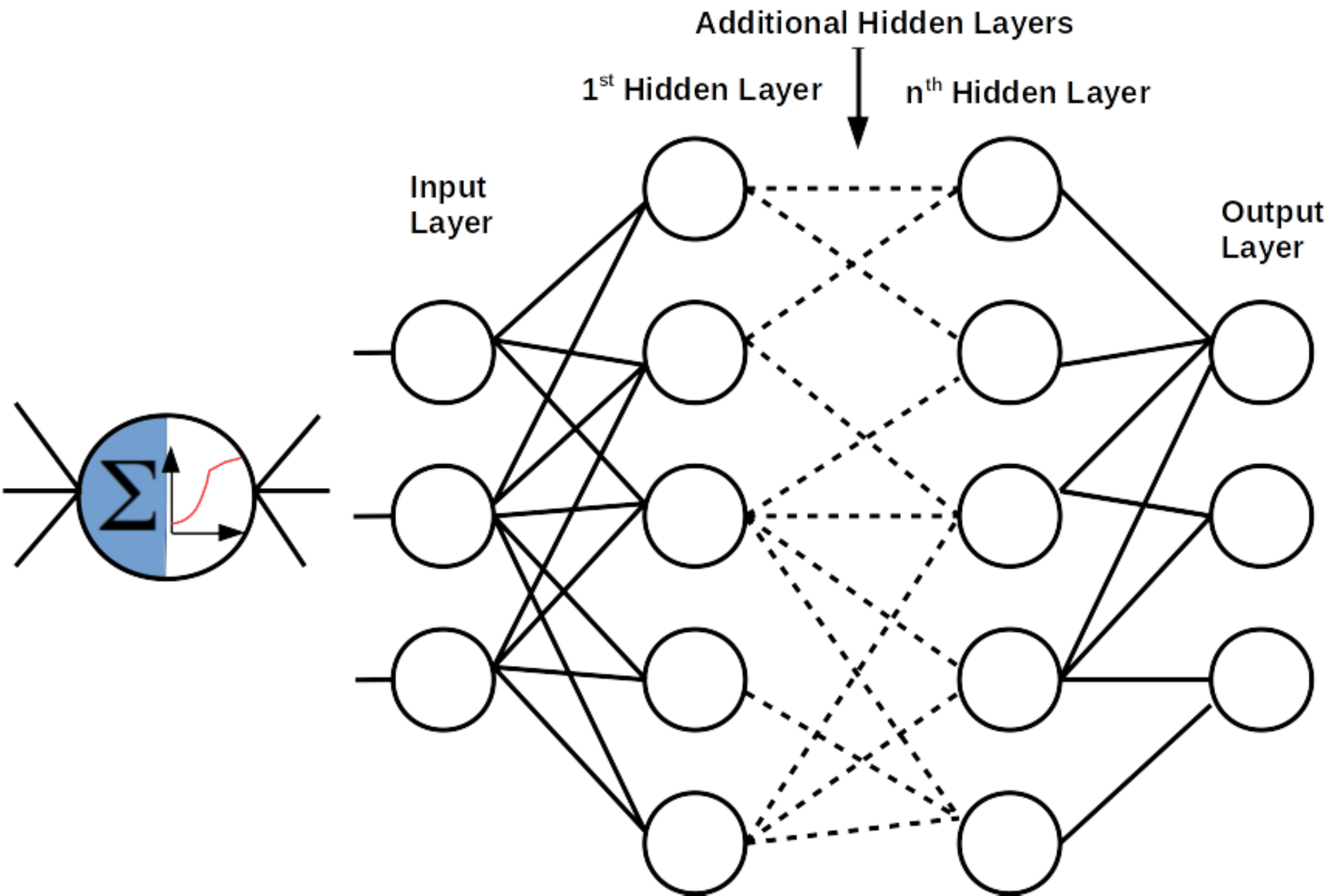}
	\end{center}
	\caption[]{(Left) Scheme of an artificial neuron. The neuron receive the input signals and builds the sum of those. This sum is used as an input function for a non-linear activation function. (Right) Graph of a feedforward artificial neural network composed of an input layer, several hidden layers, and an output layer. The connections between are the artificial synapses. Their synaptic weight is updated during the training process. Mathematically, each layer represents a vector-matrix-multiplication.}
	\label{fig:Network}
\end{figure}
artificial intelligence in software (e.g., support vector machine, deep, recurrent, or spiking neural networks - or in general, artificial neural networks (ANN)), the penalty is a massive power consumption. Although some of these algorithms adopt the principles of biological neural networks, the rigid and sequential character of the von-Neumann architecture prevents it from having the flexibility and efficiency of e.g., the human brain.\\
Neuromorphic structures are computational units (hardware) that are supposed to overcome the limitations of von-Neumann machines. They are sub-units of neural networks which are either truly hardware-based, or more often, a hybrid-system composed of software and hardware. Since any kind of neural network -- artificial as well as biological network -- is composed of neurons and synapses, it is the primary goal of neuromorphic structures to mimic the behavior of these two constituents. Thus, in the first place, neuromorphic structures might be identified as artificial synapses or neurons. However, since structural complexity is vital for neural networks as well, the term neuromorphic structures also refers to full architectures of artificial synapses and neurons, which are highly interconnected (see Figure \ref{fig:Network} as an example of a feedforward neural network). For a more detailed introduction to ANNs comparing e.g., feedforward and recurrent neural networks, the reader is referred to the book by Aggarwal \cite{Aggarwal2018}.\\
\\
\textbf{Artificial Synapses} The von-Neumann bottleneck is the physical separation of memory and computation. A synapse though is a so-called mem-computing unit, able to carry out computation and store information at the very same time. The key property enabling synapses to do so is denoted as plasticity \cite{Blitz2004}. It is the ability to tune the state of the synapse (e.g., represented by a conductivity value) depending on the history of reading/ writing of the synapse as well as the timing of the incoming spikes from the neurons. This temporal plasticity is often categorized into short- (STP, \textmu s-min regime) and long-term plasticity (LTP, in the range of minutes up to years) reflecting the different timescales of learning. STP and LTP may be further divided into synaptic depression (weakening of synaptic connection) and synaptic potentiation (strengthening of the synapse). The interplay between potentiation and depression is vital to almost any learning rule since a synapse needs to have the ability to incrementally modify its strength in any direction in order to enable efficient training of the entire synaptic network (cf. e.g., gradient ascent/descent algorithms for neural networks). However, plasticity as seen in biological neural networks is even more complex. Beyond LTP and STP, these mechanisms are referred to as higher-order plasticity, and they are essential for efficient training algorithms.
\begin{figure}[htb]
	\begin{center}
		\includegraphics[width=.99\textwidth,clip]{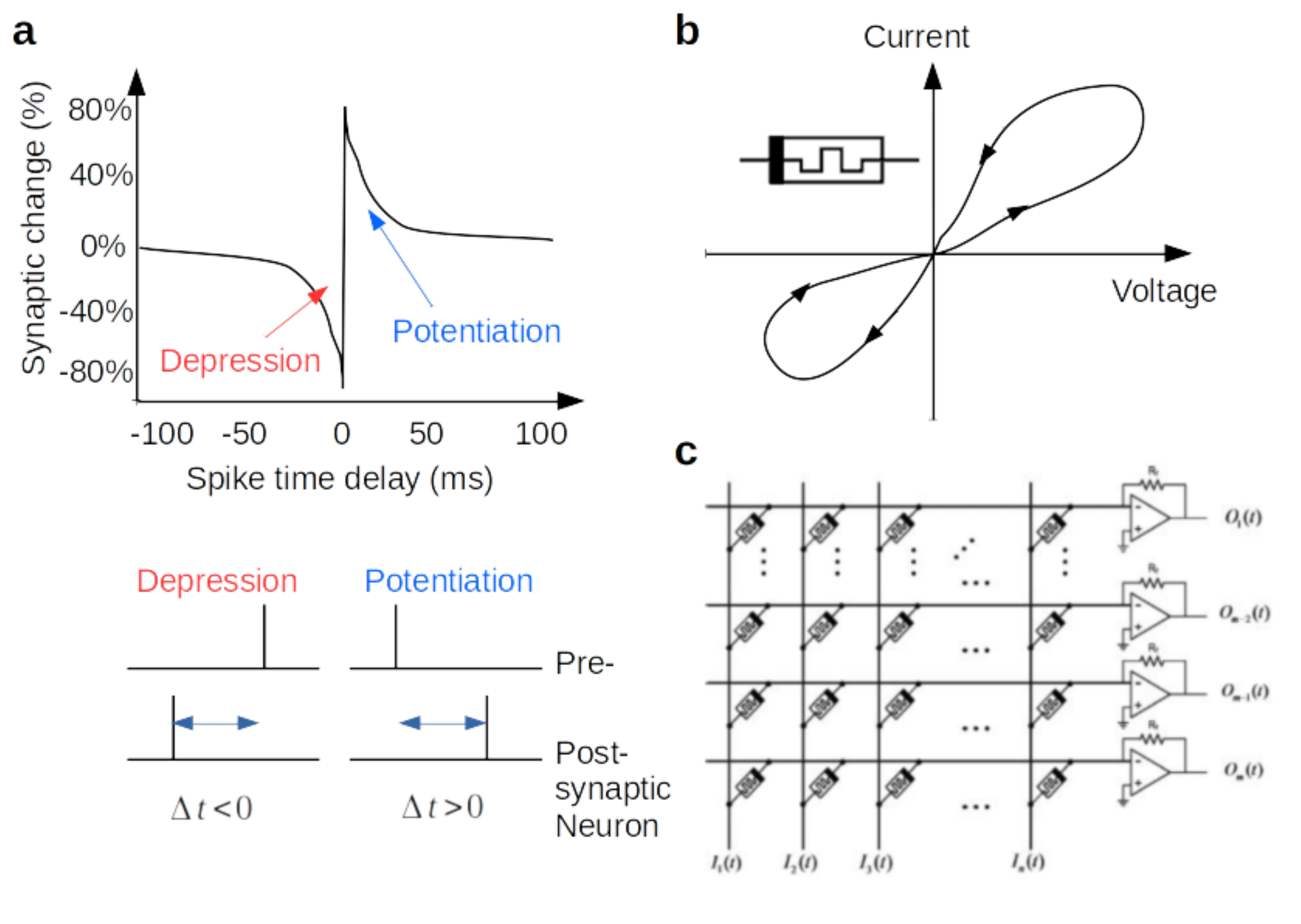}
	\end{center}
	\caption[]{\textbf{a} Functional dependency of the synaptic strength vs. the delay time of pre- and post-synaptic neuron firing (Spike-timing-dependent plasticity). The STDP-based depression and potentiation is shown. \textbf{b} Typical current-voltage curve of a memristor showing a pinched hysteresis. The inset shows the electronic symbol of an memristor. \textbf{c} Cross-bar array with a memristor at each crossing-point. The input signals I$_n$ are propagated through the memristors. At the outputs (O$_n$) the signals coming from each row are summed up. In this way, a vector-matrix-multiplication is performed.}
	\label{fig:Synapse}
\end{figure}
The best-understood type of higher-order plasticity is the so-called spike-timing-dependent plasticity (STDP) which is directly connected to the famous Hebbian learning rule. STDP refers to the synaptic strength being strongly increased if the signals from pre- and post-synaptic neuron arrive at the synapse at the same time. If the pre- or post-synaptic signals are shifted in time, the increase is weaker or even depression sets in if the post-synaptic neuron fires before the pre-synaptic neuron (see Figure \ref{fig:Synapse}a). This behavior is a direct implementation of the Hebbian learning rule - "neurons wire together if they fire together" \cite{Lowel209}. Hence, STDP is a very efficient way to train a synaptic network.\\
In the last decades, several semiconductor devices have been presented as artificial synapses. Among these devices are memristors \cite{Yao2020, Siebeneicher2012}, phase-change-memory \cite{Au2017}, and organic electro-chemical transistors \cite{Burgt2017}. In particular, memristor devices based on metal-insulator-metal structures received large attention due to the ease of integration, the compatibility with CMOS technology, and their high switching frequency. A memristor is a two-port device which is primarily characterized by the fact that it shows a pinched hysteresis in the current-voltage curve with the hysteresis reducing with increasing frequency of the voltage loop (see Figure \ref{fig:Synapse}b). The fact that the state of conductance of a memristor depends on its previous biasing condition can be understood as the synaptic plasticity of the memristor. Thus, putting a multitude of memristors together into a cross-bar array (see Figure \ref{fig:Synapse}), a neural network is realized, and vector-matrix-multiplication can be carried out fast and effectively in hardware (see Figure \ref{fig:Network}).\\
A bottleneck of metal-insulator-metal memristive devices is their device yield and the required voltage to change the state of conductance. In this regard, organic electro-chemical transistors may be an alternative. Similar to memristors, they allow to emulate synaptic plasticity but additionally, bear the advantage of low driving voltage (due to the high capacitance of the electro-chemical double-layer) \cite{Burgt2018}.\\
Another interesting aspect of the synaptic plasticity, which is usually disregarded when talking about memristors, is the ability of biological neural networks to dynamically rewire synaptic connections thereby improving the efficiency of problem-solving. This process is referred to as synaptogenesis which is an essential mechanism of the so-called homeostatic plasticity of neural networks. Since inorganic memristors are located at predefined positions in a cross-bar array and it is difficult to modify the interconnectivity during the network operation, the process of synaptogenesis is not available in such inorganic memristor networks. In OECTs though, synaptogenesis is possible, allowing for dynamic optimization of the synaptic network depending on the specific task.\\
\\
\textbf{Artificial Neurons} The second fundamental building-block of every neural network are neurons. They collect incoming signals, compute, and send out signals to other neurons. In general, the function of neurons is twofold (Figure \ref{fig:Network}). Firstly, all signals received from synapses are collected and summed up on the incoming side of the neuron. Thus, the neuron has a filter as well as an integrating function. Secondly, the integrated input signal becomes the argument of an activation function f and the value of this function is sent out from the neuron to the synapses. In principle, the activation function f is supposed to be non-linear, however, depending on the task, even simple linear functions such as step functions, the sign function, or similar might be adapted \cite{Aggarwal2018}.\\
These two functions of a neuron in combination with the synaptic plasticity are sufficient to build a first neural network for classification tasks which is only composed of one neuron and several synapses - the perceptron. Using labeled training data, the gradient of an error function is employed during the training phase to update the synaptic weights until the expected and predicted output of the neuron match. However, to increase the accuracy of the prediction and improve the fault-tolerance of the networks, more complex networks which are often composed of several layers, each containing many neurons, are required. Such systems are referred to as deep neural networks (DNN) or Deep Learning.\\
In comparison to a biological neuron though, the computational capability of the perceptron is incredibly small. The reason for that lies in the fact that artificial neurons, like in the perceptron, use predefined activation functions and the computational effort is transferred to a vector-matrix-multiplication. In contrast, biological neurons use a transformation based on a set of non-linear differential equations in order to classify data (compare e.g., the Hodgkin-Huxley-Model). More strictly speaking, the neurons transfer the input data from a low-dimensional phase space into a high-dimensional phase space which significantly simplifies classification tasks. Using such non-linear differential equations, similar data will be projected into one region of the phase space while slightly different data will go to a completely different part of the phase space. Thus, small differences in input data are amplified and translated into e.g., a different spiking frequency or amplitude which are very convenient to identify. Unfortunately, the implementation of biologically-inspired neurons into software-based ANNs is challenging since solving the set non-linear differential equations in real-time demands too much resources. However, using electronic circuits, such differential equations might be conveniently emulated and hence, brain-inspired artificial neurons are usually composed of a multitude of CMOS transistors forming e.g., a non-linear oscillator \cite{Indiveri2011}.
\subsection{Organic Electro-Chemical Transistors - Own Contributions} 
Organic electro-chemical transistors might become a powerful and versatile technology for bio-signal sensing as well as for neuromorphic computing. In particular, the combination of OECTs in a system for sensing and computation opens up the possibility to develop smart, functional sensor tags. Such systems might be used, for example, to record the heartbeat and do on-chip and real-time classification in order to detect cardiac dysfunctions. However, the development of OECT based sensors and neuromorphic systems is still at an early stage.\\
With regard to bio-sensing, often the underlying sensing mechanisms are not well understood. Furthermore, OECT-based sensors often show unfavorable cross-sensitivity to other environmental parameters which often make the readings of the sensor ambiguous. Thus, it remains an open challenge to develop OECT-based sensor platforms with high sensitivity and sufficient selectivity. With regard to neuromorphic computing, the field of OECT-based artificial neural networks has just emerged a few years ago. There are multiple challenges to overcome in this field such as 1) identifying the origins of synaptic plasticity, 2) building networks with controllable synaptic plasticity, 3) integration of OECT-based artificial synapses into complete ANNs including neurons, and 4), most importantly, to explore new approaches for hardware-based neuromorphic computing enabled by the unique properties of soft materials such as organic semiconductors.\\
Both, OECT-based bio-signal sensors as well as neuromorphic device structures are investigated in this work. These studies resulted in the following manuscripts \cite{Cucchi2021, Cucchi2021b, Tseng2021, Petrauskas2021}:
\begin{itemize}
    \item M. Cucchi, H. Kleemann, H. Tseng, G. Ciccone, A. Lee, D. Pohl, \& K. Leo, "Directed Growth of Dendritic Polymer Networks for Organic Electrochemical Transistors and Artificial Synapses", Advanced Electronic Materials, 2100586 (2021).
    \item H. Tseng, M. Cucchi, K. Leo, and H. Kleemann, "Membrane-Free, Selective Ion Sensing by Combining Organic Electrochemical Transistors and Impedance Analysis of Ionic Diffusion", ACS Applied Electronic Materials, just published  (2021).
    \item M. Cucchi, G. Gr\"uner, L. Petrauskas, P. Steiner, A. Fischer, B. Penkovsky, C. Matthus, F. Ellinger, P. Birkholz, H. Kleemann, \& K. Leo, "Reservoir computing with biocompatible organic electrochemical networks for brain-inspired biosignal classification", Science Advances \textbf{7}, eabh0693 (2021).
    \item L. Petrauskas, M. Cucchi, C. Gr\"uner, F. Ellinger, K. Leo, C. Matthus, \& H. Kleemann, "Nonlinear Behavior of Dendritic Polymer Networks for Reservoir Computing", Advanced Electronic Materials, 2100330 (2021).
\end{itemize}
\subsubsection{OECT-Based Neuromorphic Devices and Networks}
In this section, a new method for the fabrication of OECT-based neuromorphic devices and networks is discussed. This method, which is denoted as field-directed electropolymerization (FDP), allows for directed growth of polymer fibers between predefined electrodes. In this way, OECT-based neuromorphic devices can be created on-demand in a network of synapses. Furthermore, using FDP, synaptic connections within a network can be grown and dynamically modified in order to set the desired strength of synaptic plasticity. In this way, FDP emulated the process of synaptogenesis which is vital for the evolution of synaptic networks and efficient learning.\\
Beyond the level of synaptic plasticity and traditional artificial network approaches, FDP can also be employed to grow random conductive networks that show great promise for application in recurrent neural networks. In particular, the non-linear response of such networks coupled with the internal retardation of signals gives these networks an internal memory state, which allows the system (also denoted as a reservoir) to process and classify time-dependent information such as speech. 
\paragraph{Field-Directed Electropolymerization}
Polymeric materials for OECTs are typically processed by spin- or shear-coating or printing techniques \cite{Burgt2017, Worfolk14138}. For spin- or shear-coated films, the polymeric material is either patterned by photolithography and dry-etching, or more often by lift-off processes \cite{Gkoupidenis2015}. However, lift-off is a difficult process that suffers from reliability issues and it is challenging to prepare small structures.
\begin{figure}[ht]
	\begin{center}
		\includegraphics[width=.90\textwidth,clip]{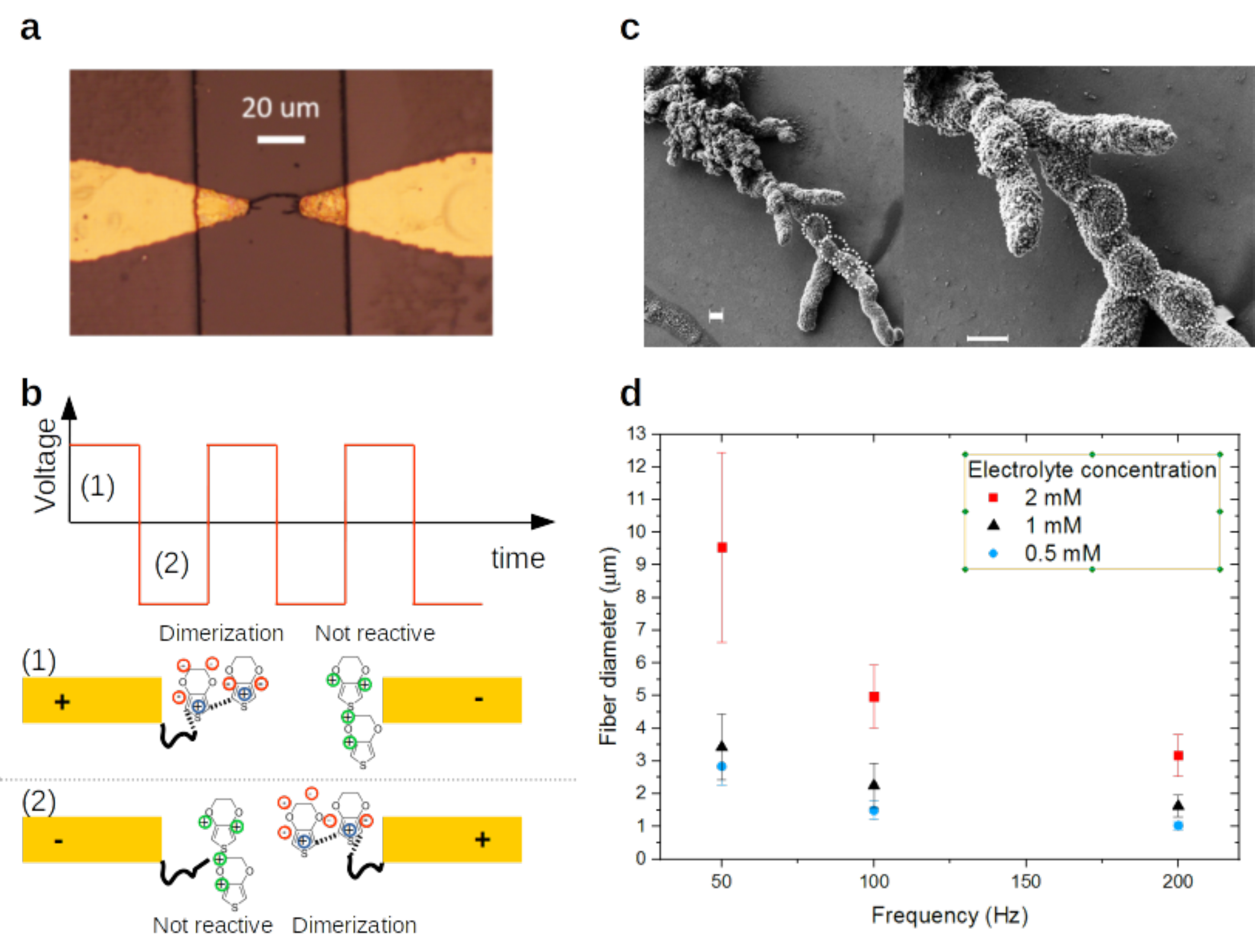}
	\end{center}
	\caption[]{\textbf{a} Photograph of a PEDOT:PF$_6^-$ fiber grown between two Au micro-electrodes (growth parameters: 1\,kHz, 2.5\,V). \textbf{b} Scheme of the FDP process. During the positive wave of the signal (1), dimerization and ultimately the growth of fibers occurs on the eletrode on the left. During the negative wave of the signal (2), the same process occurs on the other eletrode until the two fiber get in contact. \textbf{c} Scanning electron-microscopy image of a PEDOT:PF$_6^-$ fiber. The scale bar is 2\,\textmu m. Nucleation sites are visible as spherical structures. \textbf{c} Average diameter of PEDOT:PF$_6^-$ fibers grown at different AC-frequency and salt concentration. Figures adapted from \cite{Cucchi2021}.}
	\label{fig:FDP}
\end{figure}
Photolithography on materials such as PEDOT:PSS is difficult too, due to the acidity of the polymer causing undesired reactions in the photoresist. Furthermore, both techniques, lift-off and photolithography/ dry-etching can only be used for rigid structures that cannot be altered anymore after the processing. Hence, synaptogenesis cannot be emulated with spin- or shear-coated films. \\
Here a technique called field-directed electropolymerization (FDP) is proposed \cite{Cucchi2021, Cucchi2021}. This technique allows growing polymer fibers with defined properties on demand. These fibers may act as OECTs and artificial synapses, and hence, FDP enables the fabrication of synaptic networks where connections can be dynamically modified.\\
The original idea for the field-directed electropolymerization has been reported by Koizumi et al. \cite{Koizumi2016}. They have put two gold wires onto a substrate which was placed in a beaker. The beaker contained an electrolyte solution in which a monomer (EDOT) and a reducing agent was dissolved. Using large external electrodes biased with an AC-signal (42\,V), they observed the growth of thin polymeric fibers between the two fine gold wires which grow along the direction of the electric field. We modified this technique, and instead of two thin gold wires, a set of micro-electrodes is lithographically structured on the substrate. Furthermore, instead of external electrodes, the AC-signal is directly applied to the micro-electrodes. Due to these modifications, thin polymer fibers can be grown between the micro-electrodes (see Figure \ref{fig:FDP}a) at a significantly lower voltage (2 to 5\,V) and without the reducing agent. The electropolymerization process can be understood as follows (see Figure \ref{fig:FDP}b). During the positive wave of the AC-signal, the EDOT monomer forms radicals at the cathode interface. The electric field will force them to drift away and they might form a neutral complex with anions from the salt (e.g., EDOT$^+$:PF$_6^-$). In this case, the oxidized monomer is highly reactive, and in the presence of other monomer radicals an oxidative dimerization, oligomerization, and finally, polymerization will start. The long polymer chains are insoluble and deposit at the cathode. During each cycle of the AC-signal, this nucleation process occurs and the nucleation clusters are visible e.g., by scanning electron-microscopy (see Figure \ref{fig:FDP}c). During the negative wave of the AC-signal, cations accumulate to the electrode which dedopes the polymer, and hence, it is not reactive during the period. Since the same process occurs at the other electrode too, fibers periodically grow on either one or the other electrode. This process termines once the fibers from each side get in contact and form a conductive path. As a matter of fact, the growth does not stop abruptly but slowly continues until the overall conductivity of the fiber network is larger than the conductivity of the electrolyte.\\
The growth process can be understood as a diffusion-limited-aggregation. Thus, the interplay between the AC-signal frequency and the typical length scale of monomer diffusion during a period, determines the size of the nucleation centers. Hence, if the frequency is increased, the diameter of the nucleation sites gets smaller and in consequence, the diameter of the fibers too (see Figure \ref{fig:FDP}d). The same holds true for the salt concentration. A lower salt concentration results in smaller fiber diameter. The reason for this behavior is that the number of monomer radicals during each period is reduced due to a shortage of PF$_6^-$ anions.\\
The polymerization reaction always starts at the network extremities because the local field is the highest and the anion accumulation faster. If several fibers grow from one electrode, the reaction
does not occur in regions surrounded by the fibers as no net charge accumulates because of the negligible field (Faraday cage effect) as well as diffusion limited mechanisms. Frequently, bifurcations occur, resulting in higher branching degree of the network. The probability for bifurcations is increased for higher frequencies during the growth. Due to the bifurcations and the complex electrostatic situation, there is some randomness in the FDP process and same growth parameters do not necessarily translate into identical network structures. Still, using the same growth parameters, the electrical properties of the fiber networks are comparable and reproducible.\\
Overall, FDP enables the growth of polymeric connections with a size down to 1\,\textmu m of PEDOT:PF$_6$ between metallic electrodes. The connections can be grown on demand in an electrolyte solution. Since each fiber represents a fully operational OECT and an artificial synapse, FDP might be employed to mimic the process of synaptogenesis.
\paragraph{Controllable Plasticity of Synaptic Networks}
The polymer fibers of PEDOT:PF$_6$ grown by FDP form a highly conductive network. In particular, each fiber represents a fully functional OECT (see Figure \ref{fig:plasticity_1}a) and an artificial synapse. As it will be discussed, using FDP, the plasticity of each artificial synapse can be tuned over more than nine orders of magnitude from long-term to short-term and spike-timing-dependent plasticity. Furthermore, FDP enables mimicking
\begin{figure}[ht]
	\begin{center}
		\includegraphics[width=.99\textwidth,clip]{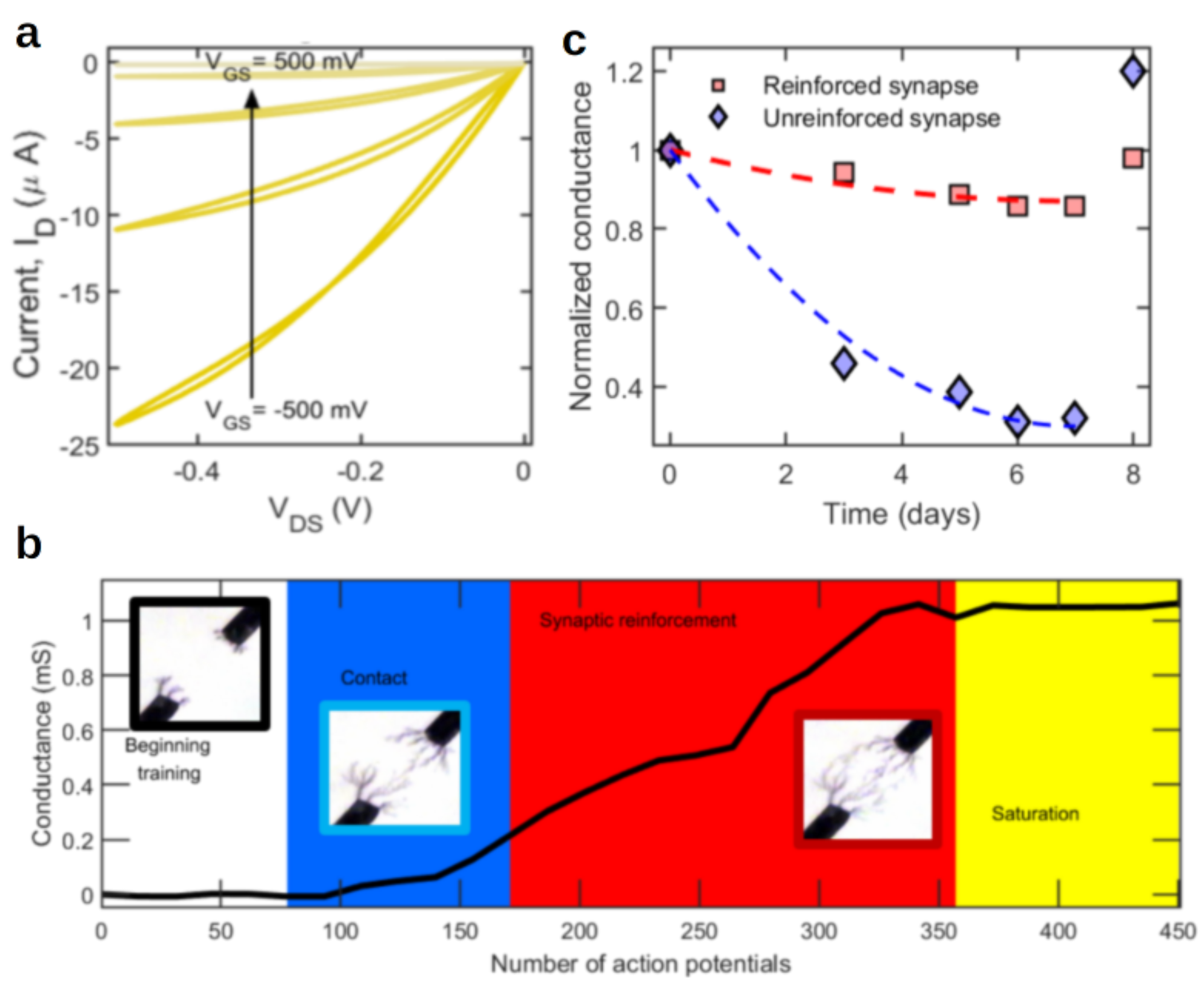}
	\end{center}
	\caption[]{\textbf{a} Output curve of an PEDOT:PF$_6$-based OECT grown by FDP. \textbf{b} Synaptic reinforcement using FDP. The sharp increase in conductance indicates gap bridging. After forming the first bridge (yellow area) more synapses/branches are formed \textbf{c} Comparison of the synaptic decay for a heavily and weakly reinforced synaptic network. Figures adapted from \cite{Cucchi2021}.}
	\label{fig:plasticity_1}
\end{figure}
learning-induced synaptogenesis. In order to visualize this mechanism, Figures \ref{fig:plasticity_1}b and c depict the process of synaptic reinforcement and depression based on synaptogenesis. Applying an AC-signal between two metal electrodes, the FDP process is initialized and fibers start growing. If the fibers have not yet bridged, the current originates exclusively from the conductivity of the electrolyte. Once the fibers bridge, a sharp increase in conductivity is obtained. The growth does not stop abruptly but rather continues until the conductivity of the fiber network is higher than the conductivity of the electrolyte, and finally saturates. This increase of the conductivity corresponds to a strengthening of the synaptic connection, which is a key biological mechanism leading to memory encoding and long-term memory retention \cite{Bailey01072015}. Furthermore, the S-shaped curve representing self-limiting learning is important for ANNs in order to avoid over-training. Since the conductance of this network is non-volatile, the synaptic reinforcement is a kind of long-term potentiation. The strength of this plasticity can be tuned by the FDP-parameters. In particular, low frequencies result in faster training and higher conductance due to larger fiber diameter (Figure \ref{fig:FDP}d).\\
However, for the brain forgetting is as vital as learning, and hence, it is mandatory to implement this function into the network of artificial synapses. One possible mechanism of forgetting of long-term memories is the synaptic decay, which can be understood as a kind of long-term depression (LTD). In the PEDOT:PF$_6$ synaptic networks, this kind of LTD occurs due to the gradual wash-out of the PF$_6^-$ from the fibers, causing a reduction of the conductivity. Using FDP, the time-constant of this decay can be controlled by the degree of synaptic
\begin{figure}[ht]
	\begin{center}
		\includegraphics[width=.99\textwidth,clip]{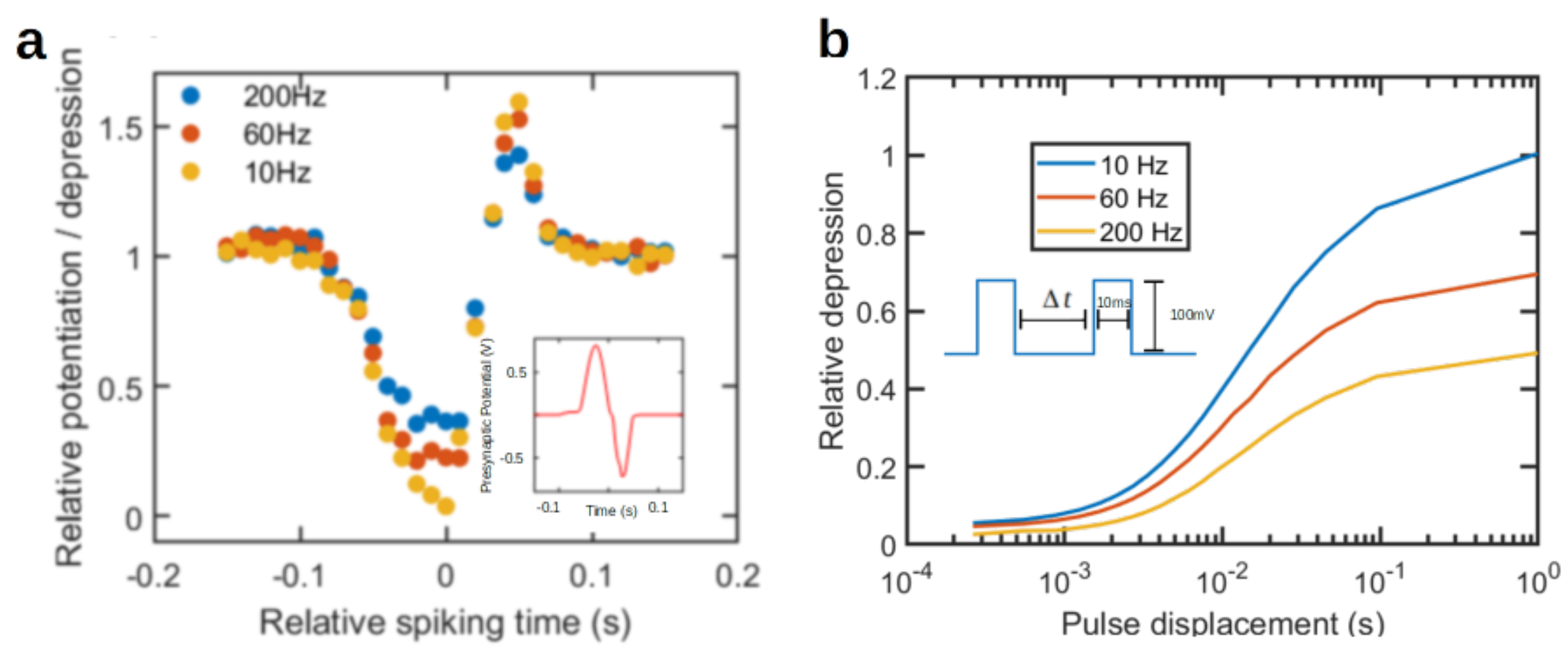}
	\end{center}
	\caption[]{\textbf{a} Spiking-timing-dependent plasticity of FDP-grown artificial synapses. The STDP behavior is shown for synapses grown at different frequencies. The inset shows the timing-diagram of the pre-synaptic neuron.  \textbf{b} Short-term depression (STD) of FDP-grown artificial synapses. The STD behavior is shown for synapses grown at different frequencies. The inset shows the timing-diagram of the pre-synaptic neuron. Figures adapted from \cite{Cucchi2021}.}
	\label{fig:plasticity_2}
\end{figure}
reinforcement. Comparing two networks, one heavily and the other weakly reinforced, a significant difference in the synaptic decay can be observed in Figure \ref{fig:plasticity_1}c. While the heavily reinforced network does not change its state of conductance within 48\,h, the weakly reinforced system losses almost 50$\%$ of its original conductivity. Thus, the process of synaptogenesis emulated by the FDP-growth is successfully used to create long-term potentiation and depression with controllable time-constants.\\
Besides long-term plasticity, also effects of short-term plasticity (typically happening in the ms-scale) can be seen in FDP-grown synaptic networks. In particular, spike-timing dependent plasticity, which is an important learning mechanism, is presented in Figure \ref{fig:plasticity_2}a. A large potentiation is observed if pre- and post-synaptic neuron fire 'almost' at the same time. A strong depression though is attained in the case the post-synaptic neuron fires before the pre-synatic neuron (acausal case). Most interestingly, the degree of
\begin{figure}[ht]
	\begin{center}
		\includegraphics[width=.99\textwidth,clip]{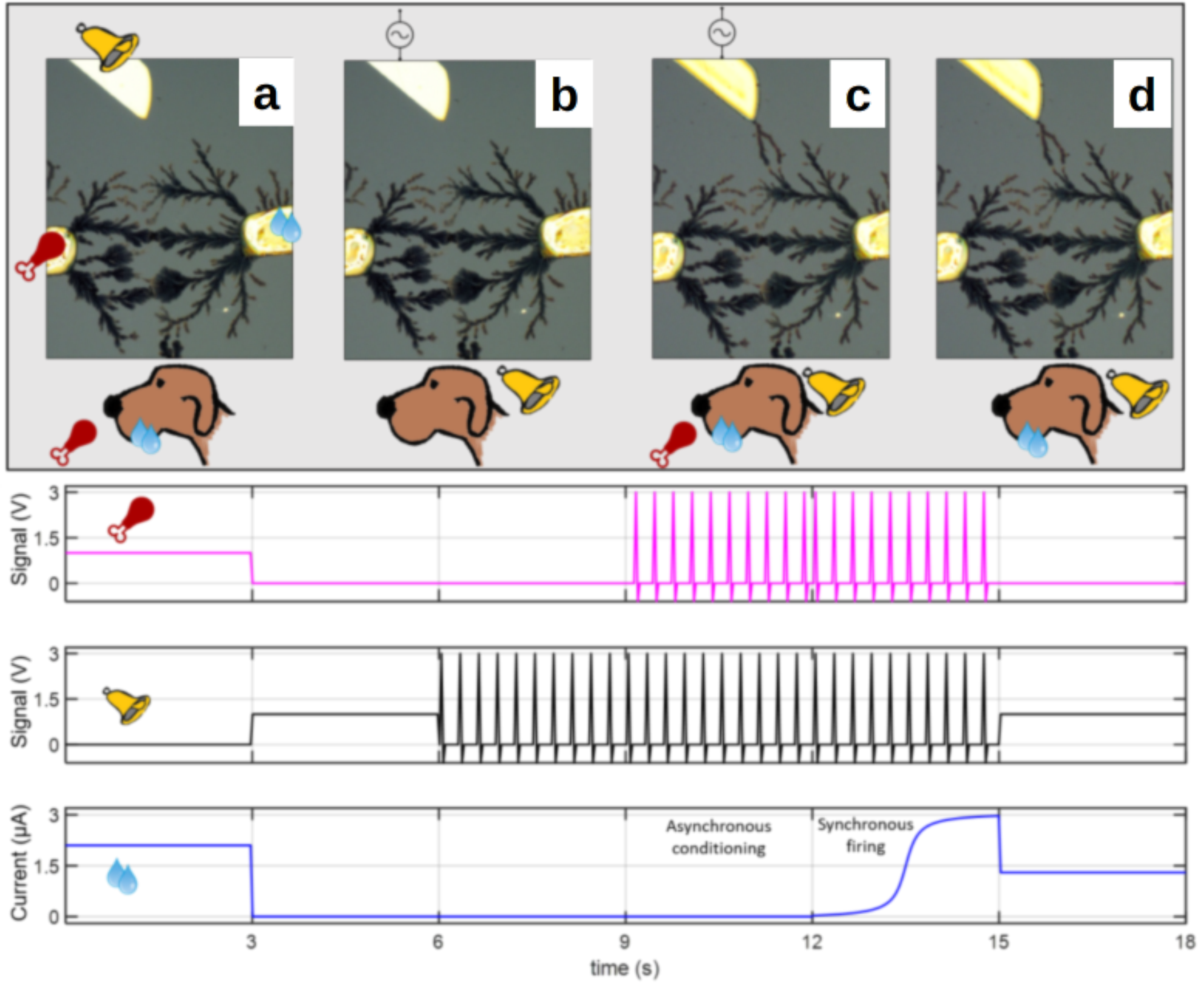}
	\end{center}
	\caption[]{Pavlovian condition using FDP - photographs of synaptic connections (top) and timing-diagram (bottom). \textbf{a} Initial condition, no connection between "bell" and "salivation" node. \textbf{b} Asynchronous firing of "bell" and "food" node - no connection between "bell" and "salivation" is grown. \textbf{c} Synchronous firing of "bell" and "food" node - conditioning: new connection is grown. \textbf{d} Dog is conditioned to salivate if the bell is ringing. Figure adapted from \cite{Cucchi2021}.}
	\label{fig:plasticity_3}
\end{figure}
synaptic depression/potentiation can be tuned by the frequency applied during FDP-growth. Similar to the LTP mechanism, synapses grown at low frequency are strongly reinforced. Due to this fact, cations are impeded from penetrating quickly in or out of the polymer fibers, and hence, a stronger depression/potentiation level is obtained.\\
Besides complex STDP, also regular short-term plasticity is seen in the polymer networks. Figure \ref{fig:plasticity_2}b displays the short-term depression of the synaptic strength measured for two consecutive pre-synaptic pulses with varying pulse displacement $\Delta$t. As can be seen, no depression is observed for pulse displacement times $\le$\,1\,ms representing the fastest response time of the system. With increasing displacement time, a strong synaptic depression is observed and its strength depends on the frequency during the FDP-growth. Again, the dominant mechanism behind this effect is the synaptic reinforcement. In strongly reinforced synapses (larger fiber diameter), the penetration of cations is impeded.\\
Overall, reliable control over synaptic properties in electropolymerized devices is given by the growth conditions. Hence, the FDP process can be effectively used to produce evolvable circuitry in electrolytic solutions.\\
As a first primitive example highlighting the potential of FDP for evolving ANNs, Pavlovian conditioning based on synaptogenesis is demonstrated in Figure \ref{fig:plasticity_3}.\\
In the beginning, there is an already existing connection between the "food" and the "salivation" node, while the "bell" node is not connected to the system (Figure \ref{fig:plasticity_3}a). In order to train the dog to salivate if the bell rings, a new connection between these two nodes needs to be created. The "bell" signal, represented here by a 3\,V, 50\,Hz action potential, is not sufficient to initialized the growth of new fibers through FDP (Figure \ref{fig:plasticity_3}b). If an additional action potential (3\,V, 50\,Hz) is emitted from the "food" node, the threshold for the FDP-growth is exceeded. However, if both signal ("bell" and "food") are emitter asynchronously, fibers do not grow (\ref{fig:plasticity_3}c). Only if the two signals are synchronous, a new connection between the "bell" node and the "salivation" node appears, and hence, the dog is ultimately conditioned to salivate if the bell rings (Figure \ref{fig:plasticity_3}d).\\
Another, more complex, example for a functional ANN composed of FDP-grown synapses is shown in Figure \ref{fig:plasticity_4}. In this particular network, 15 synaptic connections are employed for a number recognition task, using the effect of synaptogenesis as well as short term depression.
\begin{figure}[ht]
	\begin{center}
		\includegraphics[width=.90\textwidth,clip]{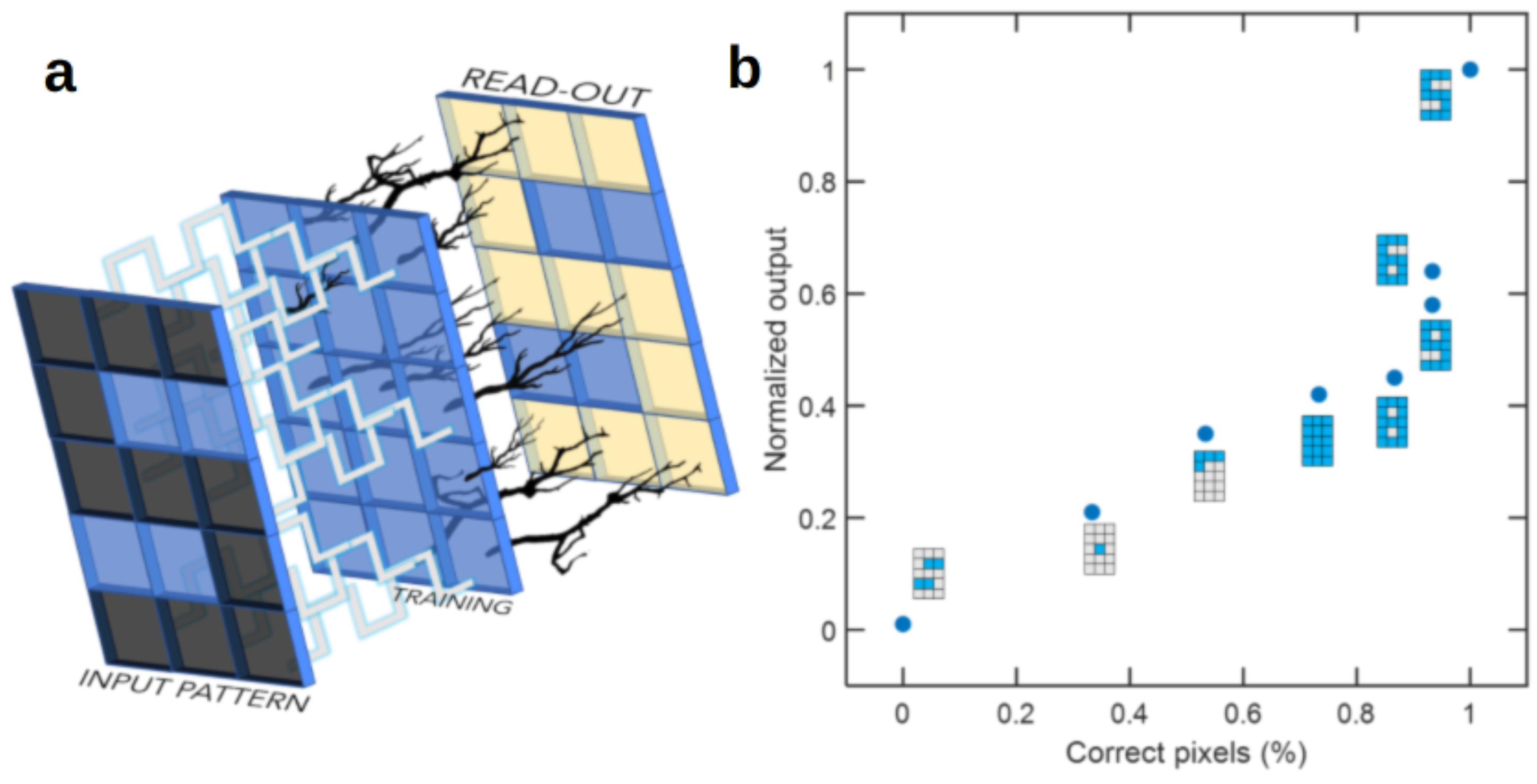}
	\end{center}
	\caption[]{Digital number recognition using 15 FDP-grown synapses. \textbf{a} Illustration of the fiber network built during the training phase. \textbf{b} Result of the readout. The network is training to recognize "5" and classify other digits based on the number of false pixels compare to "5". Figure adapted from \cite{Cucchi2021}.}
	\label{fig:plasticity_4}
\end{figure}
During the training phase, a train of pulses is applied to the network in order to build the synaptic connections. Black pixels are represented by an action potential of 5\,V, while void pixels refer to a voltage of only 2\,V. Thus, synaptic connections are only grown if the pixel is black (\ref{fig:plasticity_3}a). For the readout, the short term depression mechanism is employed. The highest current reading is recorded in case all pixels emit the same signal as during the training phase (here the pattern "5", Figure \ref{fig:plasticity_3}b). If during the readout, the "wrong" pattern is emitted, e.g. "6" (one false pixel compared to "5"), then the integrated output current is reduced since the synaptic strength of the programmed connections for "5" is weakened by the signal emitted from the false pixel which acts as a gate and causes dedoping of the PEDOT fibers. The more false pixels in the structure, the stronger the synaptic depression. Finally, for the inverted pattern of "5", the current reading is zero. Hence, the network can classify the digital numbers and the number of false pixels compare to the programmed value "5".\\
Overall, these are the first experiments towards the implementation of FDP-grown synapses into ANNs. In order to solve more complex problems such as handwritten number recognition, the synapses need to be integrated into a big cross-bar structure containing several thousand or millions of synapses. Furthermore, the fabrication of complete ANNs based on OECTs requires the development of powerful and versatile artificial neurons in the future. 
\paragraph{Reservoir Computing}
Since traditional artificial neural network approaches usually require big cross-bar arrays and a multitude of neurons, it is questionable to some extent whether OECT-based artificial synapses can compete with the inorganic semiconductor technology. In particular, silicon-based semiconductor technology is practically unbeatable in terms of device scaling, speed of computation, and fabrication of millions of perfectly equivalent devices.\\
However, there are alternative approaches for ANNs which are by far less resources-demanding and do not
\begin{figure}[ht]
	\begin{center}
		\includegraphics[width=.90\textwidth,clip]{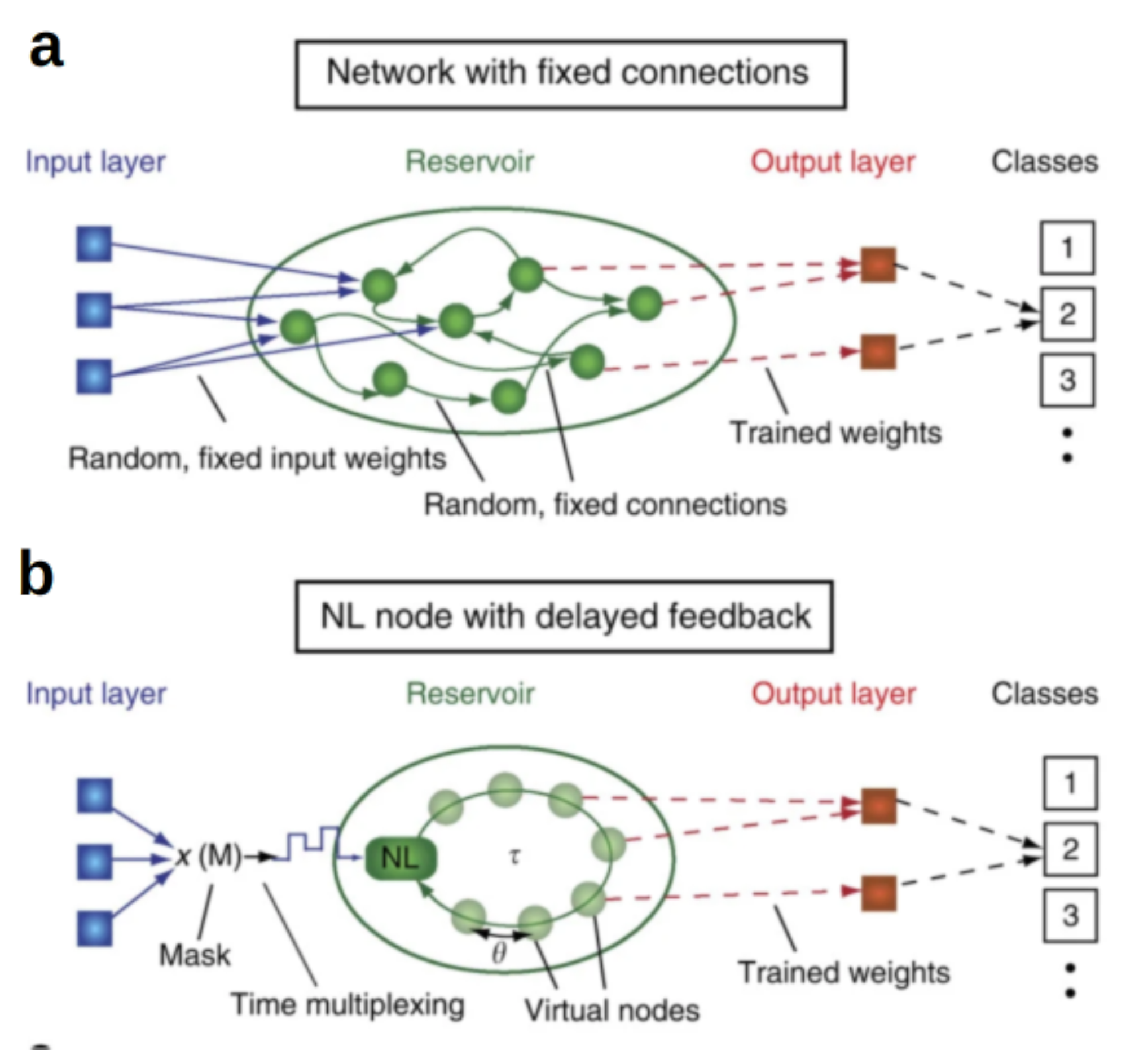}
	\end{center}
	\caption[]{\textbf{a} Scheme of classical reservoir computing. The input signal is coupled into the reservoir via fixed input weights. The connections between reservoir nodes are randomly chosen and kept fixed during training. The reservoir's response is projected during the training phase using a linear regression onto the output layer thereby defining the targeted classes. \textbf{b} Scheme of a single-node or delayed feedback reservoir. The behavior of the reservoir is emulated by the delay loop and the time multiplexing. The input states are sampled and held for a duration $\tau$, where $\tau$ is the time-constant of the delay-line. Because of the delay-line, the input data needs to be modified transformed by an input vector into a matrix where each column represents one delay step $\tau$. Figures adapted from \cite{Appeltant2011}, published under CC BY-NC-SA 3.0 license for free non-commercial reuse.}	
	\label{fig:reservoir_1}
\end{figure}
require the precision and reliability of silicon-based technology. One of these alternative approaches for ANNs is the so-called reservoir or reservoir computing, which is a special kind of a recurrent neural network. A reservoir can carry out classification tasks by projection of the input signal into a high-dimensional phase space via a non-linear transformation. While the classification of data might be difficult in low-dimensional phase space, it becomes increasingly more likely in a high-dimensional space. In this way, the function of a reservoir resembles that of a biological neuron. Like all neural networks, also a reservoir is composed of input, output, and a multitude of hidden nodes (see Figure \ref{fig:reservoir_1}a). In comparison to e.g., feedforward neural networks, the connections between the hidden layers are not trained in a reservoir during learning. Only the output of the reservoir will be trained using linear regression in order to project the output of the reservoir onto the output nodes. Thus, training of the reservoir is by far more efficient than the training of other neural networks. The reservoir itself though is composed of a set of random connections between the nodes which remain unchanged during training and operation of the network. The reservoir has two primary tasks: 1) to do the non-linear transformation, and 2) to give memory to the system. The memory of the reservoir originates from its recurrence, and it is necessary for the reservoir in order to classify a series of signals in time, e.g., for speech recognition.\\
Overall, reservoir computing is a powerful method for the classification of time-dependent signals with significantly lower training efforts than other ANNs. Due to the non-linear transformation, it is more complex to implement in software compared to other ANNs, however, it is perfectly suited for hardware-based computation since it does not require artificial synapses but rather can be composed of almost any non-linear element.
\begin{figure}[ht]
	\begin{center}
		\includegraphics[width=.90\textwidth,clip]{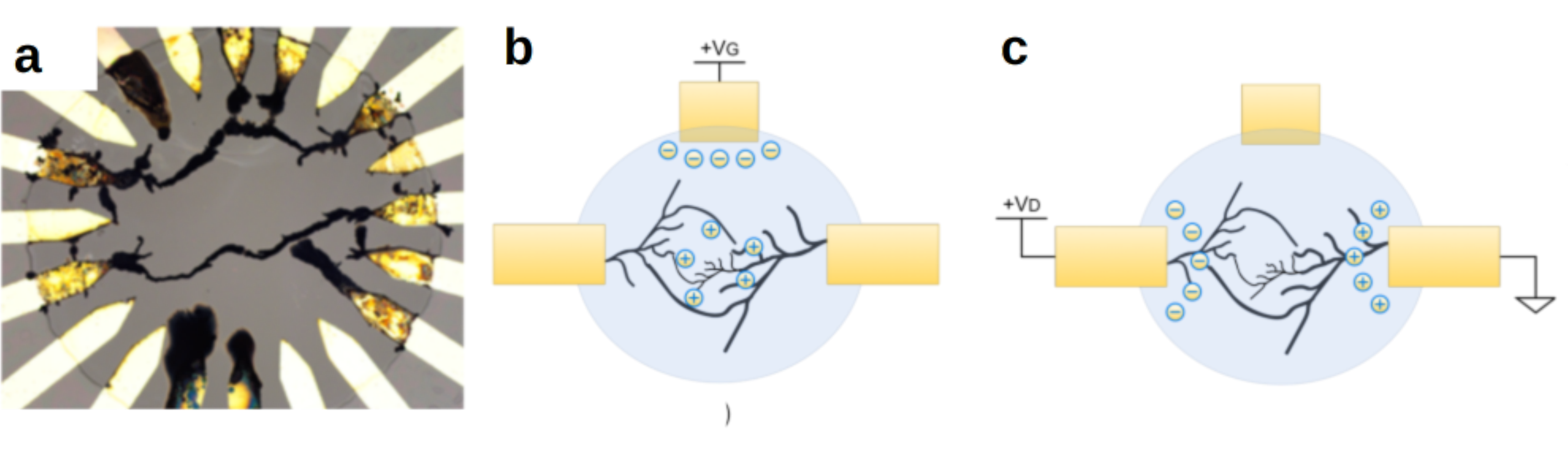}
	\end{center}
	\caption[]{\textbf{a} Photograph of an OECT-based reservoir. Input nodes are on the left, output nodes on the right. Electrodes with a polymer cover are employed as a gate electrode. \textbf{b} Illustration of the gating effect in an OECT. The strength of the gating effect depends on the position of each fiber (Faraday cage-effect). \textbf{c} Illustration of the self-gating effect in an OECT. If a time-dependent signal is applied between source and drain, the finite propagation of the signal through the electrolyte will create a local non-uniformity of the doping profile. Figure reprinted with courtesy of \cite{petrauskas2020}.}
	\label{fig:reservoir_2}
\end{figure}
One short-coming of reservoirs is that they need a large number of nodes in the hidden layer in order to be efficient and ensure sufficient recurrence. This problem though can be solved using so-called single-node or delayed-feedback reservoir systems. Instead of thousands of nodes, just a single node is used in such a system for the same computational task. The other nodes and hence the recurrence are emulated by a delayed feedback line that loops the signal back and performs multiplexing of the signal (cf. Figure \ref{fig:reservoir_1}b). For more details about reservoir computing and delayed-feedback reservoirs, the reader is referred to Penkovsky and Appeltant et al. \cite{Penkovsky2017, Appeltant2011}.\\
A reservoir composed of FDP-grown PEDOT:PF$_6$ fibers is shown in Figure \ref{fig:reservoir_2}a. Two fibers connect from the two input nodes to several output nodes. Several electrodes are also covered with the polymer and act as gate electrodes to the polymer fibers. As described above, an important property of the reservoir is its non-linear response. In this regard, four important questions arise which will be discussed in the following: 1) is the response of the network non-linear, 2) what is the origin of the non-linearity, 3) how can the level of non-linearity be tuned, and 4) which level of non-linearity is needed for powerful computation?\\
Each fiber represents an organic electro-chemical transistor. By definition, the response function of an OECT is non-linear to some extent. In particular, OECTs show a non-linear current-voltage behavior near the point of saturation and the subthreshold regime. While the non-linearity is weak close to the saturation regime (only second-order), the subthreshold regime offers higher-order non-linearity. Thus, the response function of a single-fiber OECT is non-linear and the level of nonlinearity can be set by the gate-source voltage. This description though is too simple since it only takes into account the static current-voltage behavior of an OECT and it does not reflect the complex situation in a network of fibers where local non-uniformities in the doping profiles or self-gating effects might occur. For example, if a gate voltage is applied via an external gate electrode (see Figure \ref{fig:reservoir_2}b), then the strength of the gating effect will depend on the position of each fiber since
\begin{figure}[ht]
	\begin{center}
		\includegraphics[width=.90\textwidth,clip]{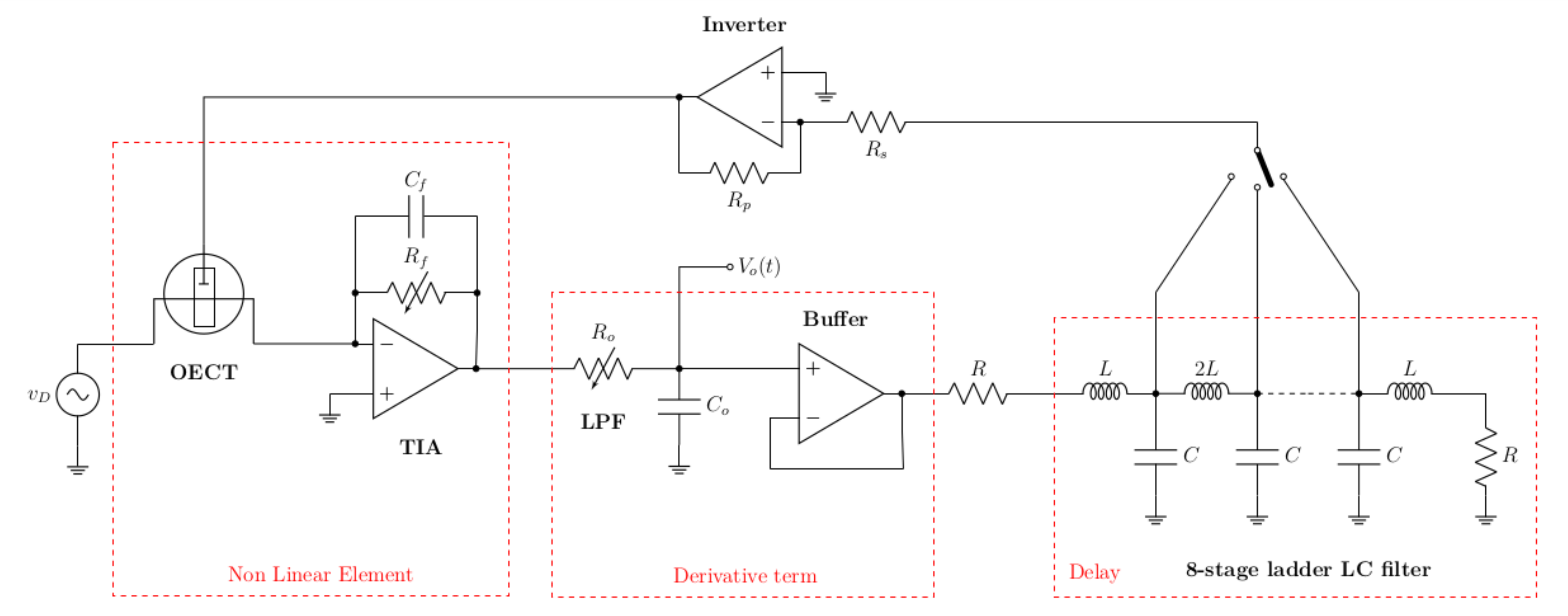}
	\end{center}
	\caption[]{Circuit diagram of the delayed-feedback line and the non-linear element. The feedback line is composed of a trans-impedance-amplifier (TIA), a low-pass filter (LPF), the delay-line, a derivative term, and an amplifier/ inverter. Figure reprinted with courtesy of \cite{petrauskas2020}.}
	\label{fig:reservoir_3}
\end{figure}
conductive fibers will screen the electrical field created by the gate electrode. Hence, each fiber-based OECT will show a different level of non-uniformity. Furthermore, having Bernards' model in mind \ref{eq:Transient}, due to these differences in the local doping profiles (caused by the screening of the electrical field), each fiber will show a different transient response. Additionally, since time-dependent signals are applied to the input and output nodes (referring to source and drain) during reservoir computing, there will also be a so-called self-gating effect which is caused by the finite propagation time of a signal through the electrolyte (see Figure \ref{fig:reservoir_2}c). In principle, this self-gating effect does not only occur at the metallic electrodes but fibers are gating each other, which in total results in a rather complex coupling of all electrical components in such a fiber-based network. Thus, the non-linear response is a property of the individual OECT which is further amplified by the coupling with the network.\\
The remaining question is whether the specific non-linearity is sufficient for reservoir computing? To answer this question, it is important to remember what the reservoir does in order to carry out a classification task - it transforms the input signal in phase space from a low to a high dimensional space. For example, if the trajectory of a 2-dimensional input signal $\Vec{\mathrm{x}}$(t) is a circle, then the trajectory of the output signal $\Vec{\mathrm{y}}$(t) is a torus or any other more complex structure. Ideally, the input signal would be transformed into a phase space with infinite dimension which is referred to as chaos. Thus, in order to prove that the non-linearity is strong enough, it is sufficient to show that the system can show chaotic behavior.\\
In order to do so, the reservoir is combined with a delayed-feedback line which is composed of a trans-impedance-amplifier (TIA), a low-pass filter (LPF), the delay-line, a derivative term, and an amplifier/ inverter (cf. Figure \ref{fig:reservoir_3}). Thus, the differential equation of the system may be written as
\begin{equation}
    \mathrm{\tau_{NL}} \dot{\Vec{\mathrm{y}}}(\mathrm{t}) + \Vec{\mathrm{y}}(\mathrm{t})=\mathrm{\lambda} f(\Vec{\mathrm{y}}(\mathrm{t-\tau}), \Vec{\mathrm{x}}(\mathrm{t}))
    \label{eq:Reservoir_1}
\end{equation}
where $\mathrm{\tau_{NL}}$ is the inherent delay-time of the non-linear system, $\mathrm{f(\cdot)}$ is the non-linear transfer function of the reservoir, $\mathrm{\tau}$ the delay-time of the delay-line, and $\mathrm{\lambda}$ is an amplification factor. In the so-called adiabatic approximation ($\mathrm{\tau_{NL}}\ll\mathrm{\tau}$), this equation breaks down to
\begin{equation}
   \Vec{\mathrm{y}}(\mathrm{t})=\mathrm{\lambda} f(\Vec{\mathrm{y}}(\mathrm{t-\tau}), \Vec{\mathrm{x}}(\mathrm{t})).
    \label{eq:Reservoir_2}
\end{equation}
Unfortunately, there is no strict formalism that one can follow in order to prove that this equation leads to chaos. Thus, the only way is to solve this equation numerically or experimentally by direct integration using the non-linear element and the proposed delay-line architecture. The numerical solution shows that even for the most simple non-linearity (second-order or logistic map, 
\begin{figure}[ht]
	\begin{center}
		\includegraphics[width=.90\textwidth,clip]{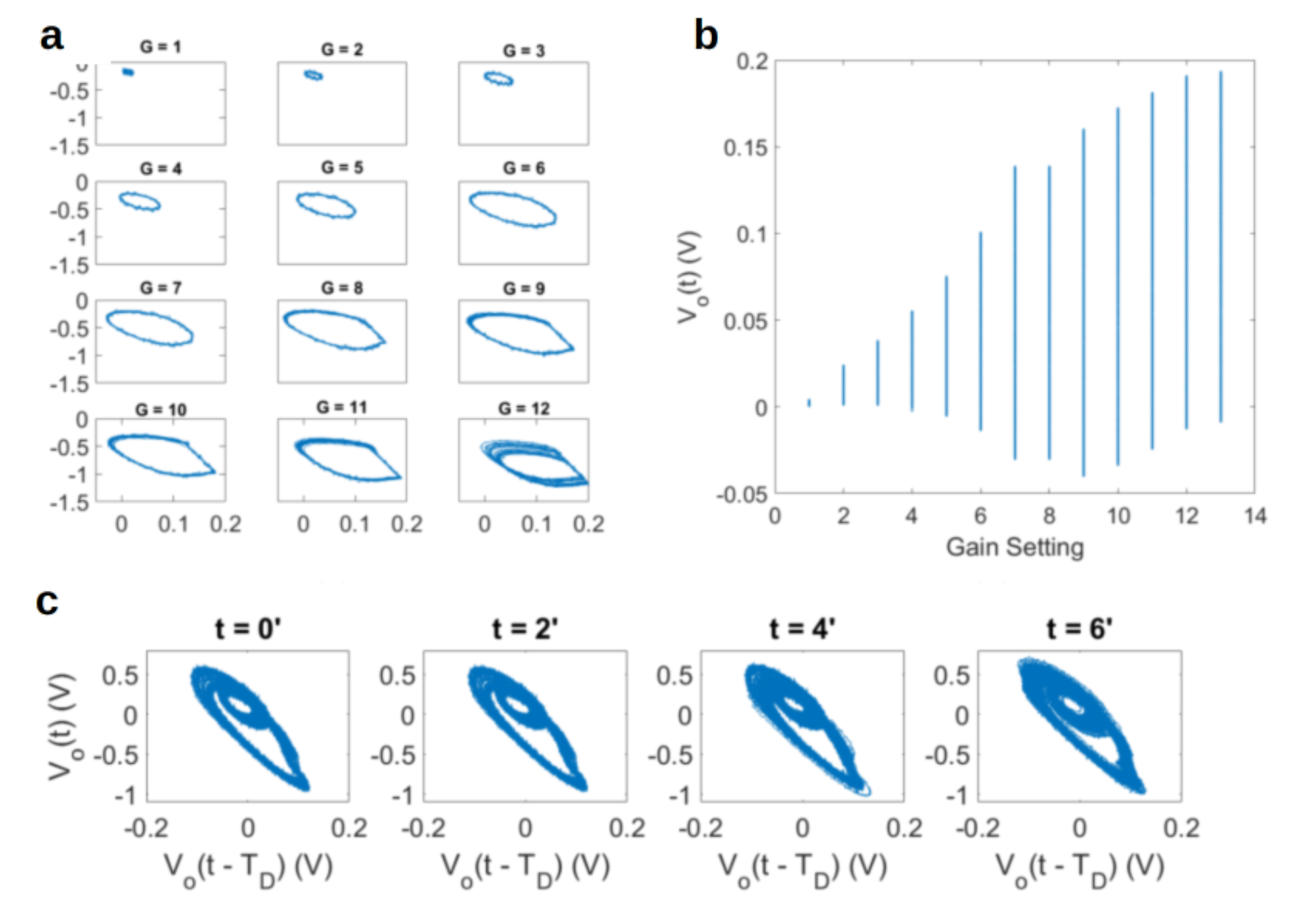}
	\end{center}
	\caption[]{\textbf{a} Phase space portraits ($\Vec{\mathrm{y}}(\mathrm{t})$ vs. $\Vec{\mathrm{y}}(\mathrm{t-\tau})$ and \textbf{b} corresponding bifurcation diagram of the experimental validation of the delayed-feedback reservoir. The gain factor g is used as a bifurcation parameter. In \textbf{b}, each vertical line (set of very dense points) represents the temporal evolution of the trajectories over time. For g\,>\,7 chaotic behavior is obtained.  \textbf{c} Repeated measurement of another non-linear network (measurement time 10\,s, repeated every 2\,min). The phase space portrait significantly changes over time suggesting instability of the system. A sinusoidal input signal with a frequency of 3\,Hz is used in all cases. Figure reprinted with courtesy of \cite{petrauskas2020}.}
	\label{fig:reservoir_4}
\end{figure}
f($\Vec{\mathrm{y}})=\mathrm{\lambda}\Vec{\mathrm{y}}(1-\Vec{\mathrm{y}})$) the system ends up in a chaotic state after multiple bifurcations (period-doubling) \cite{petrauskas2020}, which is usually presented in so-called Feigenbaum diagrams. Also in the experimental implementation of the delayed-feedback reservoir, this kind of behavior is observed. Using the gain of the feedback loop as a bifurcation parameter, several bifurcations and finally, chaotic oscillations are obtained. For gains $\le$\,7, a stable and single orbit is obtained in phase space ($\Vec{\mathrm{y}}(\mathrm{t})$ vs. $\Vec{\mathrm{y}}(\mathrm{t-\tau})$, see Figure \ref{fig:reservoir_4}a and b) representing the input sinusoidal signal. For gains >\,7, the orbit is distorted and a first bifurcation occurs. Furthermore, with increasing gain, more and more orbits occur until the system reaches a chaotic state. A direct proof of chaos is mathematically challenging, however, it might be proven indirectly by determining the fractal dimension of the trajectory in phase space. Doing so, a value of ~1.85 is obtained using the box-counting method \cite{petrauskas2020} which justifies using the term chaotic to describe the trajectory.\\
Hence, the system has been proven to be sufficiently complex to be employed for reservoir computing. However, there are still several challenges
\begin{figure}[ht]
	\begin{center}
		\includegraphics[width=.90\textwidth,clip]{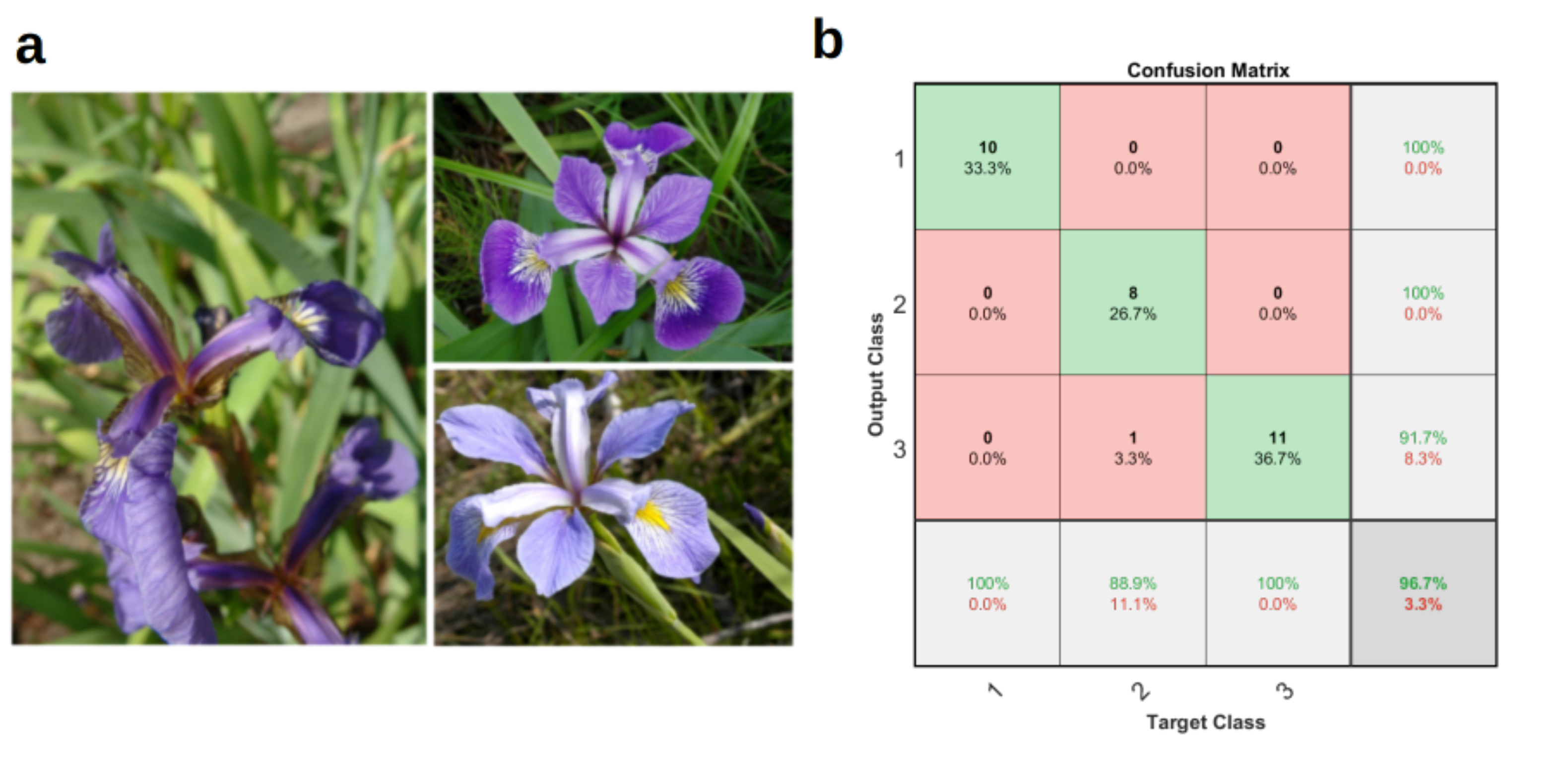}
	\end{center}
	\caption[]{\textbf{a} Photographs of the three different flowers which are part of the Iris data-set. Iris setosa (left), Iris versicolor (top right), and Iris virginica (bottom right). \textbf{b} Confusion matrix summarizing the result of the classification using the OECT-based reservoir. Figure adapted from \cite{Cucchi2021b}.}
	\label{fig:reservoir_5}
\end{figure}
to overcome before the reservoir can be used for extended classifications tasks. The most apparent issue is the stability of the non-linearity. As shown in Figure \ref{fig:reservoir_4}c, although the general shape of the trajectories is maintained, a substantial change is obtained if the reservoir is repeatedly measured (here measurement time 10\,s, repeated every two minutes). This problem arises since the electrolyte slowly evaporated over time, hence, causing a steady increase in the salt concentration. Furthermore, it turned out that the fibers connected to the fine metal electrodes (tip diameter $le$\,2$\,$\textmu m) are easily removed during the training process probably caused by the high electric field strength. In order to use the reservoir for complex classification tasks such as heartbeat classification, a training period of several tenths of minutes is required.\\
Besides these challenges, still, the potential of such OECT-based reservoirs can be demonstrated using more simple (time-independent) classification tasks. One of these tasks is given by the so-called Iris data-set where three different flowers, denoted as Iris setosa, Iris versicolor, and Iris virginica (Figure \ref{fig:reservoir_5}a, should be classified based on four properties: petal length, petal width, sepal width, and sepal length \cite{Iris}. This data-set is translated into signals that can be fed into the reservoir. In particular, all attributes are frequency encoded using sinusoidal signals with a frequency between 2 and 10\,Hz (e.g., a petal length of 1 to 6.9\,cm is linearly translated into the frequency range between 2 and 10\,Hz). Experimentally, the reservoir is composed of four input and four output nodes and a global gate electrode. While two of the input nodes are connected via PEDOT-fibers to the four output nodes, the other two input nodes serve as additional gate electrodes (covered with polymer). One of the output signals is fed back to the global gate electrode using the delayed-feedback loop. In this way, the conductivity of the fibers is modulated. For the training phase, 120 out of the 150 entries of the data-set are fed into the reservoir. Each entry is fed in for a duration of 3\,s (amplitude 1\,V) followed by a short break for 3\,s. The remaining 30 entries are used for testing the reservoir and the results are presented in the so-called confusion matrix (see Figure \ref{fig:reservoir_5}b). The first three rows show which of the flowers was classified by the reservoir, while the first three columns indicate which class was correct (total number and percentage). Thus, the diagonal shows the correctly classified flowers, while the non-diagonal elements represent the false classifications. Overall, the accuracy was about 97$\%$ which means that 29 out of 30 flowers were correctly classified. \\
In summary, neuromorphic structures based on organic electro-chemical transistors are still at an early stage of development. However, the tunability of plasticity and the possibility to use advanced growth and deposition techniques in order to build complex networks have to be seen as a big advantage of OECT-based neuromorphic structures. Furthermore, since OECT-based neuromorphic structures are compatible with large-area fabrication techniques, thousands of contact pads for the input nodes can be easily fabricated without enormous additional costs. In silicon-based application-specific integration-circuits (ASIC) for ANNs though, chip area is a cost factor and the incorporation of contact pads is expensive since in contrast to transistors and other elements, the size of the contact pads cannot be scaled down easily.
\subsubsection{Ion-Sensing using Organic Electro-Chemical Transistors}
\begin{figure}[htb]
	\begin{center}
		\includegraphics[width=.99\textwidth,clip]{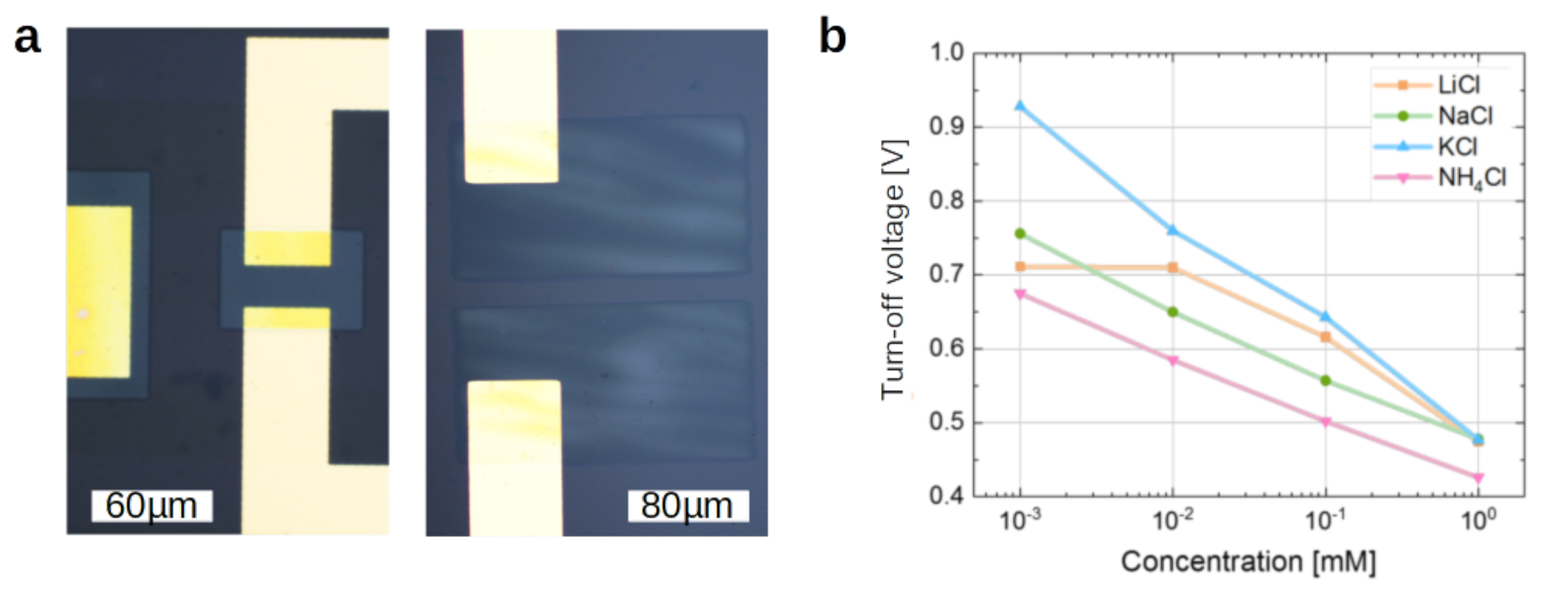}
	\end{center}
	\caption[]{\textbf{a} Photograph of the ion-sensor platform which consists of an OECT based on PEDOT:PSS and two micro-electrodes which are covered by PEDOT:PSS. \textbf{b} Determination of the turn-off voltage of the OECT-based sensor (cf. \cite{Ghittorelli2018}) versus the concentration of the electrolyte. Figures adapted from \cite{Tseng2021}.}
	\label{fig:Sensor_2}
\end{figure}

\begin{figure}[htb]
	\begin{center}
		\includegraphics[width=.99\textwidth,clip]{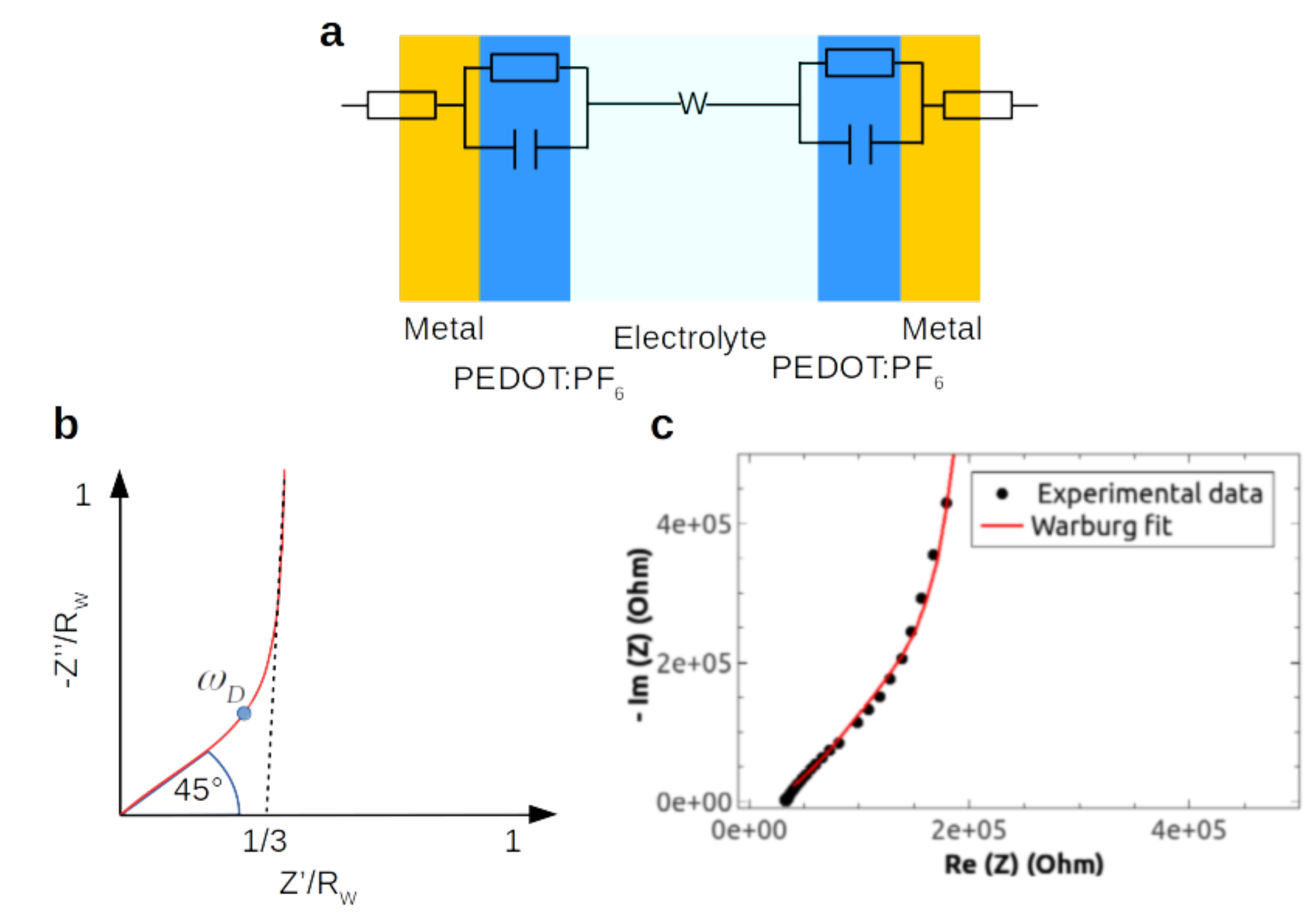}
	\end{center}
	\caption[]{\textbf{a} Cross-section of the two-electrode configuration. The series resistance R$_\mathrm{S}$ is equally distributed between both electrodes. The same holds true for the RC element describing the PEDOT:PF$_\mathrm{6}$ layer. The diffusion of cations/anions in the electrolyte is described by the Warburg element. \textbf{b} Nyquist plot of an ideal Warburg impedance assuming reflective boundary conditions. Real (Z') and imaginary part (Z'') of the impedance are normalized by R$_\mathrm{W}$. The characteristic frequency $\mathrm{\omega_D}$ marks the transition between high-frequency limit (slope 45$^\circ$) and low-frequency limit. \textbf{c} Experimental impedance data for different for a salt concentration of 0.1\,mM of NaCl and fit using the equivalent circuit shown in \textbf{a}.}
	\label{fig:Sensor_3}
\end{figure}

Organic electro-chemical transistors are devices that rely on electronic as well as ionic transport. In particular, due to the amplifying function of OECTs, small changes in the ionic conduction might be translated into large changes in the electronic conduction which can be detected and analyzed conveniently. Hence, OECTs are suited as ion-sensors as needed e.g., for environmental monitoring (water control) or the analysis of biological liquids (sudor, blood, lymph, etc.). The electrical conductivity of a PEDOT:PSS-based OECT channel is sensitive to the concentration of cations since they are used to dedope the channel. Anions though cannot be detected with PEDOT:PSS-based OECTs since additional anions from the electrolyte only weakly increase the conductivity of the channel (PEDOT is already completely doped by anions, e.g., PSS$^-$). Thus, an electron conducting polymer is required for anion detection.\\
Another aspect about ion-sensing using OECTs which should be considered is the fact that in a static transistor measurement, the conductivity depends only on the ion concentration but not on the type of ions (Equation \ref{eq:Nernst_2}). Different kinds of ions would lead to e.g., different turn-off voltages of an OECT, however, the concentration dependence of the turn-off voltage is expected to be the same for all kinds of ions. Thus, except for very large cations that might not be able to penetrate into the polymer film, OECTs are not ion-selective.\\
In order to account for these shortcomings, a sensor platform is developed that is composed of an OECT and a set of two PEDOT-covered micro-electrodes. While the OECT is used for determining the ion concentration, the set of
micro-electrodes features ion-selectivity using an impedance spectroscopy analysis. A photograph of the sensor platform is shown in Figure \ref{fig:Sensor_2}a.\\
In order to determine the ion-concentration, the turn-off voltage V$_\mathrm{TO}$ of the OECT is determined for different salt concentrations of the electrolyte as well as for different kinds of salt (see Figure \ref{fig:Sensor_2}b). As expected from Equation \ref{eq:Nernst_2}, the turn-off voltage shifts with the logarithm of
the salt concentration. In an aqueous solution, the OECT-based sensor can cover the range from mM to \textmu M - the range which is also most relevant for ion-sensing in biological systems. The sensitivity of the system is approximately 100\,mV/dec which is in agreement with previous reports and sufficient for many applications \cite{Ghittorelli2018}. In particular, the sensitivity is even better than commercial silicon sensors which usually reach a sensitivity of $\sim$70\,mV/decade \cite{Jimenez2010}. The ion-sensitivity of the OECTs might be further improved using e.g., intrinsically undoped polymers. In this case, small changes in the cation concentration might be translated into large changes of the chemical potential of the polymeric semiconductor.\\
Although different kinds of salt give different readings of V$_\mathrm{TO}$ (Figure \ref{fig:Sensor_2}b), the OECT-based sensor is not ion-selective. In particular, for an unknown electrolyte, it would be impossible to conclude from the V$_\mathrm{TO}$ reading the electrolyte concentration or kind of salt independently. In order to so, it is important to find another parameter aside the static conductivity which is linked not only to the ion concentration but also to at least one other ion-specific parameter. The most obvious choice for this parameter is the diffusion constant which might be determined e.g., by impedance analysis. However, in this case, it is important to find an adequate electrode configuration in which the diffusion can be analyzed quantitatively.\\
Here a planar two-electrode configuration of metallic micro-electrodes is proposed with a gap of 50...200\,\textmu m (Figure \ref{fig:Sensor_2}a). The micro-electrodes are covered with a layer of a highly conductive mixed conductor such as PEDOT:PF$_\mathrm{6}$ with a thickness of up to 1\,\textmu m. The PEDOT:PF$_\mathrm{6}$ layers act as transducers from ionic to electronic signals ensuring a low transition impedance between electrolyte and metallic electrode. At the interface between PEDOT:PF$_\mathrm{6}$ and the electrolyte, an electro-chemical double-layer is formed causing a gradient of ion concentration at the interface. If a small-signal voltage is applied between the two electrodes, the concentration gradient at the interface will change accordingly. In consequence, the concentration gradient will give rise to an ion diffusion current through the entire electrolyte. The most primitive equivalent circuit to describe this two-electrode configuration is shown in Figure \ref{fig:Sensor_3}a. It is composed of a series resistance accounting for the access resistance of the metal lines (R$_\mathrm{S}$), two RC-elements describing the PEDOT:PF$_\mathrm{6}$ layers and interfaces, and a so-called Warburg element. The impedance of the Warburg element is a direct solution to the diffusion equation without recombination \cite{Bisquert2002}. Depending on whether the diffusion equation is solved for absorbing or reflecting boundary conditions, the impedance can be derived as:
\begin{align}
\mathrm{Z}&=\mathrm{R_W(i\omega /\omega_D)^{-1/2}coth[(i\omega /\omega_D)^{1/2}]} &;& \quad \mathrm{reflecting\,boundary} \label{eq:Warburg_1}\\
\mathrm{Z}&=\mathrm{R_W(i\omega /\omega_D)^{-1/2}tanh[(i\omega /\omega_D)^{1/2}]}  &;& \quad \mathrm{absorbing\,boundary}
\label{eq:Warburg_2}
\end{align}
where $\mathrm{\omega_D=D/b^2}$ is a characteristic frequency (D...diffusion constant, b...distance of electrodes) and R$_\mathrm{W}$ is the diffusion resistance. The diffusion resistance is given by 
\begin{equation}
    \mathrm{R_W=\frac{k_BTb}{e^2ADc}}
\end{equation}
where T is the temperature, A the cross-section of the ion channel between the electrodes, and c the concentration of cations. Since ions cannot physically leave the electrolyte into the metal electrode/ PEDOT:PF$_\mathrm{6}$, the diffusion in the electrolyte is described by reflecting boundary conditions. In this particular case, Equation \ref{eq:Warburg_2} can be approximated in the low- and high-frequency limit as
\begin{align}
\mathrm{Z}&=\mathrm{R_W(i\omega /\omega_D)^{-1/2}} &;& \quad \mathrm{high-frequency\,limit,\,\omega\gg \omega_D} \label{eq:Warburg_3}\\
\mathrm{Z}&=\mathrm{\frac{1}{3}R_W+\frac{1}{iC_f\omega}} &;& \quad \mathrm{low-frequency\,limit,\,\omega\ll \omega_D}
\label{eq:Warburg_4}
\end{align}
with C$_\mathrm{f}$=1/(R$_\mathrm{W}\omega_D$). The Nyquist diagram of this Warburg impedance with reflecting boundaries is exemplary shown in Figure \ref{fig:Sensor_3}b.
Thus, analyzing the complex impedance of such a two-electrode configuration, the diffusion constant D might be determined by a non-linear fit of the impedance using the proposed equivalent circuit. The obvious choice for the fitting parameter would be $\mathrm{\omega_D=D/b^2}$, however, for practical reasons R$_\mathrm{W}$ is more suited. The reason for that lies in the fact, that for high ion concentrations, the magnitude of the impedance from the Warburg element and the two RC elements becomes comparable in the high-frequency limit. Hence, the two components become indistinguishable and in particular, the characteristic frequency $\mathrm{\omega_D}$ cannot be determined anymore. In the low-frequency limit though, the impedance is purely governed by the Warburg element (cf. Equation \ref{eq:Warburg_4}), and the diffusion resistance can be determined accurately. Figure \ref{fig:Sensor_3}c shows the experimental data and the fit using the proposed equivalent circuit. The experimental curves show non-ideal diffusion (exponent in Equation \ref{eq:Warburg_2} not -1/2 but rather -0.3..-0.4). Hence, Equation \ref{eq:Warburg_4} cannot be used for fitting but rather a generalized form of Equation \ref{eq:Warburg_2} having the exponent as a free parameter. Still, an equivalent diffusion resistance R$_\mathrm{W}^*$ can be extracted from the extrapolation of the impedance as shown in Figure \ref{fig:Sensor_3}b. Since the equivalent diffusion resistance contains the product of diffusion constant and ion concentration, it can be used in combination with the turn-off voltage
\begin{figure}[htb]
	\begin{center}
		\includegraphics[width=.99\textwidth,clip]{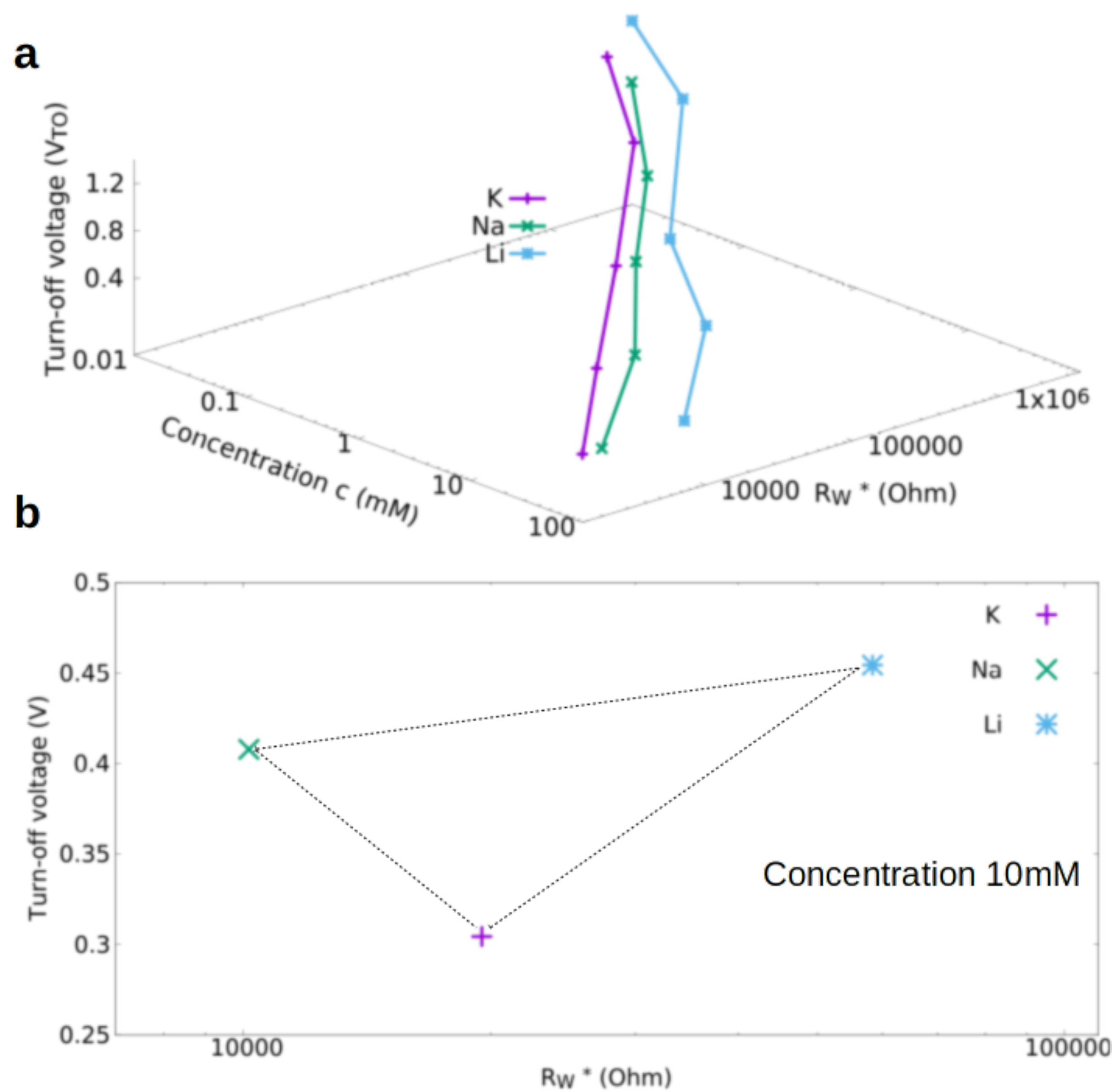}
	\end{center}
	\caption[]{\textbf{a} Turn-off voltage V$_\mathrm{TO}$ and effective diffusion resistance R$_\mathrm{W}^*$ for different salt concentrations and different kinds of salt. The lines are separable. Hence, using an unknown electrolyte, concentration and ion species can be identified independently. \textbf{b} Cut through \textbf{a} for a concentration of 10mM. For an unknown electrolyte, the composition might be determined from te position within the triangle.}
	\label{fig:Sensor_4}
\end{figure}
measurement to determine ion concentration and ion species independently. As shown in Figure \ref{fig:Sensor_4}, the measurement of equivalent diffusion resistance and turn-off voltage for different ion concentrations and species results in non-intersection curves. Consequently, ion concentration and species are linearly separable proving the ion-selectivity of the sensor.\\
As a matter of fact, for low ion concentrations, $\mathrm{\omega_D}$ can be determined unambiguously from the impedance data. Knowing the gap between the electrodes, the diffusion constant of the different ions can be extracted. We find values for the diffusion constants in the range of 10$^{-9}$\,m$^2$/s which compares well with literature values. This agreement provides indirect proof for the validity of the proposed equivalent circuit. Furthermore, it is worth mentioning that the above-described approach has not been reported in the literature since usually very large electrode distances are used (several millimeters). In this case, extremely low frequencies in the range of \textmu Hz are required to extract R$_\mathrm{W}$ correctly. Hence, large-distance electrode configurations are not suited for this analysis.

\pagestyle{fancy}
\fancyhead{}
\fancyhead[LE]{\scshape \thepage} 
\fancyhead[RE]{\scshape \rightmark}
\fancyhead[LO]{\scshape \rightmark} 
\fancyhead[RO]{\scshape \thepage}
\fancyfoot{}
\renewcommand{\headrulewidth}{1pt}
\cleardoublepage
\setcounter{section}{4}
\section*{Outlook}
\addcontentsline{toc}{section}{4 \hspace{0.1cm} Outlook}
\thispagestyle{empty}
\vspace{2cm}
\setlength{\epigraphwidth}{.5\textwidth} 
\setlength{\epigraphrule}{1pt} 
\epigraph{\large{\textit{'Denn der h\"angt von der Zukunft ab, der mit der Gegenwart nichts anzufangen weiss.'}}}{\footnotesize{Lucius Annaeus Seneca, Philosopher}}
\vspace{2cm}
The digital revolution and the internet of things (IoT) require massive amounts of application-specific integrated circuits (ASIC) for numerous applications such as sensor tags, communication units, data processing, etc. The market for high-end performance ASICs (e.g., for communication) will be dominated by silicon-based semiconductor devices. However, there is also a large market for low and medium performance circuits (smart tags \& labels, healthcare monitoring, sensor tags, etc.) where silicon-based technology is under pressure due to the high production costs (i.e., profit margin is governed by the density of circuits per area). The organic transistors developed in this work perfectly fit into this low and medium performance segment. In particular, the transition frequency of OPBTs and VOFETs is sufficiently high for near- and medium-range communication, and the supply voltage fits the requirements for portable devices. Due to the possibility to employ flexible substrates (possibly even by roll-to-roll processes), these organic semiconductor devices break with the paradigm of wafer-based processing, and allow for a considerable reduction of the fabrication costs. In order to get to this point, more steps need to be taken. Most importantly, a robust validation of the device parameters for different geometries and processes is necessary for the development of compact models. Furthermore, the performance of hero-devices needs to be demonstrated in fully integrated circuits rather than single devices. With regard to the single device performance, investigations on the contribution and origin of contact resistance in the vertical transistors are vital in order to increase the transition frequency even further. Newly developed material systems such as triclinic rubrene can boost the performance of VOFETs and OPBTs even further due to the ten times higher vertical charge carrier mobility compared to the currently used materials.\\
The vertical organic transistors developed in Chapter 2 in combination with neuromorphic structures proposed in Chapter 3 open a new perspective for next-generation smart sensor tags (e.g., in healthcare or environmental monitoring). While the neuromorphic devices may be employed for classification tasks, the vertical transistors can be utilized for data processing and transmission. Using the advantages of large-area processing, many input electrodes for the neuromorphic network can be manufactured without increasing the production costs (contact pads are costly in silicon-based technology due to the area footprint). In this way, intelligent sensor arrays with low power consumption can be realized for on-chip data classification. The impact of this approach though goes even beyond the single sensor array. For example, a substantial problem of big data analysis is the pre-processing of data in order to make it fit the requirements of the neural network. This pre-processing demands massive computational resources and often an incredible amount of data. In this regard, an intelligent sensor tag able to pre-classify data would help to significantly lower the power consumption required for big data analysis. Thus, such organic semiconductor-based artificial intelligence unit can become a valuable part of a bigger AI network composed of small units and centralized hubs for computation. In order to get to this point, the neuromorphic structures discussed in Chapter 3 need to be further investigated and optimized. In particular, it is essential to investigate the scaling behavior of such structures in order to understand the limits of the device speed and the power consumption. In this regard, the fabrication of ultra-thin fiber networks offers the possibility to reduce the Ohmic losses in such a system. Furthermore, new semiconductor materials that are intrinsically undoped may enable the fabrication of normally-off synapses, which will help to reduce the power consumption but also increase the non-linear coupling due to a stronger gating effect.  

Another substantial problem of silicon-based electronics is its enormous environmental impact. Today’s electronic device manufacturing relies on the unsustainable exploitation of natural resources creating a massive amount of electronic waste composed of hazardous and toxic substances. Moreover, disposal in backyards or improper recycling puts an additional threat to our ecosystem. Thus, electronic devices are unsustainable, and our society demands timely innovations to overcome the challenges we are all facing. In this regard, organic electronics offer the possibility to build decomposable and even fully recyclable electronic devices with a low carbon footprint.\\
This topic though is very controversial since currently organic electronic devices often employ processes and material which are as toxic as the processes used in the silicon-based industry. Still, there is a variety of preliminary work on decomposable organic electronic devices where, e.g., transistors are built from beta-carotene or shellac. Thus, there is the potential to fabricate these devices from decomposable and recyclable materials for substrate, semiconductors, metals, and insulators. Thus, it is my vision to demonstrate that organic electronic devices, as discussed in Chapters 2 and 3, can be fabricated in a truly sustainable fashion. In order to that, appropriate materials and processes for fabrication and recycling need to be explored. 
\newpage

\pagestyle{fancy}
\fancyhead{}
\fancyhead[LE]{\scshape \thepage} 
\fancyhead[RE]{\scshape \rightmark}
\fancyhead[LO]{\scshape \rightmark} 
\fancyhead[RO]{\scshape \thepage}
\fancyfoot{}
\renewcommand{\headrulewidth}{1pt}
\cleardoublepage

\pagestyle{fancy}
\fancyhead{}
\fancyhead[LE]{\scshape \thepage} 
\fancyhead[RE]{\scshape \rightmark}
\fancyhead[LO]{\scshape \rightmark} 
\fancyhead[RO]{\scshape \thepage}
\fancyfoot{}
\renewcommand{\headrulewidth}{1pt}

\pagestyle{fancy}
\fancyhead{}
\fancyhead[LE]{\scshape \thepage} 
\fancyhead[RE]{\scshape \rightmark}
\fancyhead[LO]{\scshape \rightmark} 
\fancyhead[RO]{\scshape \thepage}
\fancyfoot{}
\renewcommand{\headrulewidth}{1pt}
\pagestyle{fancy}
\fancyhead{}
\fancyhead[LE]{\scshape \thepage} 
\fancyhead[RE]{\scshape \rightmark}
\fancyhead[LO]{\scshape \rightmark} 
\fancyhead[RO]{\scshape \thepage}
\fancyfoot{}


\addcontentsline{toc}{section}{Contributions to Joint Publications}
\fancyhead{}
\fancyhead[R]{Contributions to Joint Publications} 
\fancyhead[L]{\scshape \thepage}
\cleardoublepage
\thispagestyle{empty}
\newpage
\section*{Contributions to Joint Publications}
This thesis summarizes joint publications in the field of solid-state electronics. My contributions to these publications are listed below.
\subsection*{1. Publications as Lead Author}
\begin{itemize}

\item H. Kleemann, G. Schwartz, M. Baumann, S. Zott, and M. Furno, "Megahertz operation of vertical organic transistors for ultra-high resolution active-matrix display", Flexible and Printed Electronics \textbf{5}, 1, 014009 (2020).\\
Together with G. Schwartz, I planned the experiments and did the measurements. I wrote the major part of the manuscript.

\item H. Kleemann, K. Krechan, A. Fischer, and K. Leo, "Review of Vertical Organic Transistors", Advanced Functional Materials \textbf{20}, 1907113 (2020).\\
Together with my colleagues, I have collected the literature overview and I wrote the major part of the manuscript.

\item M.I.B. Utama, H. Kleemann, W. Zhao, C. Ong, H. Felipe, D.Y. Qiu, H. Cai, H. Li, R. Kou, S. Zhao, S. Wang, K. Watanabe, T. Taniguchi, S. Tongay, A. Zettl, S.G. Louie, and F. Wang, "A dielectric-defined lateral heterojunction in a monolayer semiconductor", Nature Electronics \textbf{2}, 60-65 (2019). \cite{Utama2019}\\
M.I.B. Utama and I contributed equally to this work. We designed the experiments, fabricated the devices, carried out the transport and optical measurements and developed parts of the theory.

\item H. Kleemann, G. Schwartz, and J. Blochwitz-Nimoth, "Method for producing an organic transistor and organic transistor", US Patent US10497888B2 (2019).\\
Together with G. Schwartz, I am the main inventor and we have co-written the patent together with a lawyer.

\item H. Kleemann, G. Schwartz, "Method for producing a vertical organic field-effect transistor, and vertical organic field-effect transistor", US Patent US10170715B2 (2019).\\
Together with G. Schwartz, I am the main inventor and we have co-written the patent together with a lawyer.

\item H. Kleemann, B. L\"ussem, K. Leo, and A. G\"unther, "Method for producing an organic field effect transistor and an organic field effect transistor", US Patent US9705098B2 (2017).\\
I am the main inventor and we have co-written the patent together with a lawyer.

\item H. Kleemann, A. Zakhidov, B. L\"ussem, and K. Leo, "Method for manufacturing an organic electronic device and organic electronic device", US Patent US9837609B2 (2017).\\
I am the main inventor and we have co-written the patent together with a lawyer

\item H. Kleemann, M. Furno, G. Schwartz, and J. Blochwitz-Nimoth, "39.4 L: Late-News Paper: Vertical Organic Transistors (V-OFETs) for Truly Flexible AMOLED Displays", SID Symposium Digest of Technical Paper \textbf{46}, 597-600 (2015).\\
The work leading to this publications has been done at the Novaled GmbH Dresden. I planned the experiments, fabricated and measured the devices, interpreted the results and wrote the manuscript. I was supported by a team of technicians and other research of the team.

\end{itemize}

\subsection*{2. Publications as Principal Investigator}
The following publication were headed by me. I motivated the work, conceived experiments, supervised the students, and supported the student with the writing of the manuscripts. 
\begin{itemize}

\item E. Guo, S. Xing, F. Dollinger, R. H\"ubner, S.-J. Wang, Z. Wu, K. Leo, \& H. Kleemann, "Integrated complementary inverters and ring oscillators based on vertical-channel dual-base organic thin-film transistors", Nature Electronics \textbf{4}, 588–594 (2021).

\item E. Guo, F. Dollinger, B. Amaya, A. Fischer, \& H. Kleemann, "Organic Permeable Base Transistors–Insights and Perspectives", Advanced Optical Materials \textbf{9}, 2002058 (2021)

\item D. Kneppe, F. Talnack, B.K. Boroujeni, C. da Rocha, M. H\"oppner, A. Tahn, S. Mannsfeld, F. Ellinger, K. Leo, \& H. Kleemann, "Solution-processed pseudo-vertical organic transistors based on TIPS-pentacene", Materials Today Energy \textbf{21}, 100697 (2021).

\item   G. Darbandy, F. Dollinger, E. Guo, A. Kloes, K. Leo, \& H. Kleeman, "Unraveling Structure and Device Operation of Organic Permeable Base Transistors", Advanced Electronic Materials \textbf{6}, 2000230 (2020).

\item E. Guo, Z. Wu, G. Darbandy, S. Xing, S.J. Wang, A. Tahn, M. G\"obel, A. Kloes, K. Leo, \& H. Kleemann, "Vertical Organic Permeable Dual-Base Transistors for Logic Circuits", Nature Communications \textbf{11}, 4725 (2020).

\item E. Guo, S. Xing, F. Dollinger, Z. Wu, A. Tahn, M. L\"offler, K. Leo, \& H. Kleemann, "High‐Performance Static Induction Transistors Based on Small‐Molecule Organic Semiconductors", Advanced Materials Technologies \textbf{5}, 2000361 (2020).

\item K.G. Lim, E. Guo, A. Fischer, Q. Miao, K. Leo, and H. Kleemann, "Anodization for Simplified Processing and Efficient Charge Transport in Vertical Organic Field-Effect Transistors", Advanced Functional Materials \textbf{30}, 2001703 (2020).

\item A. Zakhidov, B. L\"ussem, K. Leo, and H. Kleemann, "Method for producing an organic field effect transistor and an organic field effect transistor", US Patent US9722196B2 (2017).

\item R. Lessmann, and H. Kleemann, "Array of several organic semiconductor components and method for the production thereof", US Patent US9240560B2 (2016).
\end{itemize}

\subsection*{3. Publications as Co-Author} 
These results were published during different stages of my PostDoc time. The majority of manuscripts (starting from 2018) was published while I was heading the ODS research group of Prof. Leo. He initiated the research, I supervised the day-to-day activities of the students, conceived experiments, analyzed and interpreted data, and co-wrote the manuscripts. Furthermore, I coordinated the funding from different funding agencies.    
\begin{itemize}
\item M. Schwarze, M.L. Tietze, F. Ortmann, H. Kleemann, \& Karl Leo, "Universal Limit for Air-Stable Molecular n-Doping in Organic Semiconductors", ACS Applied Materials \& Interfaces \textbf{12}, 40566 (2020).

\item K. Orstein, S. Hutsch, A. Hinderhofer, J. Vahland, M. Schwarze, S. Schellhammer, M. Hodas, T. Geiger, H. Kleemann, H.F. Bettinger, F. Schreiber, F. Ortmann, \& Karl Leo, "Energy Level Engineering in Organic Thin Films by Tailored Halogenation", Advanced Functional Materials \textbf{5}, 2002997 (2020).

\item	X. Shen, X. Wang, E. Guo, A. Fischer, H. Kleemann, and K. Leo, "Organic Thin-Film Red-Light Photodiodes with Tunable Spectral Response Via Selective Exciton Activation", ACS Applied Materials \& Interfaces  \textbf{12}, 11, 13061 (2020).

\item	F. Dollinger, H. Iseke, E. Guo, A. Fischer, H. Kleemann, and K. Leo, "Electrically Stable Organic Permeable Base Transistors for Display Applications", Advanced Electronic Materials  \textbf{5}, 1900576 (2019).
	
\item M. H\"{o}ppner, D. Kneppe, H. Kleemann, and K. Leo, "Precise patterning of organic semiconductors by reactive ion etching", Org. Electron. \textbf{76}, 105357 (2020).
 
 \item X. Fan, D. Kneppe, V. Sayevich, H. Kleemann, A. Tahn, K. Leo, V. Lesnyak, and A. Eychm\"{u}ller, "High-Performance Ultra-Short Channel Field-Effect Transistor Using Solution-Processable Colloidal Nanocrystals", Physcal Chemistry Letters \textbf{10}, 14, 4025 (2019).
 
 \item Z.Wu, Y. Liu, L. Yu, C. Zhao, D. Yang, X. Qiao, J. Chen, C. Yang, H. Kleemann, K. Leo, and D. Ma, "Strategic-tuning of radiative excitons for efficient and stable fluorescent white organic light-emitting diodes", Nat. Comm. \textbf{1}, 2380 (2019).
 
 \item F. Dollinger, K.G. Lim, Y. Li, E. Guo, A. H\"{u}bner, P. Formánek, A. Fischer, H. Kleemann, and K. Leo, "Vertical Organic Thin‐Film Transistors with an Anodized Permeable Base for Very Low Leakage Current", Adv. Mat. \textbf{31}, 1900917 (2019).
 
 \item C. Jin, J. Kim, I. Utama, E.C. Regan, H. Kleemann, H. Cai, Y. Shen, M. Shinner, A. Sengupta, K. Watanabe, T. Taniguchi, S. Tongay, A. Zettl, and F. Wang, "Spatial-Temporal Imaging of Pure Spin-Valley Current in Transition Metal Dichalcogenide Heterostructures", Science \textbf{360}, 893-896 (2018).
 
 \item F.M. Sawatzki, DH Doan, H. Kleemann, M. Liero, A. Glitzky, T. Koprucki, and K. Leo, "Balance of Horizontal and Vertical Charge Transport in Organic Field-Effect Transistors", Physical Review Applied \textbf{10}, 034069 (2018).
 
 \item B. L\"ussem, A. Zakhidov, H. Kleemann, and K. Leo, "Organic field effect transistor and method for producing the same", US Patent US9899616B2 (2018).
 
\item M.L. Tietze, J. Benduhn, P. Pahner, B. Nell, M. Schwarze, H. Kleemann, M. Krammer, K. Zojer, K. Vandewal, and K. Leo, "Elementary steps in electrical doping of organic semiconductors", Nature Communications \textbf{9}, 1182 (2018).

\item B. Kheradmand-Boroujeni, M.P. Klinger, A. Fischer, H. Kleemann, K. Leo, and F. Ellinger, "A Pulse- Biasing Small-Signal Measurement Technique Enabling 40 MHz Operation of Organic Transistors", Scientific Reports \textbf{8}, 7643 (2018).

\item M.P. Klinger, A. Fischer, H. Kleemann, and K. Leo, "Non-Linear Self-Heating in Organic Transistors Reaching High Power Densities", Scientific Reports \textbf{8}, 9806 (2018).

\item A. G\"unther, H. Kleemann, B. L\"ussem, K. Leo, and D. Kasemann "Method for manufacturing an organic electronic device and organic electronic device", US Patent US9620730B2 (2017).

\item T. Menke, D. Ray, H. Kleemann, K. Leo, and M. Riede, "Determining doping efficiency and mobility from conductivity and Seebeck data of n-doped C60 layers", physica status solidi b \textbf{252}, 1877-1883 (2015).

\item J. Fischer, D. Ray, H. Kleemann, P. Pahner, M. Schwarze, C. K\"{o}rner, K. Vandewal, and K. Leo, "Density of states determination in organic donor-acceptor blend layers enabled by molecular doping", Journal of Applied Physics \textbf{117}, 245501 (2015).

\item J. Fischer, J. Widmer, H. Kleemann, W. Tress, C. K\"{o}rner M. Riede, and K. Leo, "A charge carrier transport model for donor-acceptor blend layers", Journal of Applied Physics 117, 045501 (2015).

\item B. L\"{u}ssem, H. Kleemann, D. Kasemann, V. Fentsch, and K. Leo, "Organic Junction Field-Effect Transistor", Advanced Functional Materials \textbf{24}, 1011-1016 (2014).

\item J. Fischer, W. Tress, H. Kleemann, J. Widmer, K. Leo, and M. Riede, "Exploiting diffusion currents at Ohmic contacts for trap characterization in organic semiconductors", Organic Electronics \textbf{117},  2428-2432 (2014).

\item S. Nicht, H. Kleemann, A. Fischer, K. Leo, and B. L\"{u}ssem, "Functionalized p-dopants as self-assembled monolayers for enhanced charge carrier injection in organic electronic devices", Organic Electronics \textbf{117},  654-660 (2014).

\item T. Menke, D. Ray, H. Kleemann, M.P. Hein, K. Leo, and M. Riede, "Highly efficient p-dopants in amorphous hosts", Organic Electronics \textbf{15}, 365-371 (2014).

\item B. L\"{u}ssem, M. L. Tietze, H. Kleemann, C. Hossbach, J.W. Bartha, A. Zakhidov, and K. Leo, "Doped organic transistors operating in the inversion and depletion regime", Nature Communications \textbf{4}, 2775 (2013).

\item P. Pahner, H. Kleemann, L. Burtone, M.L. Tietze J. Fischer, K. Leo, and B. L\"{u}ssem, "Pentacene Schottky diodes studied by impedance spectroscopy: Doping properties and trap response", Physical Review B \textbf{88}, 195205 (2013).
 
\end{itemize}
\newpage

\bibliographystyle{frankystyle}
\bibliography{main}
\addcontentsline{toc}{section}{Bibliography}
\newpage

\pagestyle{fancy}
\fancyhead{}
\fancyhead[LE]{\scshape \thepage} 
\fancyhead[RE]{\scshape \rightmark}
\fancyhead[LO]{\scshape \rightmark} 
\fancyhead[RO]{\scshape \thepage}
\fancyfoot{}
\renewcommand{\headrulewidth}{1pt}
\newpage

\section*{List of Symbols}
\addcontentsline{toc}{section}{List of Symbols}
\fancyhead{}
\fancyhead[R]{List of Symbols} 
\fancyhead[L]{\scshape \thepage}
\begin{center}
\begin{longtable}{m{2cm} m{9.5cm}m{2cm}}
\rule{14cm}{0.5mm}\\
\textbf{~Symbol} & \textbf{Description} & \textbf{Unit} \\
\rule{14cm}{0.5mm}\\
\vspace{1cm}\\
~A & Active device area  & m$^2$\\
~A & Intrinsic voltage gain of a transistor  & $-$\\
\\
~b & Distance between micro-electrodes (ion sensor) & m \\
\\
~c$_\mathrm{d}$ & Specific capacitance per area of the double-layer capacitance & F/m$^2$\\
~c & Electrolyte concentration & mol/l\\
~$\tilde{\mathrm{C}}$ & Specific insulator capacitance (OTFT) & F/m$^\mathrm{2}$\\
~C$_\mathrm{{Ch}}$ & Channel capacitance (OTFT)& F\\
~C$_\mathrm{{f}}$ & Diffusion capacitance (Warburg element) & F\\
~C$_\mathrm{{GS}}$ & Gate-Source capacitance (OTFT)& F\\
~C$_\mathrm{{GD}}$ & Gate-Drain capacitance (OTFT)& F\\
~C$_\mathrm{{LCD}}$ & Capacitance of LCD-element in active-matrix pixel & F\\
~C$_\mathrm{{S}}$ & Storage capacitance of active-matrix pixel & F\\
~C$_\mathrm{{tot}}$ & Total device capacitance (OTFT) & F\\
\\
~d$_\mathrm{eff}$ & Lateral spreading of channel underneath the drain electrode (VOFET) & m\\
~ D & Cation diffusion constant & cm$^2$/s\\
\\
~E$_\mathrm{F}$ & Electro-Chemical Potential & eV\\
~E$_{\mathrm{Act}}$ & Activation energy of electrical conductivity & eV\\
~E$_\mathrm{F,0}$ & Standard electrode potential & eV\\
\\
~ f$_\mathrm{T}$ & Unity-gain transition frequency & $Hz$\\
~ f & Frequency & Hz\\
\\
~g$_\mathrm{m}$ & Transconductance & $\mathrm{\Omega^{-1}}$\\
\\
~h$_{21}$ & Differential current gain of a transistor & $-$\\
\\
~I$_\mathrm{C}$ & Collector current (OPBT) & A\\
~I$_\mathrm{D}$ & Drain current & A\\
~i$_\mathrm{D}$ & Small signal drain current & A\\
~I$^{\mathrm{lin}}_\mathrm{D}$ & Drain current in the linear regime & A\\
~I$^\mathrm{sat}_\mathrm{D}$ & Drain current in the saturation regime & A\\
~I$_\mathrm{G}$ & Gate current & A\\
~i$_\mathrm{G}$ & Small signal gate current & A\\
~I$_{\mathrm{off}}$ & Off-state current (OPBT) & A\\
~I$_{\mathrm{ref}}$ & Reference current for steady-state operation at T$_a$ (OPBT) & A\\
~I$_{SS}$ & Steady-state drain current (OECT) & A\\
\\
\\
\\
~L & Channel length OTFT & m\\
~L$_{\mathrm{OV}}$ & Gate-Drain or Gate-Source overlap for OTFTs & m\\
~L$_{\mathrm{OV}}$ & Overlap of source-insulator over source electrode for VOFETs & m\\
~LW & Line width of metal electrodes & m\\
~L$_{\mathrm{T}}$ & Transfer length for charge injection(OTFT) & m\\
~L$\mathrm{Pin}$ & Diameter of a pinhole (OPBT) & m\\
~l & Distance between gate electrode and semiconductor-electrolyte interface (OECT) & m\\
\\
\\
~$\mathrm{\tilde{n}}$ & Surface-density of electrons & cm$^{-2}$\\
~N$\mathrm{Pin}$ & Density of pinholes (OPBT) & 1/m$^2$\\
\\
~OL & Overlay of electrodes & m\\
\\
~p$_0$ & Initial hole density in the semiconductor (OECT, Bernard's model) & m$^{-2}$\\
\\
\\
~R$_\mathrm{Ch}$ & Channel-width normalized channel resistance (OTFT)& $\mathrm{\Omega} \mathrm{m}^{-1}$\\
~R$_{\mathrm{c}}$ & Channel-width normalized contact resistance (OTFT)& $\mathrm{\Omega} \mathrm{m}^{-1}$\\
~R$_{\mathrm{CBS}}$ & Resistance of charge-blocking-layer (VOFET)& $\mathrm{\Omega}$\\
~R$_{\mathrm{W}}$ & Diffusion resistance (Warburg model, ion sensor) & $\mathrm{\Omega}$\\
~R$_{\mathrm{W}}^*$ & Equivlent diffusion resistance (Warburg model, ion sensor) & $\mathrm{\Omega}$\\
\\
~S & Subthreshold slope & V/dec\\
\\
~T & Temperature & K\\
~T$_\mathrm{a}$ & Ambient temperature (293.15K) & K\\
~T$_\mathrm{G}$ & Glass transition temperature & K\\
~t & Gate insulator thickness (OTFT) & m\\
~t$_\mathrm{OSC}$ & Thickness of the semiconductor film (OECT) (OTFT) & m\\
\\
~V$_{\mathrm{BE}}$ & Base-Emitter voltage (OPBT) & V\\
~V$_{\mathrm{CE}}$ & Collector-Emitter voltage (OPBT) & V\\
~V$_{\mathrm{data}}$ & Voltage on the data-line in a AM display & V\\
~V$_{\mathrm{DS}}$ or V$_{\mathrm{D}}$& Drain-Source voltage & V\\
~V$_{\mathrm{GS}}$ or V$_{\mathrm{G}}$ & Gate-Source voltage & V\\
~V$_{\mathrm{P}}$ & Pinch-off voltage (OECT, Bernard's model) & V\\
~V$_{\mathrm{ref}}$ & Reference voltage for steady-state operation at T$_a$ (OPBT) & V\\
~V$_{\mathrm{sel}}$ & Voltage on the select-line in a AM display & V\\
~V$_{\mathrm{th}}$ or V$_\mathrm{Th}$ & Threshold voltage & V\\
~V$_\mathrm{TO}$ & Turn-off voltage (OECT) & V\\
\\
~W & Channel width OTFT & m\\
\\
~z & Valence of ions  & $-$\\
~$\tilde{Z}' $ & Real part of impedance & $\Omega$\\
~$\tilde{Z}'' $ & Imaginary part of impedance & $\Omega$\\
~$\tilde{Z} $ & Complex impedance & $\Omega$\\
\\
~$\alpha$ & Power law exponent describing the current-voltage curve of an OPBT & -\\
\\
\\
~$\mathrm{\epsilon}$ & Relative permittivity & $-$\\
\\
\\
\\
\\
~$\theta_\mathrm{th}$ & Thermal resistivity of the substrate & KW$^{-1}$\\ 
\\
\\
~$\mathrm{\lambda}$ & Amplification factor & $-$\\
\\
~$\mathrm{\mu}$ & Charge carrier mobility & cm$^2$/(Vs)\\
~$\mathrm{\mu_{FET}}$ & Charge carrier mobility as measurement in a field-effect transistor & cm$^2$/(Vs)\\
\\
\\
\\
~$\mathrm{\sigma}$ & Specific conductivity & $(\mathrm{\Omega}$ cm)$^{-1}$\\
~$\mathrm{\sigma}_0$ & Specific conductivity at T=0K& $(\mathrm{\Omega}$ cm)$^{-1}$\\
\\
~$\mathrm{\tau}$ & Delay-time of the delay-line & s\\
~$\mathrm{\tau_i}$ & Time-constant of ionic sub-circuit (OECT) & s\\
~$\mathrm{\tau_e}$ & Time-constant of ionic sub-circuit (OECT) & s\\
~$\mathrm{\tau_{NL}}$ & Inherent delay-time of the non-linear reservoir & s\\
\\
\\
~$\mathrm{omega}$ & Angular frequency & Hz\\
~$\mathrm{omega_D}$ & Characteristic frequency (Warburg element) & Hz\\
\rule{14cm}{0.5mm}\\
\end{longtable}
\end{center}

\fancyfoot{}
\newpage
\section*{Physical Constants}
\fancyhead{}
\fancyhead[L]{\scshape \thepage}
\fancyhead[R]{Physical Constants} 
\begin{center}
\begin{tabular}{m{5.5cm} m{3.5cm}m{4.5cm}}
\rule{14.5cm}{0.5mm}\\
\textbf{~Quantity} & \textbf{Symbol} & \textbf{Value} \\
\rule{14.5cm}{0.5mm}\\
\vspace{1cm}\\
~Electron-volt energy & $eV$ & $1eV=1.60218\times10^{-19}C$ \\
~Elementary charge & $e$ & $1.60218\times10^{-19}C$\\
~Boltzmann constant & $k_B$ & $1.38066\times10^{-23}J/K$\\
~Vacuum permittivity & $\epsilon_0$ & $8.85418\times10^{-14}F/cm$\\
~Planck constant & $h$ & $6.62606957\times10^{-34}Js$\\
\rule{14.5cm}{0.5mm}\\
\end{tabular}
\end{center}


\newpage
\section*{Acknowledgements}
\fancyfoot{}
\fancyhead{}
\fancyhead[L]{\scshape \thepage}
\fancyhead[R]{Acknowledgements} 
\vspace{2cm}
\setlength{\epigraphwidth}{.5\textwidth} 
\setlength{\epigraphrule}{1pt} 
\epigraph{\large{\textit{'Die vollends aufgeklärte Welt erstrahlt im Zeichen triumphalen Unheils.'}}}{\footnotesize{Theodor W. Adorno, philosopher.}}
\vspace{2cm}
\vspace{1cm}
\thispagestyle{empty}
At this point, I would like to express my deepest gratitude to everyone who joined me during the last couple of years and helped me during the process of the Habilitation. In particular, I thank Prof. Karl Leo for his amazing support over more than the last 10 years. His enthusiasm and attitude to push things forward are inspiring not only to me but also to all other members of the IAPP. He created an institute with a perfect experimental and scientific infrastructure and IAPP is simply a great place to be. Furthermore, I am very grateful to Prof. Dr. Bj\"orn Lüssem, Prof. Martin Knupfer, and Prof. Manfred Helm for agreeing to review this thesis.\\
I am also very thankful to all my colleges at IAPP, UC Berkeley, IHM, and University Paris Saclay who supported me during the last 5 years and contributed to a large extent to this thesis. In particular, I would like to thank all members of the ODS group at IAPP. In particular, I am very grateful to Felix, Judy, Markus, Axel, Henning, Zhongbin, and Baltasar for their work on pushing the OPBTs to their limits. Furthermore, I wish to say thank you to Cindy, Matteo, Giuseppe, Anton, Lautaro, Anton, and Veronika for their enthusiasm to develop the FDP method and explore the field of neuromorphic. Special thanks to Matteo and Veronika for being the first members of the OECT-team and for being open to explore new ground - although it was sometimes not clear where the journey will go. My deepest gratitude also to Marco, Micha, David, J\"orn, Ilia, Shu-Jen, Kevin, and Jaebok who spend countless hours in the labs to optimize the growth of rubrene and push the performance of the VOFETs. Micha, I am very grateful to you for your organizational talent and your engagement with lab etiquette and safety. Furthermore, I also would like to thank Katrin, Martin, and Frank for their excellent work on photoelectron spectroscopy and modeling in order to describe the principles of doping and charge carrier transport in organic semiconductor materials. Thanks also to Shen and Jonas for their work on the organic photodetectors. Furthermore, I am grateful to Micha and Felix for proof-reading.\\ 
There are even many colleges to mention with whom I enjoyed to work and who kept things running: Fanny, Johanna, Angelika, Annette, Peter, Kai, Thomas, Sven, Tobias, Andreas, Tina, Caroline, Sascha, Marcus, Stefan, Johannes, Freddi, Sebastian, André, Martin K., Bahman, Peter, and Peter.\\
Furthermore, I would like to thank Laurie and Bogdan from Paris Saclay who introduced me to the concept of reservoir computing. I am glad to have so knowledgeable partners on this topic.\\
My deepest gratitude also to all the members of the Wang group in Berkeley. It was an amazing time full of interest in science. The Wang group is full of brilliant researchers and it was an honor to be there.\\
Last but not least, I would like to thank my family and friend for their love, support, and patience. In particular, I thank my beloved family, Fritz, Fridolin, Janine, Romeo, Liselotte, Anke, and Tom, for just being amazing the way they are. 

\end{document}